\documentclass[12pt]{article} 

\usepackage{natbib}
\usepackage{placeins}
\usepackage{amsmath,amsfonts,esint}
\usepackage{amstext,rotating}
\usepackage{amssymb,graphics,xspace,endnotes}
\usepackage{float}
\usepackage{forest}
\usepackage{lineno}
\usepackage{epsfig}
\usepackage{srcltx}
\usepackage{subfig}
\usepackage{graphicx}
\usepackage{xcolor}	
\usepackage{verbatim}
\usepackage{tikz}
\usetikzlibrary{patterns}
\usepackage{mathabx}
\usepackage{threeparttable}
\usepackage{booktabs}
\usepackage{caption}	
\usepackage[utf8]{inputenc}
\usetikzlibrary{arrows.meta,decorations.pathreplacing}

\usepackage{multirow}
\usepackage{soul}
\usepackage[margin=1in]{geometry}
\usepackage{setspace}

\usepackage[
    colorlinks=true,
    citecolor=blue,
    linkcolor=black,
    urlcolor=black
]{hyperref}

\definecolor{intervalblue}{RGB}{30,91,156}
\definecolor{intervalgreen}{RGB}{31,125,84}
\definecolor{intervalred}{RGB}{183,50,48}
\newenvironment{figurenotes}
{\vspace{0.2cm}\footnotesize\noindent\textit{Notes: }}
{\par\vspace{0.2cm}} 
\newtheorem{prop}{Proposition}
\newtheorem{assumption}{Assumption}
\newtheorem{theorem}{Theorem}

\newtheorem{notation}{Notation}	
\newtheorem{definition}{Definition}

\newcommand{\x}{\mathbf{x}}

\setcitestyle{sort&compress}

\usepackage{etoolbox}
\AtBeginEnvironment{thebibliography}{\singlespacing}
\begin{document} 

\title{The Anatomy of Dependence in Multivariate Ordered Choice\thanks{This paper was previously circulated as ``Multivariate ordered discrete response models with two layers of dependence.'' We are grateful to  participants at Yale, Harvard-MIT, Erasmus University Rotterdam, LSE-STICERD seminars, and $EC^2$ 2023 conference  for helpful comments.}} 
\author{Tatiana Komarova\thanks{Faculty of Economics, University of Cambridge.  tk670@cam.ac.uk.}  \and
William Matcham\thanks{Department of Economics, Royal Holloway University of London.  William.Matcham@rhul.ac.uk.}} 

\date{\today \\ \vspace{.5cm}}

\maketitle
\thispagestyle{empty}


\begin{abstract}
When individuals make simultaneous decisions across multiple ordered dimensions, standard multivariate ordered choice models impose narrow bracketing, with each dimension decided as if the others did not exist. We develop a general rectangular structure model capturing broad bracketing while nesting the standard model. The framework introduces two layers of dependence, one through the threshold structure associated with decision rules and the other through the correlation of latent utilities. We provide microfoundations, prove identification, and discuss estimation. We showcase the model in applications to parental investment in children’s education, cryptocurrency familiarity and expectations, and health insurance, where the latter separates moral hazard from adverse selection.

\vspace{0.05in}

\begin{description}
\item[Keywords:] Ordered response, multiple dimensions, broad bracketing, narrow bracketing, coherency, insurance, moral hazard, adverse selection 

\item[JEL Classification:] C14, C31, C35, D9
\end{description}
\end{abstract}
\spacing{1.3}

\newpage
\setcounter{page}{1}


\section{Introduction} 
\label{intro}
\noindent 
How do people make simultaneous decisions across multiple ordered dimensions? How does a worker choose a health insurance plan and adjust healthcare
utilization? How do parents allocate time and money to their children's
education?  In these cases, each of the two decisions is shaped by the other, and the economic content of that interaction, such as complementarity and substitutability or moral hazard and adverse selection, is precisely what researchers wish to recover.

The standard approach extends univariate ordered response models dimension by dimension. Each dimension has its own latent utility and its own thresholds, and these thresholds are assumed to be functionally independent across dimensions. This is what we call the \textit{lattice model} as its threshold intersections form a lattice in the latent space. It corresponds to an agent who \textit{narrowly brackets}, making each decision as if the others did not exist  thereby prioritizing simpler choice rules over general joint utility optimization,\footnote{Functionally independent thresholds are compatible with additively separable joint utility, as shown in Section \ref{sec:microfound}. They are not, however, implied by general joint utility specifications. The lattice model should therefore be understood not as ruling out joint utility maximization altogether, but as sharply restricting the cross-dimensional interactions that joint utility can generate in the induced decision rules.} with all  dependence between decisions entering only through the correlation of latent utilities.\footnote{In an auction context, the assumption of functionally independent decision thresholds is akin to suggesting that a firm bidding in two simultaneous auctions for complementary objects would employ functionally independent equilibrium strategies in each, a notion contradicted by auction theory literature. See discussion in \cite{gentrykomarovasch2023} and \cite{gentrykomarovasch2019monotone} for more detail.} This oversimplified decision making process is illustrated in the left panel of Figure \ref{fig:ex1}. 
This paper introduces a richer class of models we call \textit{general
rectangular structures}. The key innovation is allowing each dimension's
decision threshold to depend on the outcome indices of all other
dimensions. The right panel of Figure \ref{fig:ex1} illustrates
the models we propose. The latent space remains partitioned into rectangular regions, and a clean mapping from latent utilities to observed discrete outcomes is preserved, but the boundaries of those rectangles can shift as a function of where the agent stands in other dimensions. This captures behavior akin to \textit{broad bracketing} since the agent evaluates their options jointly, and their decision rule in one dimension adjusts to reflect the constraints and opportunities in others. The lattice model is the special case with no  such adjustment. The flexibility of our new, more comprehensive, model enables us to capture more complex economic behavior than lattice models.

\begin{figure}[!t]
\centering
\begin{tikzpicture}[scale=0.2]
\draw [very thick] (-3,-9) -- (-3,9);
\draw [very thick] (3,-9) -- (3,9);

\draw [very thick] (-9,-3) -- (9,-3);
\draw [very thick] (-9,3) -- (9,3);

\node at (-3,-3) [circle,fill,inner sep=1.5pt]{};
\node at (3,3) [circle,fill,inner sep=1.5pt]{};

\node at (-3,3) [circle,fill,inner sep=1.5pt]{};
\node at (3,-3) [circle,fill,inner sep=1.5pt]{};

\draw[very thick, dotted, -> ] (-8.5,-11) -- (8.5,-11) node[below] {Latent Utility 1};
\draw[very thick, dotted, -> ] (-11,-8.5) -- (-11,8.5) node[above] {Latent Utility 2};

\end{tikzpicture}
\hskip -0.2in \begin{tikzpicture}[scale=0.2]

\draw [very thick] (-10,-7) -- (-5,-7);
\node at (-5,-7) [circle,fill,inner sep=1.5pt]{};
\node at (-5,-5) [circle,fill,inner sep=1.5pt]{};
\node at (-5,1) [circle,fill,inner sep=1.5pt]{};
\node at (-3,1) [circle,fill,inner sep=1.5pt]{};
\node at (2,1) [circle,fill,inner sep=1.5pt]{};
\node at (2,-5) [circle,fill,inner sep=1.5pt]{};
\node at (0,-5) [circle,fill,inner sep=1.5pt]{};
\node at (4,1) [circle,fill,inner sep=1.5pt]{};
\draw [very thick] (-5,-5) -- (10,-5);
\draw [very thick] (-10,1) -- (10,1);

\draw [very thick] (-5,-10) -- (-5,1);
\draw [very thick] (-3,1) -- (-3,10);
\draw [very thick] (0,-10) -- (0,-5);
\draw [very thick] (2,-5) -- (2,1);
\draw [very thick] (4,1) -- (4,10);

\draw[very thick, dotted, -> ] (-8.5,-11) -- (8.5,-11) node[below] {Latent Utility 1};
\draw[very thick, dotted, -> ] (-11,-8.5) -- (-11,8.5) node[above] {Latent Utility 2};
\end{tikzpicture}
\caption{Models with a lattice (left) and a general rectangular structure (right)}
\label{fig:ex1}
\end{figure}

Allowing thresholds to shift across dimensions introduces a logical consistency requirement that is absent from lattice models. We formalize this problem as \textit{coherency} and define it as the requirement that the rectangular regions partition the latent space. We  characterize it through a set of local  conditions on adjacent thresholds that capture choices in an immediate neighborhood. Our notion of coherency is the econometric counterpart of internal consistency of a decision maker, and it constrains the threshold structure in a way that can be directly enforced in the estimation. 

The separation of two layers of dependence is the central economic payoff
of the framework. In any multivariate ordered choice setting, observed
covariation between outcomes reflects two conceptually distinct forces. One is the \textit{threshold layer}, through which one dimension's decision
rule responds to the other, and the other one is the \textit{unobservables layer},
through which latent shocks are correlated conditional on observables. In
a lattice model, both forces are absorbed into a single dependence layer in correlation, which becomes a mixture of the two and is therefore interpretable as neither. Our framework assigns each force to a separate object, enabling the researcher to ask whether dependence between outcomes reflects a structural behavioral link or correlation of unobservables. For instance, the threshold decision rule may exhibit substitutability, while the dependence structure of the unobservables in the latent utilities may reflect complementarities. Furthermore, within the threshold decision structure itself, patterns of complementarity or substitutability can vary across different regions of the latent utility space. Another advantage of general rectangular structure models relative to lattice models is that they allow partial effects on the conditional probability of exceeding a given level to change sign across the domain, capturing more nuanced and flexible behavioral patterns. They also allow indirect effects of covariates from other processes on the conditional choice probability for a given process. For instance, a price subsidy meant to encourage health insurance enrollment can indirectly influence the probability of exceeding a given level in the pension plan dimension.

Some might view the use of sequential timing as a natural response to the limitations of lattice models. This approach involves deciding one dimension first, with the other's thresholds then free to depend on that first-stage outcome. However, restricting thresholds to a strict sequential order blocks certain feedback channels by construction. For example, if the insurance decision rule is modeled as functionally independent of utilization outcomes, individuals cannot select insurance based on their anticipated behavioral response. Since the ordering of decisions is challenging to observe, such assumptions are far from innocuous. Our general rectangular structure framework nests sequential timing as a special case, thereby eliminating the need to maintain a particular timing order as an assumption.

From a complementary conceptual perspective, the paper provides some  microfoundations for general rectangular structures. We  first introduce synergies and crowding-out effects directly into joint utility, and  illustrate this using bivariate models. In a second interpretation, we relate the general threshold system to a collection of local decision rules. Coherency ensures that these rules jointly select exactly one outcome for every latent realization.

Econometrically, we develop a semiparametric setting for our general rectangular structure models and formalize  how the models generate richer behavior than lattice specifications, especially through flexible partial effects and cross-process covariate effects mentioned above. We establish identification sequentially, beginning with parameters associated with exclusive covariates and eventually turning to the more involved identification of thresholds. The argument relies on independence between unobservables and covariates, as well as eventually on the presence of at least one relevant exclusive covariate in each process. Online Appendix C discusses semiparametric estimation, building on \cite{coppejans2007} and imposing coherency through equality restrictions on thresholds. 

We also study a parametric version in which unobservables are multivariate normal. A bivariate numerical exercise illustrates identification under weaker conditions than in the semiparametric case. We then discuss maximum likelihood estimation under the same coherency constraints and present Monte Carlo evidence comparing the proposed general rectangular structure (also called ``non-lattice'') probit estimator with standard bivariate ordered probit estimators, the latter of which effectively estimates a misspecified lattice model.

We demonstrate the practical importance of our general rectangular structures and the separation of two dependence layers in three applications. In U.S. health insurance markets, we draw on the Medical Expenditure Panel Survey and revisit the classic question of separating moral hazard from adverse selection. A dominant method since \citet{chiappori2000testing} is to estimate a bivariate probit model and interpret the sign of the latent correlation. As \citet{cohen2010} emphasize, a positive correlation conflates adverse selection with moral hazard. Separating these two forces has required either quasi-experimental variation or strong structural assumptions. Our general rectangular structure resolves the identification problem and disentangles moral hazard from selection (adverse or advantageous) differently by allowing functionally dependent thresholds that capture moral hazard in a coherent and data driven way, leaving selection to be fully captured by correlation of unobservables. We find that the asymmetric information is mostly confined to moral hazard, and  adverse selection is minimal, although the binary categorization of coverage limits what we can conclude about selection on moral hazard. These  findings are similar to recent papers in the literature that rely on context specific identification strategies or fully fledged structural economic models. Finally, we also estimate non-zero partial effects of the partner's insurance status on own utilization, which the lattice model rules out.

Our second application examines the relationship between cryptocurrency familiarity and optimism. The non-lattice approach reveals differences in how individuals form and express opinions about bitcoin's value, obscured by the lattice model. We also consider a sequential model in which cryptocurrency familiarity is determined first (thresholds functionally independent of optimism), then with  optimism thresholds allowed to depend on familiarity, and compare its findings to the general rectangular structure model. We find that, among other things,  the framework matters for index coefficient estimates whose signs may reverse due to  lattice coefficients conflating the structural effect with threshold misattribution.

The third application revisits parental investment in children's human capital using the PSID Child Development Supplement. The literature, following \citet{DelBocaFlinnWiswall2014}, typically assumes complementarity between financial and time investments in child quality production. Our model allows the data to reveal the sign of this relationship at each layer separately. At the threshold layer, higher home cognitive stimulation lowers the financial spending
threshold, confirming production complementarity in parental decision making. At the unobservables layer, however, the correlation between the two is negative, reflecting a labor supply trade-off. Thus, households in which parents work more accumulate money but lose time. The lattice model captures only the net effect and returns a moderate positive correlation, obscuring both mechanisms. 

The paper proceeds as follows. After reviewing related work and situating our contribution in the literature, Section~\ref{modelcontent} introduces general rectangular structures and coherency. Section~\ref{sec:microfound} provides microfoundations. Sections~\ref{sec:semiall} and~\ref{sec:parametric} develop identification and estimation for the semiparametric and parametric cases, respectively. Section~\ref{sec:MonteCarlo} presents Monte Carlo evidence, Section~\ref{sec:appl} presents three empirical applications, and Section~\ref{sec:conclusion} concludes.

\paragraph*{Our contributions and literature review} Our paper contributes to the literature on the economic content and econometrics of ordered choice models. The literature on \textit{univariate} ordered response models is extensive. \cite{cameron1998}, \cite{heckman1999}, and \cite{carneiro2003} develop foundational frameworks that ground thresholds in economic decision-making. \cite{cunha2007} generalizes this by allowing thresholds to depend on both observables and unobservables, capturing dynamic settings such as schooling decisions. On the semiparametric side, \cite{lewbel2000} establishes identification of binary, ordered, and multinomial choice models via a continuous special regressor that is conditionally independent of the error. \cite{lewbel2003} extends this approach to ordered models with random or sequentially determined thresholds.\footnote{\cite{lewbel2014} provides a unified treatment of the special regressor method.} Reduced-form treatments of thresholds as psychological cut-points or cost-benefit barriers without optimization foundations are surveyed in \cite{greene2010} and \cite{boes2006}.  \cite{Anderson1984} pursues a related approach in which thresholds relate to category proportions rather than absolute utility levels.\footnote{In the \cite{Anderson1984} model, ordinal categories are not tied to a single latent variable with fixed thresholds.}  \cite{bhat1998} derives thresholds from a range-based utility specification, and \cite{ApesteguiaBallester2023} proposes type-ordered random utility models in which ordered choices arise from heterogeneous preference types.

Our paper moves this body of work from univariate to \textit{multivariate} ordered decisions. The key departure is that thresholds in each dimension depend on the realized discrete outcomes of the other dimensions. This requires a joint model of all endogenous processes and introduces a flexibility in the threshold structure unavailable in any existing framework. Our contribution to the microfoundations of ordered choice (see Section~\ref{sec:microfound}) speaks to the  strand of univariate models focused on utility optimization, as we show that general rectangular structures arise naturally from joint utility specifications with complementarities/substitutabilities. We also explain it as a result of a collection of local decision rules. 

\cite{leesalanie2018} study treatment assignment
through several threshold-crossing rules, a framework
related to ours. In our model, however, several
thresholds apply to the same latent variable, and a
change in an equation's index shifts that variable
relative to all its thresholds at once. As a result, the induced vector of threshold crossings cannot in general be varied freely in all of its coordinates. The identification strategy of \citet{leesalanie2018}, by contrast, relies on an open range condition that provides sufficiently rich multidimensional variation in the vector of thresholds. That condition does not hold in our setting, and our identification results thus require a different argument.

In demand estimation, \cite{gentzkow2007} models two goods that interact in joint utility but also with correlated standalone utilities. It shows that goods that appear complementary in the raw data can be substitutes once the unobserved heterogeneity is separated out. We generalize the separation between threshold and unobservable layers in \cite{gentzkow2007} to ordered, multi-category and a  higher dimensional settings.

The economic content of threshold interdependence connects our work to the literature on \textit{choice bracketing}. The question of whether agents evaluate options jointly or dimension by dimension has several interpretations. For example, it is \textit{sequential vs.\ simultaneous choice} in \cite{simonson1992}, \textit{narrow vs. broad decision frames} in \cite{kahneman1993}, \textit{local vs. overall value functions} in \cite{heyman1996}, and \textit{isolated vs. distributed choice} in \cite{herrnstein1991}. This literature is largely theoretical and experimental \citep{tversky1981,read1999,thaler1999,rabin2009,lian2020,camara2021,zhang2021}, with only a few descriptive or structural empirical studies \citep{camerer1997,thakral2021}.\footnote{\citet{tversky1981} provides a classic example of narrow bracketing in experimental settings.}  What has been missing is a structural econometric model that simultaneously nests both behaviors and allows the data to reveal which is operative. Our framework provides this. The lattice model is the precise econometric representation of narrow bracketing, whereas the general rectangular structure is its broad bracketing generalization. The hypothesis of narrow bracketing becomes a potentially testable restriction.

A tangentially related literature is discrete choice  with \textit{strategic interactions}, in which one player's outcomes depend on the others' actions \citep{tamer2003,berry2007,ciliberto2009,honor2010,chesher2017,chesher2020,aradillas2022}.  A central challenge in that literature is that best-response correspondences can yield incoherent or incomplete outcome mappings. As we detail in the following section, by \textit{coherency} we mean logical consistency in decision making ensuring that rectangular regions representing different discrete responses do not overlap and together cover the entire latent space $\mathbb{R}^D$.\footnote{For more on coherency, see \citet{heckman1978}, \cite{lewbel2007} and \citet{tamer2003}. \citet{tamer2003} distinguishes between incoherency and incompleteness in games with strategic interactions, a distinction followed by later studies. In our setting, we use the term ``coherency'' to refer more generally to overall logical consistency.}
Our setting is conceptually distinct from models with  strategic interactions, as we have a \textit{single} agent choosing across multiple dimensions. Consequently, this decision problem is free from the strategic interaction sources of incoherency studied in that literature and is internally consistent by construction. But since general rectangular structures do not guarantee coherency, explicit threshold constraints for coherency will be required and  enforced in estimation, and characterizing these constraints is one of our theoretical contributions.

\section{Model with a general rectangular structure}
\label{modelcontent} 

We define general rectangular structures and lattice  multivariate ordered response models for an agent making decisions across $D \geq 2$ dimensions. These models map a $D$-variate latent continuous metric $(Y^{*c_1}, \ldots, Y^{*c_D})$ to a discrete metric $(Y^{c_1}, \ldots, Y^{c_D})$, with ordered responses in dimension $d$ denoted as $y^{(d)}_j$, $j=1, \ldots, M_d$, and satisfying $y^{(d)}_1 < \cdots < y^{(d)}_{M_d}$.

\begin{definition}[General rectangular structure model] 
\label{def:noinlattice} A  model has a general rectangular structure (we will also refer to it as a \textit{non-lattice} model) if
\[
(Y^{c_1}, \ldots, Y^{c_D}) = (y^{(1)}_{j_1}, \ldots, y^{(D)}_{j_D}) \quad \iff \quad (Y^{*c_1}, \ldots, Y^{*c_D}) \in R_{j_1, \ldots, j_D}, \text{ where}
\]
\begin{equation} 
\label{rectangleR}
R_{j_1, \ldots, j_D} = \displaystyle\bigtimes_{d=1}^D
\left({\alpha}^{(d)}_{j_1, \ldots, j_{d-1}, {\color{orange} j_{d} \: - \: 1}, j_{d+1}, \ldots, j_D}, {\alpha}^{(d)}_{j_1, \ldots, j_{d-1}, {\color{orange} j_d}, j_{d+1}, \ldots, j_D}\right],
\end{equation}
with thresholds ${\alpha}^{(d)}_{j_1, \ldots, j_{d-1}, {\color{orange} j_d}, j_{d+1}, \ldots, j_D}$ strictly increasing in $j_d$ for fixed other indices and normalized at the boundary as 
\[
\alpha^{(d)}_{j_1, \ldots, j_d, \ldots, j_D} = +\infty \text{ for } j_d = M_d, \quad \alpha^{(d)}_{j_1, \ldots, j_d, \ldots, j_D} = -\infty \text{ for } j_d = 0.
\]
\end{definition}
The threshold intersections in Definition \ref{def:noinlattice} do not \textit{necessarily} form a lattice in $\mathbb{R}^D$. Instead, they reflect functionally interdependent decision rules, akin to \textit{broad bracketing} in behavioral economics.\footnote{Definition \ref{def:noinlattice} outlines the intended threshold decision rules. However, for this mapping to be valid for every latent draw, the defined regions must be exhaustive and mutually exclusive. Definition \ref{def:noinlattice} alone does not guarantee this division and we discuss this below under ``Coherency''.} 

\begin{definition}[Lattice model] 
\label{def:lattice}
A \textit{lattice} model is a special case of a general rectangular structure model in which each threshold
$\alpha^{(d)}_{j_1, \ldots, j_d, \ldots, j_D} $ depends only on  index $j_d$: 
\[
(Y^{c_1}, \ldots, Y^{c_D}) = (y^{(1)}_{j_1}, \ldots, y^{(D)}_{j_D}) \quad \iff \quad Y^{*c_d} \in ( \alpha^{(d)}_{j_d-1}, \alpha^{(d)}_{j_d} ] \quad \forall d, 
\]
with $\alpha^{(d)}_{j_d} = +\infty$ for  $j_d = M_d$, and $\alpha^{(d)}_{j_d} = -\infty$ for  $j_d = 0$.
\end{definition}

In Definition \ref{def:lattice}, thresholds are functionally independent across dimensions, forming a lattice in $\mathbb{R}^D$ at their intersections. Lattice models describe decisions made dimension by dimension and will misspecify a decision maker with more complex rectangular decision rules.  

Since the general rectangular structure model allows for the lattice form, the synonym ``non-lattice'' should not be viewed as a ruling out of the possibility of a lattice structure but instead a model that does not need to have a lattice structure.

Thus, in our setting, the distinction between broad and narrow bracketing is fully captured by the functional interdependence or independence of decision rules, which is determined by the thresholds. Both lattice and non-lattice models permit correlated decisions (conditional on observables) through latent processes, but the latter distinguishes correlation in unobservables from interdependent decision rules.

\paragraph*{Coherency} The flexibility of general rectangular structure models comes at a cost, as Definition \ref{def:noinlattice} does not guarantee that the division of latent space into regions $R_{j_1, \ldots, j_D}$ is  exhaustive or mutually exclusive. As a result,  the condition that each latent profile maps to exactly one observed response (which is associated with logical consistency in decision making) and to which we refer as \textit{coherency} is not guaranteed by Definition \ref{def:noinlattice}. Since the model aims to describe the behavior of a logically consistent decision maker, it should always satisfy coherency, especially when we take it to the data. In general rectangular structure models, ensuring coherency requires constraints on the thresholds. Lattice models, by contrast, are coherent by construction. 

We examine coherency in a bivariate general rectangular structure ordered response model and provide a formal condition on thresholds for the model to be coherent. The characterization of coherency for $D>2$ is more involved and is presented in the Online Appendix. 

\begin{prop}[Coherency for $D=2$] 
\label{prop:coherencyD2}
Consider a bivariate general rectangular structure  model, defined by sets  of thresholds  $\left\{ \alpha_{j_1, j_2}^{(1)} \right\}_{j_1=1, j_2=1}^{M_1-1, M_2}$ and $ \left\{ \alpha_{j_1, j_2}^{(2)} \right\}_{j_1=1, j_2=1}^{M_1, M_2-1}
$. Given threshold normalizations at the boundary, the model is \textit{coherent} -- i.e., the latent space is partitioned into mutually exclusive and exhaustive rectangular regions
$R_{j_1, j_2} = (\alpha^{(1)}_{j_1-1, j_2}, \alpha^{(1)}_{j_1, j_2}] \times (\alpha^{(2)}_{j_1, j_2-1}, \alpha^{(2)}_{j_1, j_2}]$  each corresponding to a unique observed outcome -- iff, for all $(j_1, j_2)$, $j_1=1,\ldots, M_1-1$, $j_2=1, \ldots, M_2-1$, 
\begin{equation}
\label{eq:coherencybiv}
( \alpha^{(2)}_{j_1+1, j_2} - \alpha^{(2)}_{j_1, j_2} ) \cdot ( \alpha^{(1)}_{j_1, j_2+1} - \alpha^{(1)}_{j_1, j_2} ) = 0.
\end{equation}
\end{prop}

\begin{figure}[!t]
\centering
\begin{tikzpicture}[scale=0.2]

\fill [opacity = 0.75, pattern=crosshatch, pattern color=orange] (2,-1) rectangle (-5,3);
\fill [opacity = 0.75, pattern=crosshatch, pattern color=orange] (2,-1) rectangle (6,5);
\fill [opacity = 0.75, pattern=crosshatch, pattern color=orange] (0,-1) rectangle (-7,-5);
\fill [opacity = 0.75, pattern=crosshatch, pattern color=orange] (0,-1) rectangle (6,-5);

\draw [very thick] (-10,-7) -- (-7,-7);
\draw [very thick] (-7,-5) -- (6,-5);
\draw [very thick] (6,-4) -- (10,-4);

\draw [very thick] (-10,-1) -- (6,-1);
\draw [very thick] (6,2) -- (10,2);

\draw [very thick] (-10,3) -- (2,3);
\draw [very thick] (2,5) -- (10,5);

\draw [very thick] (-7,-10) -- (-7,-1);
\draw [very thick] (-5,-1) -- (-5,3);
\draw [very thick] (-3,3) -- (-3,10);

\draw [very thick] (-2,-10) -- (-2,-5);
\draw [very thick] (0,-5) -- (0,-1);
\draw [very thick] (2,-1) -- (2,10);

\draw [very thick] (6,-10) -- (6,5);

\draw [very thick] (8,5) -- (8,10);

\end{tikzpicture}
\caption{Intuition for a non-lattice model being coherent in the bivariate case}
\label{fig:ex1cont}
\end{figure}

In other words, for the model to be coherent, the thresholds must satisfy a local condition for each $2 \times 2$ block of adjacent cells. Specifically, within each block, at least one of the dimensions must have constant thresholds across that block. One  economic interpretation of that is when an agent faces a choice within a $2 \times 2$ block, they make their decision sequentially: first along one dimension, where the thresholds remain fixed, and then along the other. An illustration of a local problem and the coherency requirement is in Figure \ref{fig:ex1cont}. Thus, in every local decision problem one dimension is leading.   Leading dimensions may differ across different parts of the domain (e.g., when one considers health insurance levels vs retirement contribution level, it may be the case that for lower levels the insurance decision is the leading one whereas for higher levels of both the leading decision is the retirement contributions as at those levels long-run financial planning may be of more relevance).

\section{Utility rationalization and local decision rules} 
\label{sec:microfound}
One way to justify a general rectangular structure model from a microeconomic foundations perspective, is to  directly model  complementarities and substitutabilities in the joint utility across different pairings of options. We illustrate how it can be done in a simple bivariate model with two discrete options (1 and 2) in each dimension. The left panel in Figure \ref{fig:bivariate} shows substitutability in the decision structure reflected in a larger threshold in the second dimension when $Y^{c_1}=2$. It shows that choosing a higher level in dimension 1  makes it harder to choose a higher level in the other dimension, for example, due to resource constraints. The right panel in Figure \ref{fig:bivariate} shows complementarity  as choosing a higher level in dimension 1  facilitates a higher level in the other dimension through a lower threshold.

Consider the following utilities across four pairs of discrete choices: for  constant $v<0$, 
\begin{align} U(1,1) & =0, \; \; \;
U(2,1) = Y^{*c_1}, \; \; \;
U(1,2) =Y^{*c_2},  \; \notag \\ 
U(2,2)& =D(Y^{*c_1} + Y^{*c_2}-v)+ (1-D)(Y^{*c_1} + Y^{*c_2}), \label{U22}
\end{align}
 where $D=\mathbf{1}(Y^{*c_1} > 0 )\mathbf{1}(Y^{*c_2} > v)$. Then $\text{argmax}_{j_1,j_2} U(j_1,j_2)$ is (i) (1,1) when $Y^{*c_1}, Y^{*c_2}\leq 0$; (ii) $(1,2)$ when $Y^{*c_1} \leq 0, Y^{*c_2} > 0$; (iii) (2,1) when $Y^{*c_1} >  0, Y^{*c_2} \leq v$; and (iv) (2,2) when  $Y^{*c_1} > 0$, $Y^{*c_2} > v$. The lower threshold $v<0$ facilitates choosing $ Y^{c_2} = 2 $ when $ Y^{c_1} = 2 $, suggesting complementarity (as in the right panel of Figure \ref{fig:bivariate}). The utility boost $-v > 0 $ in $ U(2,2) $ when $ D = 1 $ acts like a synergy term, incentivizing (2,2) over (2,1) or (1,2) when propensities are sufficiently high. This reflects scenarios where choosing one high option reduces the marginal cost of the other, such as  economies of scale, shared fixed cost, learning spillovers, subsidies, or other mechanisms through which one high-level choice raises the marginal value of the other.

 To obtain substitutes in the decision structure (as in the left panel of Figure \ref{fig:bivariate}),  consider $w>0$ and define $D=\mathbf{1}(Y^{*c_1}>0)\mathbf{1}(Y^{*c_2} \leq w)$, $U(1,1)=0$, 
$U(1,2)=Y^{*c_2}$, 
$U(2,1)=Y^{*c_1}+wD$, 
$U(2,2)=Y^{*c_1}+Y^{*c_2}$. This utility specification delivers the desired rectangular decision rule, with a unique argmax away from threshold boundaries. When $Y^{*c_1}>0$ and $Y^{*c_2}<w$, $D=1$ and $U(2,1)-U(2,2)=w-Y^{*c_2}>0$, so $(2,1)$ is preferred to $(2,2)$. When $Y^{*c_2}>w$, $D=0$ and $U(2,2)$ strictly exceeds both $U(2,1)$ and $U(1,2)$ because $Y^{*c_2}>0$ and $Y^{*c_1}>0$. The remaining regions follow analogously. 
The  term  $wD$ is a specialization premium which  captures substitutability: when the  high-level choice in the first dimension is attractive, but the second propensity remains below $w$, the agent receives an additional payoff for choosing the high option in the first dimension alone rather than choosing both high options jointly. Economically, this reflects competition for a common resource such as time, effort, attention, or budget. Once $Y^{*c_2}>w$, the value of the second high-level choice dominates the specialization premium, and the joint alternative $(2,2)$ becomes optimal. Analogous constructs can be done for any number of ordered choices in each dimension.

\begin{figure}[!t]
\centering
\begin{tikzpicture}[scale=0.15]
\draw [very thick, dotted] (-12,0) -- (-10,0);
\draw [very thick, dotted] (10,3)  -- (12,3);
\draw [very thick, dotted] (0,6)   -- (0,8);
\draw [very thick, dotted] (0,-6)  -- (0,-8);

\draw [very thick] (0,-6) -- (0,6);     
\draw [very thick] (-10,0) -- (0,0);    
\draw [very thick] (0,3) -- (10,3);     

\node at (0,0) [circle,fill,inner sep=1.5pt]{};
\node at (0,3) [circle,fill,inner sep=1.5pt]{};

\node at (0,-10) {\small $0$};
\node at (-13,0) {\small $0$};
\node at (13,3)  {\small $w$};

\draw [font=\large] (-6,-3) node {$(1,1)$};
\draw [font=\large] (-6,4)  node {$(1,2)$};
\draw [font=\large] (6,-2)  node {$(2,1)$};
\draw [font=\large] (6,5)   node {$(2,2)$};

\draw[very thick,->] (-14,-12) -- (15,-12) node[right] {$Y^{*c_1}$};
\draw[very thick,->] (-14,-12) -- (-14,9)  node[above] {$Y^{*c_2}$};
\end{tikzpicture}
\hspace{0.6cm}
\begin{tikzpicture}[scale=0.15]
\draw [very thick, dotted] (-12,0)  -- (-10,0);
\draw [very thick, dotted] (10,-3)  -- (12,-3);
\draw [very thick, dotted] (0,6)    -- (0,8);
\draw [very thick, dotted] (0,-6)   -- (0,-8);

\draw [very thick] (0,-6) -- (0,6);     
\draw [very thick] (-10,0) -- (0,0);    
\draw [very thick] (0,-3) -- (10,-3);   

\node at (0,0)  [circle,fill,inner sep=1.5pt]{};
\node at (0,-3) [circle,fill,inner sep=1.5pt]{};

\node at (0,-10) {\small $0$};
\node at (-13,0) {\small $0$};
\node at (13,-3) {\small $v$};

\draw [font=\large] (-6,-3) node {$(1,1)$};
\draw [font=\large] (-6,4)  node {$(1,2)$};
\draw [font=\large] (6,-5)  node {$(2,1)$};
\draw [font=\large] (6,2)   node {$(2,2)$};

\draw[very thick,->] (-14,-12) -- (15,-12) node[right] {$Y^{*c_1}$};
\draw[very thick,->] (-14,-12) -- (-14,9)  node[above] {$Y^{*c_2}$};
\end{tikzpicture}
\caption{Illustration of first microfoundation in the bivariate case with two discrete options in each dimension}
\label{fig:bivariate}
\end{figure}

 Another way to motivate the general rectangular structure is through  representing local decision rules (which may be associated with discrete analogues of first-order conditions). In this interpretation the agent evaluates whether one-step movements in each dimension are locally desirable, holding the other dimension fixed. Consider, for example, the bivariate case. For a candidate outcome \((y^{(1)}_{j_1},y^{(2)}_{j_2})\), local optimality requires that moving upward to this outcome is desirable relative to the immediately lower category in each dimension, while moving one step further upward is not desirable. We formalize these local comparisons by 
 \[ Y^{*c_1}-\alpha^{(1)}_{j_1-1,j_2}>0, \; Y^{*c_1}-\alpha^{(1)}_{j_1,j_2}\leq 0, \;  \text{ and } \; Y^{*c_2}-\alpha^{(2)}_{j_1,j_2-1}>0, \; Y^{*c_2}-\alpha^{(2)}_{j_1,j_2}\leq 0.  \]
 Equivalently, \(\alpha^{(1)}_{j_1-1,j_2} < Y^{*c_1} \leq \alpha^{(1)}_{j_1,j_2},\) \(\alpha^{(2)}_{j_1,j_2-1} < Y^{*c_2} \leq \alpha^{(2)}_{j_1,j_2}. \) Thus, the candidate outcome \((y^{(1)}_{j_1},y^{(2)}_{j_2})\) is selected if and only if \( (Y^{*c_1},Y^{*c_2})\in R_{j_1,j_2}. \) When the threshold system is coherent, these local discrete first-order conditions assign each realization of \((Y^{*c_1},Y^{*c_2})\) to a unique observed outcome.
 
\section{Semiparametric specification, partial effects, identification} 
\label{sec:semiall}

To take our model to the data, we adopt the standard approach in the discrete response literature, specifying each $d$-th continuous latent process as a linear index:
\begin{equation}
Y^{*c_d} = x_d \beta_d + \varepsilon_d, \quad d=1, \ldots, D,
\label{index_structure}
\end{equation}
where $x_d$ is a row vector of observable covariates, $\beta_d$ is a $k_{d}$-dimensional column vector of unknown parameters, and $\varepsilon_d$ is an unobservable error term. This structure interprets $x_d \beta_d$ as the systematic component of an agent's latent propensity to choose an ordered category in dimension $d$, with $\varepsilon_d$ capturing random shocks to the latent utility. The shocks $\varepsilon_1, \ldots, \varepsilon_D$ may be dependent, allowing latent processes $Y^{*c_d}$ to be correlated conditional on covariates. This linear index, which is standard in discrete choice models, supports the estimation of threshold-based interdependence in general rectangular structure  models while maintaining parsimony. We maintain linear indices throughout the identification analysis. Nonlinear systematic components can be identified in other discrete choice frameworks under suitable restrictions,  see e.g. \cite{Matzkin1993}. Extending the present results to a specified nonlinear class would require additional identification arguments.

Without a loss of generality, $x_d\beta_d$, $d=1,\ldots, D$, does not contain an intercept term. 

Given the complexity of the model, particularly with regard to the two-layer dependence structure, we should be prepared for fairly stringent requirements on the data to ensure identification of the following objects of interest: $\beta_d$, $d=1,...,D$, the joint c.d.f of $\boldsymbol{\varepsilon}=(\varepsilon_1, \ldots, \varepsilon_D)'$, and the thresholds. We start by imposing Assumption \ref{assn:errors}, which relates to the independence of $\varepsilon$ from index covariates, and a useful notation. 
\begin{assumption}\label{assn:errors} $\boldsymbol{\varepsilon}=(\varepsilon_1, \ldots, \varepsilon_D)'$ is independent of $x=(x_1,\ldots, x_D)$ and has a convex support in $\mathbb{R}^D$.  Suppose the boundary of this support has probability zero. 
\end{assumption} 
\begin{notation} For any $d=1, \ldots, D$, let $\kappa_d $ denote either the $\leq$ or $>$ sign. Denote
$F_{\kappa_1, \kappa_2, \ldots, \kappa_D}(t_1,...,t_D)=P\left(\bigcap_{d=1}^D(\varepsilon_d \, \kappa_d \, t_d )\right) $ 
for any $\kappa_d \in \{\leq, >\}$, $d=1,..,D$. Functions $F_{\leq, ..., \leq}$ and  $F_{>, ..., >}$  are the joint c.d.f. and the joint survival function of $\boldsymbol{\varepsilon}$, respectively, and for simplicity we will interchangeably denote them as $F$ and $\overline{F}$, respectively. All the other cases correspond to hybrid  forms of c.d.f.s and survival functions. Similar notations apply to subsets of dimensions. 

Also denote 
$F_{d,\kappa_d}(t_d) = P(\varepsilon_d \, \kappa_d \, t_d)$ for $\kappa_d \in \{\leq, >\}$. Thus, $F_{d,\leq}$ is the marginal c.d.f.  and $F_{d,>}$ is the marginal survival function of $\varepsilon_d$.

\end{notation}

In our related paper \cite{km2025lattice}, we outline the increasing degree of restrictions required to ensure identification in semiparametric \textit{lattice} models. There, we discuss why identification of parameters and thresholds can rely on a weaker version of Assumption \ref{assn:errors} -- namely, the independence of $\varepsilon_d$ from $x_d$ for each $d=1, \ldots, D$. However, as we argue there, the identification of the joint distribution of $\boldsymbol{\varepsilon}$ \textit{does} largely rely on Assumption \ref{assn:errors} even in lattice models. Thus, when one of the objectives is to identify the joint distribution of errors in our general rectangular structure models (motivation for this is discussed later), our Assumption \ref{assn:errors} is similarly restrictive to what is required in  lattice models.

\subsection{Advantages of non-lattice models over lattice models} 
\label{sec:semiadvantages}

Before turning to identification in the described semiparametric contexts, we formalize the  advantages of general rectangular structure models over lattice models for capturing interdependencies among discrete choices.  Even in the simplest possible  case of a bivariate model with 2 discrete choices in each dimension  we can talk about at least four different empirical structures. These arise from the interactions between complementarity or substitutability in decision thresholds (the first layer) and complementarity or substitutability in unobservables conditional on observables (the second layer), the latter captured through positive or negative dependence that reflects whether shocks reinforce or offset one another.

In many applications, substitutability/complementarity relationships at each layer may be unknown \textit{a priori} and need to be identified from the available data. We now describe two advantages of general rectangular models relating to richer partial effects.

\begin{flushleft}
\textbf{Advantage 1. Cross-partial effects: Partial effects in one dimension can depend on covariates exclusive to other processes.}
\end{flushleft}
Under Assumption \ref{assn:errors}, models with general rectangular structures  allow $P(Y^{c_d} \geq y^{(d)}_j | x)$, which is the likelihood that an individual selects at least a certain level of  commitment in one decision area, given all relevant personal and contextual factors, to depend on covariates exclusive to other processes, enabling analysis of cross-partial effects $\frac{\partial P(Y^{c_d} \geq y^{(d)}_j | x)}{\partial x_{h,m}}$. This effect is \textit{indirect} as covariates exclusive to other processes affect the  probabilities of decision in a given dimension \textit{indirectly} through their influence on latent utilities in other dimensions. In lattice models under  Assumption \ref{assn:errors} such partial effects are absent (that is, zero).

For example, consider high school students choosing academic effort, measured by study hours, and extracurricular participation, measured by involvement in sports or clubs. These choices are interdependent at the level of the decision structure, but the direction of that interdependence is ambiguous a priori as high academic effort may limit time for extracurriculars, while extensive extracurricular involvement may encourage greater academic effort for college applications. Our model allows the data to determine the form and strength of this interdependence. Some covariates may be dimension-specific, such as parental education or tutoring access for academic effort, and school resources or peer involvement for extracurriculars. Others, such as socioeconomic status or school quality, may be shared by both processes. A general rectangular structure model allows covariates entering only the extracurricular process to affect the probability of high academic effort through the indirect channel, which is informative for resource allocation policies. Furthermore, a lattice model rules out such indirect effects and would therefore miss it. 

To illustrate this theoretically, take the model  in Figure \ref{fig:bivariate2} and note that 
\begin{multline*} P(Y^{c_1} \leq 1 | x) = P(Y^{c_1} \leq 1, Y^{c_2} =1 | x) +P(Y^{c_1} \leq 1, Y^{c_2} = 2 | x) \\ 
= F\left( \alpha^{(1)}_{1,1} -x_1\beta_1,   \alpha^{(2)} -x_2\beta_2\right)+F_{1, \leq}\left( \alpha^{(1)}_{1,2} -x_1 \beta_1\right) -
F\left( \alpha^{(1)}_{1,2} -x_1\beta_1,   \alpha^{(2)} -x_2\beta_2\right)
\end{multline*}
With the lattice structure ($\alpha^{(1)}_{1,1}= \alpha^{(1)}_{1,2}$ ), only the middle term in this representation is left. Therefore, with the lattice structure, $\frac{\partial P(Y^{c_1} \leq 1 | x)}{\partial x_{2,m}}=0$, where $x_{2,m}$ is a covariate exclusive to the process $Y^{*c_2}$. Thus, cross-partial effects are not possible in the lattice model. 

Under the general rectangular structure, assume that the distribution of $\varepsilon$ admits a density and let $e_{2} = \alpha^{(2)}-x_{2}\beta_{2}$. Then  
{\small$$\frac{\partial P(Y^{c_1} \leq 1 | x)}{\partial x_{2,m}}=-\beta_{2,m} \frac{\partial F\left( \alpha^{(1)}_{1,1} -x_1\beta_1,   \alpha^{(2)} -x_2\beta_2\right)}{\partial e_2}+\beta_{2,m} \frac{\partial F\left( \alpha^{(1)}_{1,2} -x_1\beta_1,   \alpha^{(2)} -x_2\beta_2\right)}{\partial e_2}$$}is not necessarily 0 because $\alpha^{(1)}_{1,1}\neq \alpha^{(1)}_{1,2}$. In contrast,  under Assumption \ref{assn:errors}, lattice models restrict $P(Y^{c_d} \geq y^{(d)}_j | x) = P(Y^{c_d} \geq y^{(d)}_j | x_d)$, ignoring cross-process effects.

\begin{flushleft}
\textbf{Advantage 2. Partial effects can vary in sign across the domain.}
\end{flushleft}

\begin{figure}[!t]
\centering
\begin{tikzpicture}[scale=0.18]
\draw [very thick, dotted ] (15.2,-4) -- (28.2,-4);
\draw [very thick, dotted ] (-8.2,-4) -- (-12,-4);

\draw [very thick, dotted ] (1,-10.2) -- (1,-12);

\draw [very thick, dotted ] (7,0) -- (7,3);

\draw [very thick ] (-10,-4) -- (28,-4);
\draw [very thick ] (1,-10) -- (1,-4);
\draw [very thick ] (7,-4) -- (7,1.5);
\draw [very thick ] (17,-4) -- (17,1.5);
\draw [very thick, dotted ] (17,0) -- (17,3);
\draw [very thick ] (20,-10) -- (20,-4);
\draw [very thick, dotted ] (20,-10.2) -- (20,-12);

\node at (1,-4) [circle,fill,inner sep=1.5pt, ]{};
\node at (7,-4) [circle,fill,inner sep=1.5pt, ]{};
\node at (17,-4) [circle,fill,inner sep=1.5pt, ]{};
\node at (20,-4) [circle,fill,inner sep=1.5pt, ]{};

\draw [, font = \large] (-7,-9) node (00) {$(1,1)$};
\draw [, font = \large] (7,-9) node (20) {$(2,1)$};
\draw [, font = \large] (25,-9) node (31) {$(3,1)$};
\draw [, font = \large, below] (-2,0.5) node (01) {$(1,2)$};
\draw [, font = \large, below] (12,0.5) node (21) {$(2,2)$};
\draw [, font = \large, below] (22,0.5) node (32) {$(3,2)$};

\draw [, ] (1, -12.5)  node[] {$\alpha^{(1)}_{1,1}$}; 
\draw [, ] (20, -12.5)  node[] {$\alpha^{(1)}_{2,1}$}; 
\draw [, ] (7, 5)  node[] {$\alpha^{(1)}_{1,2}$}; 
\draw [, ] (17, 5)  node[] {$\alpha^{(1)}_{2,2}$}; 

\draw[very thick,-> ] (-12,-14.75) -- (30,-14.75) node[right] {$Y^{*c_1}$};
\draw[very thick,-> ] (-14,-16) -- (-14,4) node[above] {$Y^{*c_2}$};

\end{tikzpicture}
\caption{Example of a  $3 \times 2$ bivariate model}
\label{fig:bivariate2}
\end{figure}

Let us first illustrate this property for cross-partial effects. Using the model in Figure \ref{fig:bivariate2} and the formula for $\frac{\partial P(Y^{c_1} \leq 1 | x)}{\partial x_{2,m}}$ above, we can see that this partial effect will be positive with a positive probability (and non-negative a.e.)  if and only if  $\beta_{2,m} ( \alpha^{(1)}_{1,2} - \alpha^{(1)}_{1,1})>0$. In a completely analogous way, we can consider $\frac{\partial P(Y^{c_1} \leq 2 | x)}{\partial x_{2,m}}$ and obtain that this partial effect will be positive with a positive probability (and non-negative a.e.)  if and only if  $\beta_{2,m} ( \alpha^{(1)}_{2,2} - \alpha^{(1)}_{2,1})>0$. Since $\alpha^{(1)}_{2,2} - \alpha^{(1)}_{2,1}<0$ and $\alpha^{(1)}_{1,2} - \alpha^{(1)}_{1,1}>0$, partial effects $\frac{\partial P(Y^{c_1} \leq 1 | x)}{\partial x_{2,m}}$ and $\frac{\partial P(Y^{c_1} \leq 2 | x)}{\partial x_{2,m}}$ will have different signs. In contrast, in lattice models the sign of such cross-partial effect would remain consistent across the whole domain.

Let us now look at the partial effects with respect to own covariates. If $x_{d.m}$ is exclusive to $Y^{*c_d}$, then the own partial effect $\frac{\partial P(Y^{c_d} \leq y^{(d)}_j | x)}{\partial x_{d,m}}$ will have the same sign  in a non-lattice model for any $y^{(d)}_j$, $j=1,\ldots, M_d -1$. The situation is different if $x_{d,m}$ is shared with another latent process. Using our model in Figure \ref{fig:bivariate2}, suppose $x_{1,m}=x_{2,m_2}$ and obtain that in this case,

{\small\begin{multline*}
\frac{\partial P(Y^{c_1} \leq y^{(1)}_j | x)}{\partial x_{1,m}}=-\beta_{1,m} \frac{\partial F( \alpha^{(1)}_{j,1} -x_1\beta_1,   \alpha^{(2)} -x_2\beta_2)}{\partial e_1} - \beta_{2,m_2} \frac{\partial F( \alpha^{(1)}_{j,1} -x_1\beta_1,   \alpha^{(2)} -x_2\beta_2)}{\partial e_2} \\
-\beta_{1,m}f_1(\alpha^{(1)}_{j,2} -x_1\beta_1)
+\beta_{1,m} \frac{\partial F( \alpha^{(1)}_{j,2} -x_1\beta_1,   \alpha^{(2)} -x_2\beta_2)}{\partial e_1} + \beta_{2,m_2} \frac{\partial F( \alpha^{(1)}_{j,2} -x_1\beta_1,   \alpha^{(2)} -x_2\beta_2)}{\partial e_2},
\end{multline*}}where $f_d$ denotes the p.d.f. of $\varepsilon_d$. 

As an example, consider a special case of independent  $\varepsilon_1$ and $\varepsilon_2$. Then 
{\small\begin{multline*}
\frac{\partial P(Y^{c_1} \leq y^{(1)}_j | x)}{\partial x_{1,m}}=\beta_{2,m_2} f_2(\alpha^{(2)} -x_2\beta_2)( F_1( \alpha^{(1)}_{j,2} -x_1\beta_1)-F_1( \alpha^{(1)}_{j,1} -x_1\beta_1)) \\ -\beta_{1,m}( f_1( \alpha^{(1)}_{j,1} -x_1\beta_1)F_2(\alpha^{(2)} -x_2\beta_2)  + f_1( \alpha^{(1)}_{j,2} -x_1\beta_1)(1-F_2(\alpha^{(2)} -x_2\beta_2))).
\end{multline*}}
If $\beta_{2,m_2}(\alpha^{(1)}_{j,2}-\alpha^{(1)}_{j,1})$ and $\beta_{1,m}$ have the same sign, then we have the difference of either two positive or two negative terms. Each of them can potentially dominate the other depending on the values of indices (and, hence, $x_1$ and $x_2$), thresholds, as well as $\beta_{1,m}$ and $\beta_{2,m_2}$. 

The feature of own partial effects with respect to a shared covariate or cross-partial effects not having the same sign across the domain complicates the identification analysis that follows.

\subsection{Identification in the semiparametric model}
\label{sec:semiid}

We now proceed to establishing identification in semiparametric models with general rectangular structures. The knowledge of index parameters and both layers of dependence, which is embedded in threshold parameters and the joint c.d.f., is central to counterfactual analysis and policy design involving joint outcomes such as household decisions on, say, healthcare and education investments. For example, the double-layer dependence structure will determine whether bundled interventions reinforce or crowd out each other.

The approach in \cite{km2025lattice} for the identification of index parameters and threshold differences in lattice models relies on the ability to isolate different dimensions and consider one dimension at a time. This will not work in non-lattice models as we cannot isolate different dimensions. For example, from  Figure \ref{fig:bivariate2} one can see that $P\left(Y^{(1)}\leq y_j^{(1)}|x\right)$ for $j=1,2$, cannot be expressed just in terms of the index $x_1\beta_1$ and the marginal c.d.f. $F_{1,\leq}$.  General rectangular structures therefore require a different approach to identification. Intuitively, the identification of parameters $\beta_d$ and the threshold structure in these models should be more demanding on the data compared to lattice models, especially given an unknown dependence structure of unobservables. This is indeed the case until we get to the stage of identifying the joint c.d.f. of unobservables. At that stage, as follows from \cite{km2025lattice}, both lattice and general rectangular models become similarly demanding on the data, which is worthy of note.  

To prove identification, we introduce some definitions and notations:

\begin{definition}
Covariate $x_{d,\ell}$ is exclusive to process $d$ if  $ x_{d,\ell}  \, | x_{-d}$ has a non-degenerate distribution almost everywhere for $x_{-d} \equiv  (x_1, \ldots, x_{d-1}, x_{d+1}, \ldots, x_D)$.
\end{definition}
Intuitively, an exclusive covariate is one that contains information unique to the 
$d$-th process $Y^{*c_d}$ and cannot be perfectly predicted from the covariates associated with other processes.\footnote{The support conditions in subsequent theorems will specify the additional variation on exclusive covariates required for identification.}   It is without a loss of generality that these exclusive covariates are arranged to be the first covariates within $x_d$: 

\begin{notation} For each $d=1, \ldots, D$, let $x_{d, 1:L_d}$  denote the subvector of $x_d$ that consists of all the covariates in $x_d$ that are exclusive to the process $Y^{*c_d}$.
\end{notation} 	
Throughout, references to effective covariates count each covariate shared across processes only once.

We establish identification through the following four steps:
\begin{itemize} \itemsep0em
\item[1st] step: identification of parameters corresponding to exclusive covariates in each process (Theorem \ref{th:nonlattice_semiparametric}).

\item[2nd] step: identification of parameters corresponding to shared covariates (Theorem \ref{th:nonlattice_semiparametric2}).

\item[3rd] step: identification of the joint c.d.f. of $\varepsilon$ (Theorem \ref{th:nonlattice_semiparametric3}).

\item[4th] step: identification of the thresholds $\alpha$  (Theorem \ref{th:nonlattice_semiparametric4}).
\end{itemize}

We now proceed to the identification of parameters corresponding to exclusive covariates,  

\begin{theorem} 
\label{th:nonlattice_semiparametric}
Consider a $D$-variate ordered discrete response model with the index
structure in  (\ref{index_structure}). Suppose Assumption  \ref{assn:errors}  holds and the model has a coherent general rectangular structure.  Suppose that, for each
$d=1,\ldots,D$, the following conditions are satisfied:

\begin{enumerate}
\item[(a)] $L_d\geq 1$.

\item[(b)] The coefficient $\beta_{d,1}$ corresponding to $x_{d,1}$
in $x_d\beta_d$ is normalized to $1$.

\item[(c)] There exist a nonempty interval $(\underline x_{d,1},\overline x_{d,1})$,  a set $S_{d,2:L_d}\subseteq\mathbb R^{L_d-1} $ with full affine dimension $L_d-1$, and a set $S_{d,-}$ of values of the remaining effective covariates $(x_{d,L_d+1:k_d},x_{-d})$, having positive probability, such that, for almost every fixed value in $S_{d,-}$, the set  $(\underline x_{d,1},\overline x_{d,1}) \times S_{d,2:L_d}$  is contained in the conditional support of $(x_{d,1},x_{d,2:L_d})$ given the remaining effective covariates.\footnote{When $L_d=1$, the  requirement on full affine dimension is understood to hold vacuously.}  Let $q(x)$ denote the conditional probability of a  fixed ``corner''
outcome of the rectangular partition.\footnote{E.g., $q(x) = P( Y^{c_1}=y^{(1)}_1,\ldots,Y^{c_D}=y^{(D)}_1 | x )$ for the lower ``corner''. More generally, each $Y^{c_d}$ should take either the first or the last discrete value. For the theorem, it does not matter which corner is chosen as long as it satisfies theorem's conditions.}  For the same fixed values of the remaining effective covariates: 
\begin{enumerate} 
\item[(i)] for every $x_{d,2:L_d}\in S_{d,2:L_d}$, the function  $x_{d,1}\mapsto q(x)$  is continuous and strictly monotone on $(\underline x_{d,1},\overline x_{d,1})$, with the direction of monotonicity determined by the chosen corner, holding all other
covariates fixed;
\item[(ii)] there exists $p\in(0,1)$, which may depend on the fixed remaining covariates, such that  $p\in \bigcap_{x_{d,2:L_d}\in S_{d,2:L_d}} \left\{ q(x): x_{d,1}\in (\underline x_{d,1},\overline x_{d,1}) \right\}$.  \end{enumerate} \end{enumerate} 
Then the parameters  $\beta_{d,1:L_d}$,  $d=1,\ldots,D$,  corresponding to the exclusive covariates in each process are identified.
\end{theorem}

\noindent Condition (a) states that each process has at least one exclusive covariate, and  condition (b) effectively states that the first (and potentially only) exclusive covariate in process $Y^{*c_d}$ has a non-zero coefficient. (b) further normalizes this coefficient to 1 (alternatively, could be normalized to $-1$ if the impact is negative). Such normalization restrictions are standard in semiparametric models where  parameter vectors generally can only be identified up to scale.  These normalizations can be different across $d$ (some normalized to 1, some to $-1$). 

Condition (c) is a version of the usual rank condition adapted to the multivariate setting. It requires that, on an informative part of the
covariate support, the exclusive covariates $x_{d,1:L_d}$ provide
sufficiently rich variation while the remaining covariates
$(x_{d,L_d+1:k_d},x_{-d})$ are held fixed. The normalized exclusive covariate $x_{d,1}$ varies continuously over an interval, while the remaining exclusive covariates $x_{d,2:L_d}$ vary over a set of full affine dimension.

For the ``corner'' probability $q(x)$ used in the condition, increasing $x_{d,1}$ affects only the corresponding inequality in dimension $d$ (this probability is automatically weakly monotone in $x_{d,1}$).
The strictly monotonic requirement rules out locally uninformative
regions in which variation in the exclusive index does not change the
probability. 

Condition (c) also requires sufficient overlap of the
probability ranges generated at different values of $x_{d,2:L_d}$.  The remaining
exclusive covariates, however, may be discrete. 

Our next result identifies the coefficients on shared regressors,
building on the identification of the coefficients of exclusive covariates
 in Theorem \ref{th:nonlattice_semiparametric}.
Theorem \ref{th:nonlattice_semiparametric2} strengthens the support
conditions on exclusive covariates. The argument starts with the probability that all outcomes take specified extreme categories, such as their lowest categories. To study one outcome, we vary the exclusive covariates of the other processes until those other outcomes take their specified categories with probability approaching one. The joint probability then approaches the probability of the outcome we are studying. We next compare different values of the shared covariates and ask how much the studied process’s exclusive covariate must change to keep this probability fixed. These compensating changes identify the coefficients on the shared covariates.

This argument requires enough variation in the exclusive covariates to make the other outcomes approach their specified extreme categories with probability one. Depending on the support of the unobservables, a sufficiently wide bounded range of covariate values may be sufficient. Otherwise, unbounded support may be needed. Such large enough  support requirements are common in the semiparametric literature, in particular, in semiparametric univariate ordered response models (see e.g., \citealp{manski1985,manski1988, horowitzbook, lewbel2000, lewbel2003}). Large support and exclusion type restrictions are common in semiparametric models 
of complex choice environments. For example, \cite{BarseghyanMolinariThirkettle2021}
use common and alternative specific excluded regressors to identify preferences and
limited consideration mechanisms. In their framework, excluded regressors with
large support play the role of special regressors. Related exclusion and support requirements appear in \cite{Matzkin1993}.

For the next two theorems, fix a ``corner'' outcome
$(j_1,\ldots,j_D)$, where $j_d\in\{1,M_d\}$, and let
$q(x)=P(
    \bigcap_{d=1}^D\{Y^{c_d}=y^{(d)}_{j_d}\}
    \,|\,x
)$. 
Write $a_d=
\alpha^{(d)}_{j_1,\ldots,j_{d-1},r_d(j_d),j_{d+1},\ldots,j_D}$,
with 
$r_d(1)=1$, $r_d(M_d)=M_d-1$. Take
$\kappa_d$ as $\leq$ if $j_d=1$ and $\kappa_d$ as $>$ if $j_d=M_d$. Let $w$ collect the effective covariates other than
$x_{1,1},\ldots,x_{D,1}$. By exclusivity, $x_{d,-1}$ is
determined by $w$. Denote the marginal support endpoints of
$\varepsilon_d$ by $\underline e_d$ and $\overline e_d$,
allowing infinite endpoints.

\begin{theorem}
\label{th:nonlattice_semiparametric2}
Suppose all conditions of Theorem
\ref{th:nonlattice_semiparametric} hold and the joint distribution of $\boldsymbol{\varepsilon}$
is absolutely continuous with respect to Lebesgue measure
on $\mathbb R^D$ and $M_d\geq2$ for every $d$. 
Also assume the following: 
\begin{enumerate}
\item[(a)]
There is a measurable set $S$ of values of $w$, with
$P(w\in S)>0$, such that, for almost every $w\in S$, the
conditional support of $(x_{1,1},\ldots,x_{D,1})$ contains $\prod_{d=1}^D
(\underline x_{d,1},\overline x_{d,1})$, where the intervals are nonempty and their endpoints may
be infinite.

For each $d$ with $L_d<k_d$, the conditional distribution
of $x_{d,L_d+1:k_d}$ given $w\in S$ is not concentrated on
a proper affine hyperplane of $\mathbb R^{k_d-L_d}$.

\item[(b)] For each $d$, for a.e. $w\in S$, and for every
$x_{d,1}\in(\underline x_{d,1},\overline x_{d,1})$, we have $P(
    \bigcap_{h\neq d}\{Y^{c_h}=y^{(h)}_{j_h}\}
    \,|\,x)
\rightarrow 1$ 
as, jointly for all $h\neq d$,
$x_{h,1}\downarrow\underline x_{h,1}$ if $j_h=1$
and $x_{h,1}\uparrow\overline x_{h,1}$ if $j_h=M_h$,
holding $x_{d,1}$ and $w$ fixed.\footnote{This holds automatically if for every $h$,
$\underline x_{h,1}=-\infty$ if $j_h=1$, and
$\overline x_{h,1}=+\infty$ if $j_h=M_h$. Finite endpoints may also suffice in case the support of $\varepsilon$ is bounded.}

\item[(c)]
For each $d$, take the limit of $q(x)$ as, for every
$h\neq d$, $x_{h,1}\downarrow\underline x_{h,1}$ if $j_h=1$
and $x_{h,1}\uparrow\overline x_{h,1}$ if $j_h=M_h$,
holding $x_{d,1}$ and $w$ fixed. Denote this limiting
probability by $q_d(x_d)$.

For each $d$ with $L_d<k_d$, there exists $p_d\in(0,1)$
such that, for almost every $w\in S$,
\[
p_d\in
\left\{
q_d(x_d):
x_{d,1}\in(\underline x_{d,1},\overline x_{d,1})
\right\}.
\]
\end{enumerate}

 Then all $\beta_d$,
$d=1,\ldots,D$, are identified.
\end{theorem}

Since $\beta_{d,1:L_d}$ has been identified by Theorem \ref{th:nonlattice_semiparametric}, the  additional rank condition in (a) concerns only the shared regressors
$x_{d,L_d+1:k_d}$. When $L_d=k_d$, the entire coefficient
vector has already been identified by Theorem \ref{th:nonlattice_semiparametric}. The Cartesian conditional support in (a) allows the joint
limits in (c), while (b) requires the other outcomes to  jointly approach their selected corner categories with probability one.  Condition (c) gives  the remaining
overlap requirement in requiring  a probability level common across
the values of $w$ used to identify the shared coefficients.\footnote{
The overlap condition in Theorem
\ref{th:nonlattice_semiparametric} does not imply this as 
its probability level may depend on the fixed remaining
covariates, including shared regressors, and it concerns
the original ``corner'' probability rather than its boundary
limit.}

Our third result is on the identification of the joint distribution of $\boldsymbol{\varepsilon}$. It is enough to identify one function $F_{\kappa_1 \ldots \kappa_D}$ to fully characterize this distribution. We can identify the distribution from one of the ``corners'' in our partitioning that gives us enough variation in probabilities.   

\begin{theorem}
\label{th:nonlattice_semiparametric3} 
Suppose all conditions of Theorem
\ref{th:nonlattice_semiparametric2} hold for the chosen
``corner'' and set $S$. Suppose additionally that, for every
$d=1,\ldots,D$ and almost every $w\in S$, 
$(0,1)\subseteq
\left\{
q_d(x_d):
x_{d,1}\in(\underline x_{d,1},\overline x_{d,1})
\right\},$ 
holding $w$ fixed.

Under the location normalizations $F_{d,\leq}(e_{0d})=c_{0d}$,
$d=1,\ldots,D$, where $e_{0d}$ and $c_{0d}\in(0,1)$ are known, the joint
distribution of $\boldsymbol{\varepsilon}$ and the
thresholds $a_d$, $d=1,\ldots,D$, defining the chosen
``corner'' are identified.
\end{theorem}

The additional condition strengthens Theorem
\ref{th:nonlattice_semiparametric2}'s overlap requirement
to coverage of all probability levels in $(0,1)$ for
every equation. This identifies the marginal error
distributions up to location shifts, which the
normalizations determine along with the chosen corner
thresholds. The inherited Cartesian support condition
then identifies the joint error distribution.

As we have seen, conditions have strengthened across Theorems~\ref{th:nonlattice_semiparametric}--\ref{th:nonlattice_semiparametric3}, reflecting the increasing demand on data as we move from exclusive covariate parameters to shared parameters and then to the joint distribution of unobservables. 


Our final result is on the identification of threshold parameters. This result will require us to strengthen some conditions further but at the same time will allow us to distinguish whether the threshold structure is
consistent with broad or narrow bracketing.  Threshold identification, as is clear from Theorem  \ref{th:nonlattice_semiparametric4}, requires index
variation that distinguishes the remaining rectangular
regions. The  theorem states sufficient
conditions using boundary limits, locally observed
probability boundaries, and integration over unrestricted
index coordinates.

\begin{theorem}\label{th:nonlattice_semiparametric4}
Suppose the conditions of Theorem
\ref{th:nonlattice_semiparametric3} hold.
For each region $R_{\mathbf j}$, let $A=\{d:1<j_d<M_d\}$ and $B=\{1,\ldots, D\}\setminus A$. Partition $A$ into $U=\{d\in A:\operatorname{supp}(\varepsilon_d)=\mathbb R\}$ and $C=A\setminus U$.
Suppose the support of the joint index $t=(x_1\beta_1, \ldots, x_D\beta_D)$ contains an open product
$\prod_d I_d$, possibly depending on $\mathbf j$,
with the following properties: 

\begin{enumerate}
\item[(a)]
For $K=B$ and $K=B\setminus\{h\}$, $h\in B$,
sending $t_d$, $d\in K$, to the lower endpoint of $I_d$
if $j_d=1$ and to its upper endpoint if $j_d=M_d$,
holding the other indices fixed, gives $P(Y^{c_d}=y^{(d)}_{j_d}\text{ for all }d\in K\mid t)
\rightarrow1$.

Denote the corresponding limits of $P(Y^{c_d}=y^{(d)}_{j_d},\ d=1,\ldots,D| t)$
by $q(t_A)$ when $K=B$, and by $q_h(t_A,t_h)$
when $K=B\setminus\{h\}$.

\item[(b)] For $d\in U$, $I_d=\mathbb R$. If $C\ne\varnothing$,
the set
$\left\{
t_C\in\prod_{d\in C}I_d:r(t_C)>0
\right\}$,
 where 
$r(t_C)=\int_{\mathbb R^{|U|}}q(t_A)\, dt_U$,
has $|C|$ supporting hyperplanes with linearly
independent outward normals and contact points inside
$\prod_{d\in C}I_d$.

Each normal has a strictly positive $d$th component
if $\operatorname{supp}(\varepsilon_d)$ is bounded below, and  strictly negative component otherwise.

\item[(c)]
For every $h\in B$, at some point in
$\prod_{d\in A}I_d\times I_h$, it holds that  $0<q_h(t_A,t_h)<q(t_A)$.

\end{enumerate}

Then all thresholds are identified.\footnote{Empty collections of limits or integrations leave
the function unchanged. When $A=\varnothing$, $q=1$.}
\end{theorem}
Condition (a) specifies restrictions for interior
outcomes in $A$ by making the selected boundary outcomes in $B$ occur
with probability approaching one. It holds automatically when the relevant endpoints
are $-\infty$ for first categories and $+\infty$
for last categories.

For an interior coordinate whose marginal error support
is $\mathbb R$, condition (b) requires the corresponding
index to range over $\mathbb R$. 
This allows the proof to integrate the observed
probabilities over that index. For the remaining
coordinates, the data must reveal where the probability
becomes zero. These boundary locations depend on the
unknown thresholds and the already identified error
support. Observing them in enough different directions
is exactly what will allow  the thresholds to be determined separately. 

Condition (c) identifies each remaining boundary
threshold from a conditional probability strictly between zero and one.

 \begin{figure}[!t]
\centering

\begin{tikzpicture}[scale=0.2]

\draw [very thick] (-10,-7) -- (-7,-7);

\draw [very thick] (-10,3) -- (-3,3);

\draw [very thick] (8,5) -- (10,5);

\draw [very thick] (6,-4) -- (10,-4);

\draw [very thick] (-7,-10) -- (-7,-7);

\draw [very thick] (-3,3) -- (-3,10);

\draw [very thick] (6,-10) -- (6,-4);

\draw [very thick] (8,5) -- (8,10);

\draw[very thick, dotted, -> ] (-8.5,-11) -- (8.5,-11) node[right] {$Y^{*c_1}$};
\draw[very thick, dotted, -> ] (-11,-8.5) -- (-11,8.5) node[above] {$Y^{*c_2}$};
\node [below=3cm, align=flush center,text width=4cm]
        {
            Panel A: Stage 1
        };
\end{tikzpicture}
\hskip 0.1in 
\begin{tikzpicture}[scale=0.2]
\draw [very thick] (-10,-7) -- (-7,-7);
\draw [very thick, dotted] (-10,-1) -- (-8,-1);
\draw [very thick, dotted] (8,2) -- (10,2);
\draw [very thick] (-10,3) -- (-3,3);
\draw [very thick] (8,5) -- (10,5);

\draw [very thick] (6,-4) -- (10,-4);

\draw [very thick] (-7,-10) -- (-7,-7);
\draw [very thick] (-3,3) -- (-3,10);
\draw [very thick, dotted] (-2,-10) -- (-2,-8);
\draw [very thick, dotted] (2,8) -- (2,10);

\draw [very thick] (6,-10) -- (6,-4);

\draw [very thick] (8,5) -- (8,10);

\draw[very thick, dotted, -> ] (-8.5,-11) -- (8.5,-11) node[right] {$Y^{*c_1}$};
\draw[very thick, dotted, -> ] (-11,-8.5) -- (-11,8.5) node[above] {$Y^{*c_2}$};
\node [below=3cm, align=flush center,text width=4cm]
        {
            Panel B: Stage 2, part 1
        };
\end{tikzpicture}

\begin{tikzpicture}[scale=0.2]

\draw [very thick] (-10,-7) -- (-7,-7);
\draw [very thick] (-10,-1) -- (-5,-1);
\draw [very thick] (6,2) -- (10,2);
\draw [very thick] (-10,3) -- (2,3);
\draw [very thick] (2,5) -- (10,5);
\draw [very thick] (6,-4) -- (10,-4);
\draw [very thick] (-7,-5) -- (6,-5);

\draw [very thick] (-7,-10) -- (-7,-1);

\draw [very thick] (-3,3) -- (-3,10);
\draw [very thick] (-2,-10) -- (-2,-5);
\draw [very thick] (2,3) -- (2,10);

\draw [very thick] (-5,-1) -- (-5,3);

\draw [very thick] (6,-10) -- (6,5);
\draw [very thick] (8,5) -- (8,10);

\draw[very thick, dotted, -> ] (-8.5,-11) -- (8.5,-11) node[right] {$Y^{*c_1}$};
\draw[very thick, dotted, -> ] (-11,-8.5) -- (-11,8.5) node[above] {$Y^{*c_2}$};

\node [below=3cm, align=flush center,text width=4cm]
        {
            Panel C: Stage 2, part 2
        };
\end{tikzpicture}
\hskip 0.1in 
\begin{tikzpicture}[scale=0.2]
\draw [very thick] (-10,-7) -- (-7,-7);
\draw [very thick] (-7,-5) -- (6,-5);
\draw [very thick] (6,-4) -- (10,-4);
\draw [very thick] (-10,-1) -- (-5,-1);
\draw [very thick, blue] (-5,-1) -- (6,-1);   
\draw [very thick] (6,2) -- (10,2);
\draw [very thick] (-10,3) -- (2,3);
\draw [very thick] (2,5) -- (10,5);
\draw [very thick] (-7,-10) -- (-7,-1);
\draw [very thick] (-5,-1) -- (-5,3);    
\draw [very thick] (-3,3) -- (-3,10);
\draw [very thick] (-2,-10) -- (-2,-5);
\draw [very thick, blue] (0,-5) -- (0,-1);     
\draw [very thick, blue] (2,-1) -- (2,3);     
\draw [very thick] (2,3) -- (2,10);     
\draw [very thick] (6,-10) -- (6,5);
\draw [very thick] (8,5) -- (8,10);
\draw[very thick, dotted, -> ] (-8.5,-11) -- (8.5,-11) node[right] {$Y^{*c_1}$};
\draw[very thick, dotted, -> ] (-11,-8.5) -- (-11,8.5) node[above] {$Y^{*c_2}$};
\node [below=3cm, align=flush center,text width=2cm]
        {
            Panel D: Stage 3
        };
\end{tikzpicture}
\caption{Stages of threshold system identification in the bivariate case.}
\label{fig:D2illustrationthreshold}
\end{figure}

 We prove the identification of thresholds in Theorem \ref{th:nonlattice_semiparametric4} sequentially in a manner somewhat consistent with solving a puzzle and it is best illustrated in the bivariate case in Figure \ref{fig:D2illustrationthreshold}.  
 
 In Stage 1, thresholds are identified that define
``corner'' regions $R_{q_1,\ldots,q_D}$ with
$q_d\in\{1,M_d\}$. The result of this step is given
in Panel A. Stage 2, part 1  identifies thresholds in the dimension corresponding to the region’s interior outcome coordinate and with  the other outcomes 
at their boundary categories. The thresholds identified after Stage 2, part 1 are in Panel B in Figure \ref{fig:D2illustrationthreshold} as dotted lines (dotted because their extent is not known). Stage 2, part 2 continues to consider border regions for each  $d=1,\ldots, D$  and identifies the remaining thresholds shaping those border rectangles. In the bivariate case the thresholds identified after Stage 2, part 2 are in Panel C in Figure \ref{fig:D2illustrationthreshold} (the thresholds from Stage 2, part 1 are now in solid lines as their extents are known). Stage 3 identifies the thresholds of regions with two interior outcome coordinates, shown in Panel D.  Each stage builds on the results of the previous stages, and regions within each stage are considered sequentially. Using thresholds identified in previous stages or for previously considered regions within the same stage ensures that, at each step, at most $D$ of the $2D$ thresholds defining a region remain unknown. These are identified using the common argument at the beginning of the proof.

The identification results also motivate semiparametric estimation, with the candidate sieve likelihood procedure outlined in Online Appendix \ref{sec:semiest} (without establishing formal asymptotic theory).  In estimation in this main text, we implement the parametric specification developed in the next section.

\section{Parametric specification}
\label{sec:parametric}

Our semiparametric framework offers an approach to achieving identification results despite model complexity and, as mentioned,  suggests a spline estimation procedure following \cite{coppejans2007} (see Online Appendix \ref{sec:semiest}). In practice, researchers may opt for a rich parametric family for the joint distribution of unobserved $\boldsymbol{\varepsilon}$, in part for computational convenience. This choice eliminates the need for nonparametric modeling of the joint distribution  and typically reduces the data requirements for identifying all unknown parameters. These parametric families must be sufficiently flexible to accommodate various negative and positive dependencies among unobservables in the latent processes. Following established traditions in statistics and econometrics, natural choices for the joint distribution of unobservables are the multivariate normal distribution and a multivariate extension of the logistic distribution. 

\subsection{Numerical illustration of identifiability}
Although parametric versions of the model will be identified under the conditions outlined in Section \ref{sec:semiid}, intuitively, weaker conditions ensuring sufficient variation (potentially with discrete and/or no exclusive covariates) should suffice for identification in the parametric case. A useful analogy is the comparison between a univariate single index model with an unknown monotone link function and the probit model. The probit model requires only a finite number of points satisfying a rank condition, whereas the single index model demands richer variation, often guaranteed by the presence of a continuous covariate not needed in probit.


Finding clear-cut weaker conditions that do not involve exclusive or continuous covariates and are sufficient to identify all model parameters in a parametric case, is theoretically challenging even though, of course, they may exist in theory. This difficulty mirrors the lack of straightforward identification conditions in multinomial probit or logit models that allow unknown correlations among different latent processes. We instead illustrate parametric identification with only shared covariates through a numerical exercise. The exercise also clarifies the source of identification in this setting -- the full rectangular cell structure can contain information absent from any single region.

Consider $D=2$. Let \(\boldsymbol{\varepsilon}=(\varepsilon_1,\varepsilon_2)'\) be jointly normal with mean \((0,0)'\), unit variances, and correlation \(\rho\).\footnote{As usual, we normalize means and variances because shift and scale changes yield observationally equivalent parameter vectors.} Denote its bivariate normal cumulative distribution function by \(\Phi_2(\cdot,\cdot;\rho)\).  Write
$P_{j_1 j_2}(x)\equiv P(Y^{c_1}=y^{(1)}_{j_1},Y^{c_2}=y^{(2)}_{j_2}| x)$ and
collect the lower ``corner'' parameters in
$\theta=(\beta_1,\beta_2,\alpha^{(1)}_{1,1},\alpha^{(2)}_{1,1},\rho)$. A single ``corner'' case does not identify these parameters when the same covariate
enters both indices as the lower ``corner'' probability is
\begin{equation}
\label{eq:normal}
P_{11}(x)=\Phi_2\!\big(\alpha^{(1)}_{1,1}-x_1\beta_1,\
\alpha^{(2)}_{1,1}-x_2\beta_2;\rho\big), 
\end{equation}
and the exchangeability of the standardized bivariate normal c.d.f. implies  $\Phi_2(u,v;\rho)=\Phi_2(v,u;\rho)$. Hence, whenever $x_1=x_2=x$,
$P_{11}(x)$ is invariant under the reflection $(\beta_1,\alpha^{(1)}_{1,1})
\leftrightarrow (\beta_2,\alpha^{(2)}_{1,1})$. The ``corner'' probability therefore cannot distinguish $\theta$ from $\tilde\theta=(\beta_2,\beta_1,\alpha^{(2)}_{1,1},
\alpha^{(1)}_{1,1},\rho)$ and two parameter vectors are observationally equivalent in the ``corner'' cell for every value of the shared covariate. Thus, the ``corner'' cell on its own suffers from a global identification challenge as a result of the aforementioned symmetry. At the value $\theta_0=(1.2,-0.8,0.5,-0.2,0.4)$ used below, the reflection is
$\tilde\theta_0=(-0.8,1.2,-0.2,0.5,0.4)$.

The full $2\times2$ cell structure rules out this reflection. Embed the ``corner'' in
a coherent $2\times2$ structure with potentially different thresholds
$\alpha^{(1)}_{1,1}$ and $\alpha^{(1)}_{1,2}$ in dimension~1, and the same $\alpha^{(2)}_{2,1}=\alpha^{(2)}_{1,1}$ in dimension~2. With $x_1=x_2=x$, the  probabilities of rectangles are
\begin{align*}
P_{11}(x)&=\Phi_2\!\big(\alpha^{(1)}_{1,1}-x\beta_1,\
\alpha^{(2)}_{1,1}-x\beta_2;\rho\big), \quad 
P_{21}(x)=\Phi\!\big(\alpha^{(2)}_{1,1}-x\beta_2\big)-P_{11}(x),\\
P_{12}(x)&=\Phi\!\big(\alpha^{(1)}_{1,2}-x\beta_1\big)
-\Phi_2\!\big(\alpha^{(1)}_{1,2}-x\beta_1,\
\alpha^{(2)}_{1,1}-x\beta_2;\rho\big),\quad 
P_{22}(x)=1-P_{11}(x)-P_{21}(x)-P_{12}(x).
\end{align*}
Although the reflection above leaves $P_{11}(x)$ unchanged, the same is not true for all other rectangle probabilities. The mechanism is especially
transparent from the fact that $P_{11}(x)+P_{21}(x)
=\Phi\!\big(\alpha^{(2)}_{1,1}-x\beta_2\big)$. Thus, once the rectangle with $P_{21}(x)$ is used together with $P_{11}(x)$, the second index is separately revealed through its marginal probability and the exchangeability of $\Phi_2$ no longer makes the two indices interchangeable. Likewise, $P_{12}(x)$ depends on the completing threshold
$\alpha^{(1)}_{1,2}$ and the first index.

To make the example concrete, take $\theta_0$ as above with completing threshold
$\alpha^{(1)}_{1,2}=1.5$, a single shared covariate, and $T=10$ design points
drawn from $[-2,2]$. We compute the cell probabilities
$P^{\mathrm{obs}}_{j_1j_2}$ at $\theta_0$ and define the full-cell objective
\[
Q_T(\theta)=\sum_{i=1}^{T}\sum_{j_1=1}^{2}\sum_{j_2=1}^{2}
\Big(P^{\mathrm{obs}}_{j_1j_2}(x^{(i)})
-P_{j_1j_2}(x^{(i)};\theta)\Big)^2.
\]
Let $Q_T^{\mathrm{corner}} =\sum_{i=1}^{T} \left(
P_{1,1}^{\mathrm{obs}}(x^{(i)})
-
\Phi_2\big(\alpha^{(1)}_{1,1} - x_{1}^{(i)}\beta_1,\;
          \alpha^{(2)}_{1,1} - x_{2}^{(i)}\beta_2;\;
          \rho\big)
\right)^2$ denote the analogous objective constructed from
$P_{11}$ alone. To isolate the reflection described above, we hold
$\alpha^{(1)}_{1,2}$ at its true value, so that both objectives are functions of
the same five parameters $\theta$. 

For the numerical example, we reparametrize $\rho$ as
$z=\operatorname{atanh}(\rho)$. For each radius $r$ around $\theta_0$ in this
transformed parameter space, we evaluate the objectives along 5,000 randomly
sampled directions and report the smallest value found.\footnote{We also include the
known reflection direction explicitly so that this particular direction is not
missed by random sampling.} 

\begin{figure}[!t]
\begin{center}
  \includegraphics[width=0.95\textwidth]{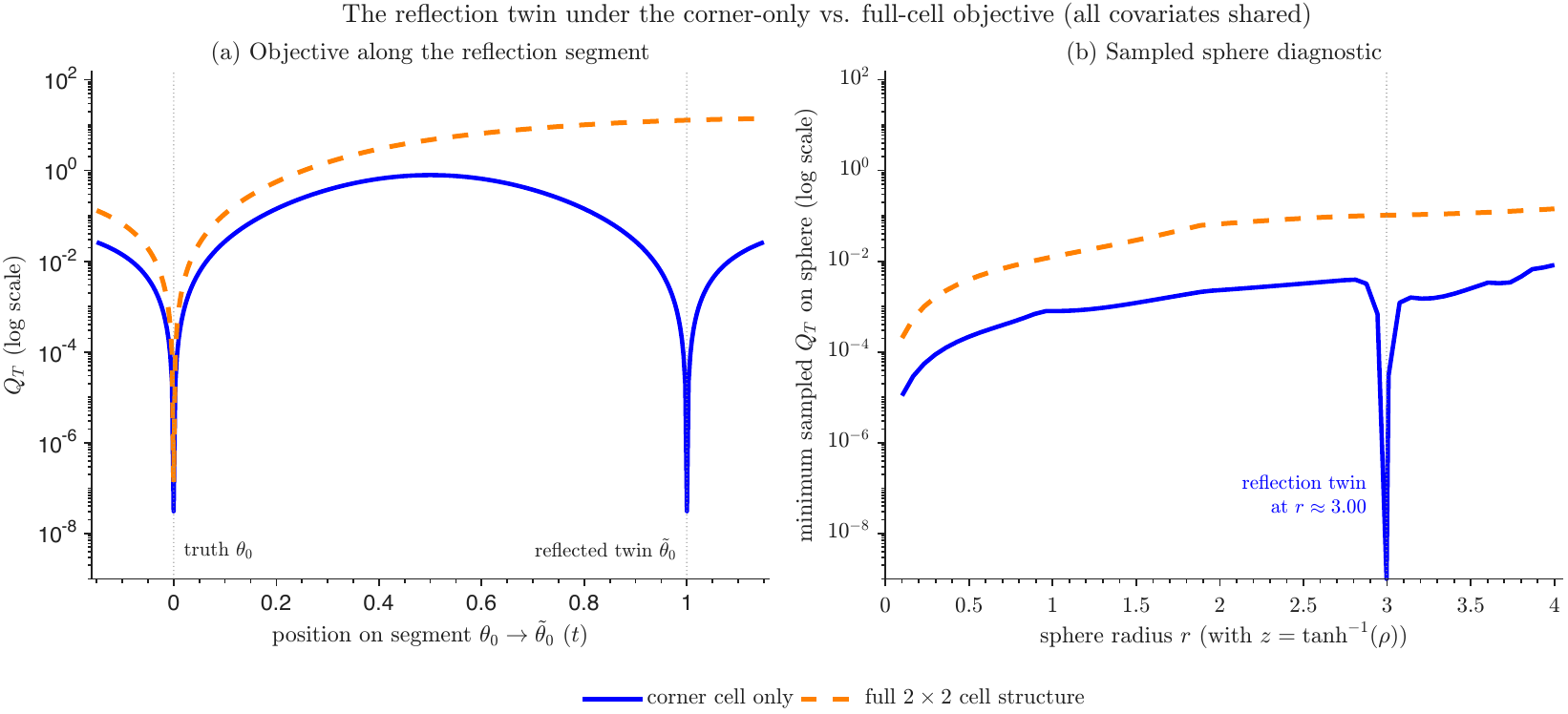}
\end{center}
\caption{Numerical illustration for the bivariate normal model with a single shared covariate.}
\caption*{\footnotesize Notes: The objective $Q_T$ (log scale) compares the lower
corner cell exercise (solid) with the full $2\times2$ cell structure (dashed).
\emph{Panel (a)}: $Q_T$ along the segment from the true value $\theta_0$
($t=0$) to its reflection $\tilde\theta_0$ ($t=1$). The  objective for the corner only 
vanishes at both endpoints, whereas the objective using all cells is strictly positive
at the reflected point. \emph{Panel (b)}: for each radius $r$ about $\theta_0$
in the $z=\operatorname{atanh}(\rho)$ parametrization, the figure reports the
smallest $Q_T$ among 5,000 randomly sampled directions together with the known
reflection direction. The corner-only objective returns to zero at the
reflection radius $r\approx3$, whereas the full-cell objective remains bounded
away from zero there. The completing threshold $\alpha^{(1)}_{1,2}=1.5$ is held
at its true value, and $T=10$ shared-covariate design points are drawn from
$[-2,2]$.}
\label{fig:identGaussian}
\end{figure}

Figure~\ref{fig:identGaussian} shows the results. Panel~(a) evaluates the two
objectives along the segment from $\theta_0$ to $\tilde\theta_0$ (moving the proportion $t$ along the vector between the two points). The
corner-only objective falls to zero at both points, displaying the reflected
global solution implied by exchangeability. By contrast, the full-cell
objective is zero at $\theta_0$ but is strictly positive at
$\tilde\theta_0$, so the additional cells break this reflection. In the
$z=\operatorname{atanh}(\rho)$ parametrization, the reflected point lies at
distance approximately $3$ from $\theta_0$. Panel~(b) gives the corresponding
sampled sphere diagnostic. The objective based on a single ``corner'' returns to zero at the
reflection radius, while among the sampled directions the full-cell objective
stays bounded away from zero at that radius and does not give an analogous 
second zero.

\subsection{Estimation} To outline an estimation approach in the parametric case, we continue with $D=2$ and $\boldsymbol{\varepsilon}=(\varepsilon_1,\varepsilon_2)'$ being jointly normal with zero mean, unit variances, and correlation \(\rho\). Given a random sample $\left\{(y^{(1)(i)}, y^{(2)(i)},x_1^{(i)},x_2^{(i)}) \right\}_{i=1}^N$ and  collecting  $\beta_1,\beta_2,\rho$ and all the thresholds in $\alpha$ in the parameter vector $\theta$, consider the log-likelihood function 
\begin{eqnarray*}
\mathcal{L}(\theta) &=& \frac{1}{N} \sum_{i=1}^{N}\sum_{j_{1}=1}^{M_{1}}\sum_{j_{2}=1}^{M_{2}} 1\left[(y^{(1)(i)},y^{(2)(i)})=(y_{j_{1}}^{(1)},y_{j_{2}}^{(2)})\right]\log(\ell^{(i)}_{j_{1},j_{2}}(\theta)) = \frac{1}{N}\sum_{i=1}^{N}\log(\ell^{(i)}(\theta)), 
\end{eqnarray*} 
$$\text{ with } \quad \ell^{(i)}_{j_{1},j_{2}} = 
\sum_{t_1=0}^{1} \sum_{t_2=0}^{1} (-1)^{t_1 + t_2} \Phi_{2}\left( \alpha^{(1)}_{j_1 - t_1, j_2} - x_1^{(i)}\beta_1, \alpha^{(2)}_{j_1, j_2 - t_2} - x_2^{(i)}\beta_2; \rho \right).$$
This log-likelihood needs to be maximized subject to linear inequality constraints that describe ordering of thresholds in each dimension, normalization constraints on the thresholds, and non-linear equality constraints $q(\theta)=0$ that collect coherency constraints (\ref{eq:coherencybiv}) across all the local models.

The constrained maximum likelihood estimator (MLE) $\hat{\theta}$ solves the optimization problem $\max_{\theta} \mathcal{L}(\theta)$  subject to the described coherency constraints on thresholds. Assuming the true parameters lie at a regular point of the constraint set (such that the Jacobian of the active coherency constraints has full row rank), the coherency constraints are differentiable and we can invoke typical MLE regularity conditions (e.g., \cite{newey1994}) to obtain the asymptotic result $\sqrt{N}(\hat{\theta}-\theta_{0}) \overset{d}{\longrightarrow} \mathcal{N}(0,V)$, where $V = J^{-1} - J^{-1}Q'(QJ^{-1}Q')^{-1}QJ^{-1}$ for terms $Q = \dfrac{\partial q(\theta_{0})}{\partial \theta'}$ and $J=\mathbb{E}\left[\frac{\partial \log(\ell^{(i)}(\theta_0))}{\partial \theta}\frac{\partial \log(\ell^{(i)}(\theta_0))}{\partial \theta'}\right]$.\footnote{Singularity arises when both threshold
differences in a coherency restriction vanish for
some $(j_1,j_2)$. The covariance formula above does
not apply at such configurations. The model can then
be re-estimated with both equalities imposed explicitly
(thus, maintaining the local lattice structure) and
using a nonredundant set of restrictions for inference. 
This inference treats the imposed equalities as given, that is, it will be conditional on that locally imposed structure.} The natural plug-in sample analogue estimator of $V$ provides a consistent estimator for the variance-covariance matrix. 

Alternatively, coherency can be imposed approximately by adding the penalty
\begin{equation}
\label{penalty}
-\lambda_N \sum_{j_1=1}^{M_1-1}\sum_{j_2=1}^{M_2-1}
\left(\alpha_{j_1+1,j_2}^{(2)}-\alpha_{j_1,j_2}^{(2)}\right)^2
\left(\alpha_{j_1,j_2+1}^{(1)}-\alpha_{j_1,j_2}^{(1)}\right)^2
\end{equation}
to the objective. With sufficiently small constraint violations and estimates stable to further increases in $\lambda_N$, differences from explicitly constrained estimation should be negligible in practice. Both approaches involve numerical tolerances, and the choice is primarily computational. We use penalized estimation in the simulations and applications below.

\section{Monte Carlo experiments}
\label{sec:MonteCarlo} 

We examine Monte Carlo simulations for the parametric case with bivariate normal errors, as outlined in Section~\ref{sec:parametric}. We compare the constrained MLE for general rectangular structure models (\textit{non-lattice probit}) with the standard bivariate ordered probit, which estimates a misspecified lattice model. The baseline model is $$Y^{*c_1} = x\beta_{1} + w_{1}\gamma_{1} + \varepsilon_{1}, \qquad
Y^{*c_2} = x\beta_{2} + w_{2}\gamma_{2} + \varepsilon_{2},$$
where unobservables are independent of regressors and jointly normal with zero means and unit variances. We explore scenarios with no exclusive covariates ($\gamma_1=\gamma_2=0$) and separately with an exclusive covariate in one latent process. Each simulation design uses 250 independent random samples of size $N=5,000$, with the penalty term in (\ref{penalty}) set to $N$.\footnote{Alternative $N$ and $\lambda_N$ yield similar results, but we favor a large sample size so that any issues with the lattice or non-lattice performance cannot be attributed to a small sample size.}

Two findings stand out. First, non-lattice model parameters are estimated with negligible bias, even in the absence of exclusive covariates. Second, imposing a lattice structure on a truly non-lattice data-generating process yields severely inconsistent estimators for \textit{all} parameters, not merely the thresholds. The degree of inconsistency depends on how well lattice models approximate the true non-lattice. The most dramatic distortion concerns the correlation parameter $\rho$. For example, in Design~1, where the truth is $\rho=0.33$, the lattice estimator consistently returns values near $-0.93$ to $-0.97$ across all covariate distributions (standard deviations of 0.01--0.02). This is a near-deterministic sign reversal of magnitude greater than one on a parameter bounded in $[-1,1]$. The mechanism for this is intuitive. Namely, the true non-lattice structure exhibits substitutability among outcomes via the threshold layer. The lattice estimator, lacking any
threshold channel, must attribute this  to $\rho$, forcing it toward large negative values. The result is that the sign of $\hat{\rho}$ under a lattice model cannot be taken as evidence of the sign of the true unobservables correlation.
This has direct implications for any empirical study that uses bivariate ordered probit, bivariate probit, or correlated ordered logit to draw conclusions about adverse selection, complementarity, or the direction of latent covariation.

\subsection*{Design 1: 2$\times$2 structure, no excluded covariates} 
First, we set $\gamma_1=\gamma_2=0$, thereby removing $w_1$ and $w_2$. We also set $\beta_1=1$, $\beta_2=0.5$, $\rho=0.33$, and use a $2\times2$ non-lattice structure with thresholds $\alpha^{(2)}_{1,1}=\alpha^{(2)}_{2,1}=1$, $\alpha_{1,1}^{(1)}=-2$, and $\alpha_{1,2}^{(1)}=1.5$. We consider three distributions for the common regressor $x$: uniform $[-5,5]$ (Design 1A),  discrete on 10 points $\{\pm 5, \pm 3.5, \pm 2.5, \pm 1.5, 0, 0.5\}$ with equal probabilities (Design 1B), and  discrete on only three points $\{\pm 2.5, 0\}$ with equal probabilities (Design 1C).

Table~\ref{tabSIM1new} reports the  means across simulations and standard deviations for the estimated parameters. Across all sub-designs A-C, the non-lattice method estimates all parameters with minimal bias, even with the minimal variation in $x$ as in Design 1C. The bivariate lattice ordered probit method performs especially poorly on $\beta_2$, $\rho$, and the thresholds. The very limited variation in the common regressor results in Design 1C exhibiting higher standard deviations than designs 1A and 1B.


\subsection*{Design 2: 4$\times$3 with one excluded covariate}
\label{subsection:des2}
We now increase the number of discrete values, $M_{d}$, in both dimensions. The discrete dependent variables $Y^{c_1}$ and $Y^{c_2}$ can take four and three values, respectively. This yields a 4$\times$3 non-lattice structure, illustrated in Figure \ref{figSimDesign2}. The common covariate $x$ follows a uniform $[-3,3]$ distribution (though continuity is not necessary here). The covariate $w_{1}$ takes values $\{-2.5, -1.5, \pm 0.5\}$ with equal probability. We set $\gamma_2=0$, thus removing $w_{2}$ in the second equation. The parameter values are $\beta_{1} = 1.5, \gamma_{1} = -4, \beta_{2} = 3$ and $\rho = 0.5$. 

Table \ref{tabSIMDesign2} reports the  means across simulations and standard deviations of the estimated index parameters and the correlation coefficient. Table \ref{tabSIM1CUT} in the Online Appendix provides the estimated threshold values. The non-lattice bivariate ordered probit method estimates all the parameters with almost no bias. By contrast, the lattice bivariate ordered probit method estimates all parameters with a relatively large bias. The mean squared errors in the non-lattice method are far lower than those in the lattice method for all the parameters. Working with a lattice structure makes estimating the correlation parameter $\rho$ markedly difficult, with the method failing to estimate the correct sign for $\rho$, let alone a value close to the true one. The intuition is similar to that in Design 1.

\begin{table}[t]
\caption{Simulation results Design 1} \label{tabSIM1new}
\centering
\scriptsize
\begin{threeparttable}
\begin{tabular}{lccccccc}
\toprule
\textbf{Parameter}  & \multicolumn{2}{c}{Design 1A} & \multicolumn{2}{c}{Design 1B} & \multicolumn{2}{c}{Design 1C}\\
\cmidrule(lr){2-3} \cmidrule(lr){4-5} \cmidrule(lr){6-7}
                   &                 \textbf{Non-latt} & \textbf{Latt}    & \textbf{Non-latt} & \textbf{Latt}    & \textbf{Non-latt} & \textbf{Latt}    \\
\hline
$\beta_{1}$ = 1    & 1.00 (0.03) & 0.77 (0.02) & 1.00 (0.03) & 0.75 (0.02)  & 1.00 (0.03) & 0.84 (0.02)  \\
$\beta_{2}$ = 0.5  & 0.50 (0.02) & 0.01 (0.01)  & 0.50 (0.02) & 0.04 (0.01)  & 0.50 (0.05) & -0.18 (0.01)  \\
$\rho$      = 0.33 & 0.33 (0.12) & -0.93 (0.02) & 0.34 (0.12) & -0.94 (0.01) & 0.33 (0.14) & -0.97 (0.01) \\
\midrule
\multirow{2}{*}{$\alpha^{(2)}_{1,1}=\alpha^{(2)}_{2,1}$=1}
   & 1.00 (0.04) & \multirow{2}{*}{0.72 (0.04)} & 1.00 (0.04) & \multirow{2}{*}{0.68 (0.03)} & 1.00 (0.05) & \multirow{2}{*}{0.64 (0.03)} \\
& 1.00 (0.04) &      & 1.00 (0.04) &            & 1.00 (0.05) &                              \\
$\alpha_{1,1}^{(1)}$ = -2  & -1.99 (0.07) & \multirow{2}{*}{-0.42 (0.02)} & -1.99 (0.07) & \multirow{2}{*}{-0.40 (0.02)} & -1.99 (0.12) & \multirow{2}{*}{-0.53 (0.02)} \\
$\alpha_{1,2}^{(1)}$ = 1.5 &  1.50 (0.08) &                               &  1.50 (0.08) &                               &  1.49 (0.15) &                               \\
\bottomrule
\end{tabular} \vspace*{-0.1cm}
\begin{tablenotes}[flushleft]
\footnotesize
\item Notes: This table reports the sample mean and sample standard deviations (in parentheses) of the estimates of the Design 1 parameters, over 250 samples. The ``Non-latt'' columns provide estimates from using the newly proposed non-lattice bivariate ordered probit model. The ``Latt'' columns assume a lattice structure.
\end{tablenotes}
\end{threeparttable}
\end{table}

\section{Applications}
\label{sec:appl}

Throughout this section, for non-lattice models, the reported standard errors are obtained by bootstrap using penalized likelihood. For lattice and sequential timing models, the standard errors are computed using the inverse Hessian.

\subsection{Identifying moral hazard and adverse selection in insurance markets}

\noindent \textit{Motivation}. The dominant empirical framework for estimating asymmetric information in insurance markets, due to \citet{chiappori2000testing}, tests for adverse selection by estimating a bivariate probit and interpreting the latent correlation as evidence of asymmetric information. As \citet{cohen2010} emphasize, this correlation conflates adverse selection with moral hazard, since both forces contribute to the positive covariation between coverage and utilization.\footnote{Despite these advances, the method remains popular in practice, with recent applications including \cite{matcham2026}, \cite{crawford2018}, and \cite{benetton2026}.} Separating them has required either quasi-experimental variation or strong structural assumptions \citep{Handel2013AER,Einav_etal2013AERselectiononmoralhazard,HackmannKolstadKowalski2015}.\footnote{An important case is
\citet{cardonhendel2001}, which jointly models health insurance choice and medical
expenditure to test for asymmetric information using the discrete-continuous
framework of \citet{dubinmcfadden1984}, and who likewise find little evidence of
selection on unobservables.}

Our general rectangular structure resolves this limitation without quasi-experimental variation, as moral hazard enters through coverage-dependent utilization thresholds, leaving the unobservables correlation to capture selection alone. Our Monte Carlo results further motivate this approach since we saw the lattice estimator's $\hat{\rho}$ being severely distorted when threshold-layer dependence is present, so studies interpreting the sign of a bivariate probit's latent correlation as evidence for or against adverse selection may be reading an artifact of model misspecification rather than a structural parameter.  Moreover, our general rectangular structure model permits two-way dependence of thresholds, so not only can utilization threshold  depend on insurance coverage, but insurance-choice thresholds can also depend on anticipated behavioral responses to insurance (particularly, with a larger number of discrete options in each dimension), accommodating what \cite{Einav_etal2013AERselectiononmoralhazard} and others term ``selection on moral hazard.'' Of course, coherency still needs to be satisfied. \\

\begin{figure}[t]
\centering
\begin{tikzpicture}[scale=0.4]

\draw [very thick] (-5,-4) -- (-3.25,-4);
\draw [very thick] (-3.25,-3) -- (8,-3);
\draw [very thick] (-5,0.5) -- (8,0.5);
\draw [very thick] (8,-1) -- (9,-1);
\draw [very thick] (8,4) -- (9,4);

\draw [very thick] (-3.25,-5) -- (-3.25,0.5);
\draw [very thick] (-0.5,0.5) -- (-0.5,5);
\draw [very thick] (0.5,-5) -- (0.5,-3);
\draw [very thick] (1.5,-3) -- (1.5,0.5);
\draw [very thick] (5,0.5) -- (5,5);
\draw [very thick] (8,-5) -- (8,5);

\draw [very thick, dotted ] (-3.25,-5) -- (-3.25,-6);
\draw [very thick, dotted ] (0.5,-5) -- (0.5,-6);
\draw [very thick, dotted ] (8,-5) -- (8,-6);

\draw [very thick, dotted ] (9,-1) -- (10,-1);
\draw [very thick, dotted ] (9,4) -- (10,4);

\draw [very thick, dotted ] (8,5) -- (8,6);
\draw [very thick, dotted ] (5,5) -- (5,6);
\draw [very thick, dotted ] (-0.5,5) -- (-0.5,6);

\draw [very thick, dotted ] (-5,-4) -- (-6,-4);
\draw [very thick, dotted ] (-5,0.5) -- (-6,0.5);

\node at (-3.25,-4) [circle,fill,inner sep=1.5pt]{};
\node at (-3.25,-3) [circle,fill,inner sep=1.5pt]{};
\node at (-3.25,0.5) [circle,fill,inner sep=1.5pt]{};
\node at (-0.5,0.5) [circle,fill,inner sep=1.5pt]{};

\node at (0.5,-3) [circle,fill,inner sep=1.5pt]{};
\node at (1.5,-3) [circle,fill,inner sep=1.5pt]{};
\node at (1.5,0.5) [circle,fill,inner sep=1.5pt]{};
\node at (5,0.5) [circle,fill,inner sep=1.5pt]{};

\node at (8,-3) [circle,fill,inner sep=1.5pt]{};
\node at (8,-1) [circle,fill,inner sep=1.5pt]{};
\node at (8,0.5) [circle,fill,inner sep=1.5pt]{};
\node at (8,4) [circle,fill,inner sep=1.5pt]{};

\node at (-3.25,-6) [circle,fill,inner sep=1.5pt]{};
\node at (0.5,-6) [circle,fill,inner sep=1.5pt]{};
\node at (8,-6) [circle,fill,inner sep=1.5pt]{};

\node at (-6,-6) [circle,fill,inner sep=1.5pt]{};
\node at (-6,-4) [circle,fill,inner sep=1.5pt]{};
\node at (-6,0.5) [circle,fill,inner sep=1.5pt]{};
\node at (-6,6) [circle,fill,inner sep=1.5pt]{};

\node at (-0.5,6) [circle,fill,inner sep=1.5pt]{};
\node at (5,6) [circle,fill,inner sep=1.5pt]{};
\node at (8,6) [circle,fill,inner sep=1.5pt]{};
\node at (10,6) [circle,fill,inner sep=1.5pt]{};

\node at (10,4) [circle,fill,inner sep=1.5pt]{};
\node at (10,-1) [circle,fill,inner sep=1.5pt]{};
\node at (10,-6) [circle,fill,inner sep=1.5pt]{};

\node [below, thick] at (-3.25,-6) {\small $-3.25$};
\node [above, thick] at (-0.5,6) {\small $ -0.5$};
\node [below, thick] at (0.5,-6) {\small $  0.5$};
\node [above, thick] at (1.5,0.5) {\small $ 1$};
\node [above, thick] at (5,6) {\small $ 5$};
\node [above, thick] at (8,6) {\small $ 8$};

\node [right, thick] at (-3.25,-4) {\small $ -4$};
\node [right, thick] at (10,-3) {\small $ -2$};
\node [right, thick] at (10,-1) {\small $0$};
\node [right, thick] at (10,0.5) {\small $ 0.5$};

\node [right, thick] at (10,4) {\small $4$};

\draw[very thick,-> ] (-6,-7.5) -- (10,-7.5) node[right] {$Y^{*c_1}$};
\draw[very thick,-> ] (-8,-6) -- (-8,6) node[above] {$Y^{*c_2}$};

\draw [very thick, dotted ] (-6,-6) -- (-6,6);
\draw [very thick, dotted ] (-6,-6) -- (10,-6);
\draw [very thick, dotted ] (10,-6) -- (10,6);
\draw [very thick, dotted ] (10,6) -- (-6,6);

\end{tikzpicture}
\caption{Latent variable space for two equations: Design 2 simulation}
\label{figSimDesign2}
\end{figure}

\noindent \textit{Data and Model Specification.} We apply the general rectangular structure to U.S. health insurance markets using the Medical Expenditure Panel Survey (MEPS), which offers nationally representative data on insurance coverage and healthcare utilization. The quality of the MEPS data has been verified in \cite{Zuvekas2009}. Our sample includes approximately 60,000 individuals from 2005 to 2010, before the Affordable Care Act. Following the standard framework, we specify latent insurance and utilization equations as 
\begin{equation}
Y_{\text{ins}}^{*} = x_{\text{ins}}\beta_{\text{ins}} + \varepsilon_{\text{ins}}, 
\quad 
Y_{\text{use}}^{*} = x_{\text{use}}\beta_{\text{use}} + \varepsilon_{\text{use}}. 
\end{equation}
 Common covariates include demographics (logged income and its square, dividend payments, family size, age, education, gender, marital status, race, region, and year) and pre-existing conditions (diabetes, asthma, high blood pressure, high cholesterol, angina, heart attack, stroke, emphysema, and arthritis). An exclusive covariate for insurance is partner's coverage provided by their job. In this and the two other empirical applications to follow, we assume a bivariate normal distribution for the vector of unobservables, with zero mean, unit variances, and an unknown correlation $\rho$.

 \begin{table}[t]
\caption{Simulation results for Design 2} \label{tabSIMDesign2}
\centering
\begin{threeparttable}
\begin{tabular}{lccccccccc}
\toprule 
\textbf{Parameter}  &&& \multicolumn{1}{c}{\textbf{Non-latt}}  &&& \multicolumn{1}{c}{\textbf{Latt}} \\
\hline
$\beta_{1} = 1.5$   &&& 1.50 (0.04) &&& 0.61 (0.01)  \\
$\gamma_{1} = -4$   &&& -4.01 (0.09) &&& -2.51 (0.04) \\
$\beta_{2} = 3$     &&& 3.00 (0.09) &&& 1.64 (0.03) \\
$\rho = 0.5$       &&& 0.50 (0.06) &&& -0.60 (0.03) \\
\bottomrule 
\end{tabular}
\vspace*{-0.1cm}
\begin{tablenotes}[flushleft]
\footnotesize
\item Notes: Sample means and sample standard deviations (in parentheses) of the estimates of the model parameters, over 250 repeated samples. 
\end{tablenotes}
\end{threeparttable}
\end{table}

The insurance outcome $Y_{\text{ins}}$ equals 1 if the individual holds private health insurance in January and is 0 otherwise. Utilization $Y_{\text{use}}$ is  measured categorically (0 for no charges, 1 for below-median charges, 2 for above-median charges).\footnote{There could be a concern that insured and uninsured individuals pay different prices, implying that charges may not reflect actual usage on an equivalent scale for the insured and uninsured. However, our utilization is constructed from MEPS total expenditures, defined as the sum of all payments received by the provider across patient out-of-pocket, private insurance, and public payers, rather than from billed charges. This is the variable the MEPS data programme designed precisely to be payer-neutral at the transaction level. Furthermore, it is also possible instead to use the number of billable events; results are similar and available on request.}

Moral hazard is isolated by allowing utilization thresholds to depend on coverage status: $Y_{\text{ins}}=1$ lowers the thresholds for utilization if individuals respond to coverage by consuming more care. Thresholds in the insurance dimension  may also vary with utilization status, accommodating ``selection on moral hazard''  \citep{Einav_etal2013AERselectiononmoralhazard}, which is the possibility that individuals select coverage partly based on their anticipated behavioral response. We estimate the model subject only to coherency constraints on the thresholds, thus remaining agnostic on both questions and letting the data determine whether either threshold interdependence is present.\footnote{In this, and all empirical examples, we use a penalty of $\lambda_{N} = N^2$, since the eventual sample size in the empirical applications are generally smaller than in the Monte Carlo simulations.} Once the behavioral effect is accounted for, the remaining correlation between $\varepsilon_{\text{ins}}$ and $\varepsilon_{\text{use}}$ can be interpreted as evidence of adverse or advantageous selection, given the assumptions of the model.\\ 

\noindent \textit{Estimated Threshold Structure.} 
 Figure~\ref{fig:thresholds} presents the threshold results for the non-lattice model. First, the model reveals moral hazard as  utilization thresholds shift downward when
$Y_{\text{ins}}=1$, with the low-to-medium usage threshold falling from $-0.02$ (0.04) to
$-0.45$ (0.05) and the medium-to-high threshold from $0.68$ (0.04) to $0.46$ (0.05), where here and in what follows, standard errors are in parentheses.

The estimated structure has coverage-dependent utilization thresholds and an invariant insurance threshold. Since both utilization cutoffs vary with coverage, coherency requires the insurance threshold to remain constant across utilization categories. The data therefore favor this direction of threshold dependence. Even though it is consistent with the absence of selection on moral hazard, a richer categorization of coverage would be better suited to studying that question. Importantly, this would be a finding of the
estimation, not a maintained assumption. A researcher who imposed sequential timing by
assumption would obtain the same insurance threshold structure but have no way of knowing
whether the data supported it.




Third, once moral hazard is absorbed into the threshold layer, the residual unobservables correlation $\hat{\rho}$ falls from $0.21$ (0.01) in the lattice model to $0.04$ (0.02) in the non-lattice model, suggesting minimal adverse selection, which is consistent with the broader literature that finds limited selection once moral hazard is accounted for
\citep{Einav_etal2013AERselectiononmoralhazard,Handel2013AER}. 

Our approach outperforms traditional lattice models and, since it can be applied with standard survey or administrative data,  provides a powerful way to revisit much of the empirical literature built on lattice approaches or ordered probit models.\\
 
\noindent \textit{Partial Effects.}  
As we explained in Section \ref{sec:semiadvantages}, the non-lattice model allows for indirect effects of exclusive covariates on outcome variables in other dimensions. In the insurance case, the insurance status of one's partner may indirectly affect one's own utilization. We estimate these partial effects and examine their magnitude. 

First, we calculate average partial effects, taking each individual in the dataset at their observed covariates and evaluating the impact on the probability of exceeding a certain value of utilization when the individual has a partner with insurance, and when they do not. Specifically, we calculate $$APE_{k} = \frac{1}{N}\sum_{i=1}^{N}\left[ P(Y_{\text{use}} \geq k|x_{i,\text{partner}=1}) - P(Y_{\text{use}}\geq k|x_{i,\text{partner}=0})  \right]$$ for $k=1$ (utilization but below the median) and $k=2$ (utilization above the median). We estimate $APE_{1}$ as 0.055 (0.0034) and $APE_{2}$ as 0.034 (0.0043), implying that having a partner with insurance raises the probability of high utilization by three percentage points on average, and the probability of non-zero utilization by $5.5$  percentage points on average.

\begin{figure}[tbp]
\centering
\begin{tikzpicture}[scale=0.25]
\draw [very thick, dotted ] (-8.2,0.5) -- (-12,0.5);
\draw [very thick, dotted ] (-8.2,-7) -- (-12,-7);
\draw [very thick, dotted ] (15.2,-3) -- (13.2,-3);
\draw [very thick, dotted ] (15.2,-10) -- (13.2,-10);
\draw [very thick, dotted ] (3,-10.2) -- (3,-12);
\draw [very thick, dotted ] (3,0) -- (3,3);
\draw [very thick, dotted ] (-12,-12) -- (-12,3);
\draw [very thick, dotted ] (15,-12) -- (15,3);
\draw [very thick, dotted ] (-12,3) -- (15,3);
\draw [very thick, dotted ] (15,-12) -- (-12,-12);
\draw [very thick ] (-10,0.5) -- (3,0.5);
\draw [very thick ] (-10,-7) -- (3,-7);
\draw [very thick ] (3,-3) -- (13,-3);
\draw [very thick ] (3,-10) -- (13,-10);
\draw [very thick ] (3,-11) -- (3,1.5);
\node at (3,3) [circle,fill,inner sep=1.5pt, ]{};
\node at (3,-3) [circle,fill,inner sep=1.5pt, ]{};
\node at (3,-7) [circle,fill,inner sep=1.5pt, ]{};
\node at (-12,-7) [circle,fill,inner sep=1.5pt, ]{};
\node at (15,-3) [circle,fill,inner sep=1.5pt, ]{};
\node at (3,0.5) [circle,fill,inner sep=1.5pt, ]{};
\node at (3,-10) [circle,fill,inner sep=1.5pt, ]{};
\node at (3,-12) [circle,fill,inner sep=1.5pt, ]{};
\node at (-12,0.5) [circle,fill,inner sep=1.5pt, ]{};
\node at (15,-10) [circle,fill,inner sep=1.5pt, ]{};
\node at (-12,3) [circle,fill,inner sep=1.5pt, ]{};
\node at (15,-12) [circle,fill,inner sep=1.5pt, ]{};
\node at (-12,-12) [circle,fill,inner sep=1.5pt, ]{};
\node at (15,3) [circle,fill,inner sep=1.5pt, ]{};
\node [thick ] at (3,-14.25) {\large $2.02 \: (0.06)$};
\node [right,thick ] at (3.7,0.5) {\large $0.68 \: (0.04)$};
\node [right,thick ] at (3.7,-7) {\large $-0.02 \: (0.04)$};
\node [right,thick ] at (15.75,-3) {\large $0.46 \: (0.05)$};
\node [right,thick ] at (15.75,-10) {\large $-0.45 \: (0.05)$};
\draw[very thick,-> ] (-12,-15.75) -- (15,-15.75) node[right] {$Y^{*}_{\text{ins}}$};
\draw[very thick,-> ] (-14,-12) -- (-14,4) node[above] {$Y^{*}_{\text{use}}$};
\end{tikzpicture}
\caption{Estimated thresholds for insurance coverage ($Y_{\text{ins}}^*$) and healthcare utilization ($Y_{\text{use}}^*$) (standard errors in parentheses)}
\label{fig:thresholds}
\end{figure}

Second, we calculate partial effects at a specific vector of covariates (typically the average) that is, $$PEA_k = P(Y_{\text{use}} \geq k|x_{i,\text{partner}=1},\bar{x}_{i,-\text{partner}}) - P(Y_{\text{use}}\geq k|x_{i,\text{partner}=0},\bar{x}_{i,-\text{partner}}) $$ For $\bar{x}_{i,-\text{partner}}$, we take numerical covariates (logged income, dividend payments, family size, years of education, and age) at their means, use the modal values of the binary variables: marital status (married), race (white), year (2009), metropolitan statistical area (yes), and region. Since the two other binary variables, male and pre-existing conditions, have means close to 0.5, we calculate PEA for the four possibilities. We estimate $PEA_{2}$ between 0.03 and 0.04 (0.01) across all four cases, whereas $PEA_{1}$ varies from 0.08 (0.01) for males without pre-existing conditions to 0.03 (0.005) for females with pre-existing conditions. Hence, the impact of partner's insurance on non-zero utilization depends on other covariates in the model.

\subsection{Cryptocurrency Familiarity and Optimism} 

\noindent \textit{Motivation, Data and Model Specification}.  In the second application, we use data from the Survey of Consumer Payment Choice (SCPC) \citep{foster2021} to study opinions on future movements in cryptocurrency prices.\footnote{See also \cite{benetton2022} and \cite{kahn2016}.} Conducted annually by the Federal Reserve Banks of Atlanta, Boston, and San Francisco, the SCPC tracks U.S. consumers' adoption of payment methods, and has recently noted a shift toward online and mobile payments due to the COVID-19 pandemic. Our sample includes around 4,600 individuals, with data on demographics (e.g., income, age, gender, education), payment method use (e.g., credit cards, cryptocurrencies, mobile platforms like Google Pay), and perceptions of safety, convenience, and cost. Additional data cover fraud exposure, FICO score ranges, and household financial roles. See \citet{foster2021} for details.

We examine whether opinions about bitcoin's future value are interdependent with cryptocurrency familiarity.  $Y^{c_1}$ is an ordered variable for familiarity with bitcoin (-1: not familiar, 0: slightly familiar, 1: somewhat familiar, 2: moderately/extremely familiar).\footnote{Moderate and extremely familiar are combined due to few respondents reporting extreme familiarity.} For $Y^{c_2}$, we use an ordered variable for bitcoin's expected value in one year (-1: decrease, 0: no change, 1: increase).\\

\noindent \textit{Estimated Threshold Structure and Coefficients}.  Figure \ref{fig:bitnonlatt} shows the estimated threshold structure for a non-lattice model. It reveals varied thresholds. For example, individuals with low familiarity expect no change (the blue dotted region for no change) at a lower threshold in the opinion dimension (i.e., lower $Y^{*Opinion}$ values yield a ``no change'' opinion compared to other familiarity groups). Conversely, those with high familiarity have closely spaced thresholds (the orange crosshatch region for no change), thereby defining a narrow ``no change'' region. For this group, most $Y^{*Opinion}$ values reflect strong opinions, either decreasing or increasing. These findings describe decision making structures and not probabilistic outcomes. 

Table \ref{tabREGbit} provides estimates of $\beta$ and $\rho$. The correlation $\rho$ ranges from 0.09 (lattice model, $s.e.=0.03$) to 0.75 (non-lattice model, $s.e.=0.14$), where the latter is large and reflects most plausibly a general disposition toward novel financial products that simultaneously raises both latent familiarity and latent optimism. Some index coefficient estimates differ by over 10\% in magnitude. Notably, the coefficient on male changes sign, being  negative in the lattice model and positive in the non-lattice model. The coefficients of the lattice model thus suggest that males are more pessimistic about bitcoin's value as the effect in this model would lead to $P(Y^{Optimism} \geq j|x_{-male}, x_{male}=1)-
P(Y^{Optimism} \geq j|x_{-male}, x_{male}=0)$ being negative for any level $j$ and any $x_{-male}$. As discussed earlier, a non-lattice model allows for changes in the signs of partial effects across the domain, therefore the positive coefficient for  males obtained there does not directly imply that this model suggests males are more optimistic for any level $j$ and any $x_{-male}$. Additional post-estimation analysis we have conducted does, however, confirm that given the estimated thresholds  the non-lattice model yields $P(Y^{Optimism} \geq j|x_{-male}, x_{male}=1)- P(Y^{Optimism} \geq j|x_{-male}, x_{male}=0)$ as positive for any level $j$ and any $x_{-male}$ in the data, thus, giving a stable sign of this partial effect across the domain.\footnote{A potentially different estimated threshold structure could have resulted in the switching of signs for this partial effect.}.  

{\tiny\begin{figure}[!t]
\centering
\begin{tikzpicture}[scale=0.3]

\fill[pattern=dots, pattern color=blue]        (-12,-11) rectangle (-6,-3);   
\fill[pattern=crosshatch, pattern color=orange] (11,1)   rectangle (15,5);    

\draw [very thick, dotted ] (-12,-14) -- (-12,7);
\draw [very thick, dotted ] (15,-14) -- (15,7);
\draw [very thick, dotted ] (-12,7) -- (15,7);
\draw [very thick, dotted ] (15,-14) -- (-12,-14);

\draw [very thick ] (2,-12) -- (2,6);
\draw [very thick, dotted ] (2,-12) -- (2,-14);
\draw [very thick, dotted ] (2,6) -- (2,7);

\draw [very thick ] (-10,-3) -- (2,-3);
\draw [very thick, dotted ] (-10,-3) -- (-12,-3);
\draw [very thick ] (-6,-12) -- (-6,-3);
\draw [very thick, dotted ] (-6,-12) -- (-6,-14);
\draw [very thick ] (-10,-11) -- (-6,-11);
\draw [very thick, dotted ] (-10,-11) -- (-12,-11);
\draw [very thick ] (-6,-7) -- (2,-7);
\draw [very thick ] (-2,-3) -- (-2,6);
\draw [very thick, dotted ] (-2,6) -- (-2,7);

\draw [very thick ] (2,1) -- (13,1);
\draw [very thick, dotted ] (13,1) -- (15,1);
\draw [very thick ] (7,-12) -- (7,1);
\draw [very thick, dotted ] (7,-12) -- (7,-14);
\draw [very thick ] (11,1) -- (11,6);
\draw [very thick, dotted ] (11,6) -- (11,7);
\draw [very thick ] (2,4) -- (11,4);
\draw [very thick ] (11,5) -- (13,5);
\draw [very thick, dotted ] (13,5) -- (15,5);

\node at (-6,-11) [circle,fill,inner sep=1.5pt]{};
\node at (-6,-7)  [circle,fill,inner sep=1.5pt]{};
\node at (-6,-3)  [circle,fill,inner sep=1.5pt]{};
\node at (-2,-3)  [circle,fill,inner sep=1.5pt]{};
\node at (2,-3)   [circle,fill,inner sep=1.5pt]{};
\node at (2,-7)   [circle,fill,inner sep=1.5pt]{};
\node at (2,1)    [circle,fill,inner sep=1.5pt]{};
\node at (2,4)    [circle,fill,inner sep=1.5pt]{};
\node at (7,1)    [circle,fill,inner sep=1.5pt]{};
\node at (11,1)   [circle,fill,inner sep=1.5pt]{};
\node at (11,4)   [circle,fill,inner sep=1.5pt]{};
\node at (11,5)   [circle,fill,inner sep=1.5pt]{};
\node at (-12,-11) [circle,fill,inner sep=1.5pt]{};
\node at (-12,-3)  [circle,fill,inner sep=1.5pt]{};
\node at (15,1)    [circle,fill,inner sep=1.5pt]{};
\node at (15,5)    [circle,fill,inner sep=1.5pt]{};
\node at (-6,-14)  [circle,fill,inner sep=1.5pt]{};
\node at (2,-14)   [circle,fill,inner sep=1.5pt]{};
\node at (7,-14)   [circle,fill,inner sep=1.5pt]{};
\node at (-2,7)    [circle,fill,inner sep=1.5pt]{};
\node at (2,7)     [circle,fill,inner sep=1.5pt]{};
\node at (11,7)    [circle,fill,inner sep=1.5pt]{};

\draw[very thick,-> ] (-14,-17.75) -- (15,-17.75) node[right] {$Y^{*\text{Familiarity}}$};
\draw[very thick,-> ] (-14,-14) -- (-14,7) node[above] {$Y^{*\text{Opinion}}$};

\node [below, thick ] at (-6,-14)  {$-0.80 \: (0.15)$};
\node [below, thick ] at (2,-15.5) {$0.36 \: (0.12)$};
\node [below, thick ] at (7,-14)   {$0.73 \: (0.14)$};
\node [above, thick ] at (-2,7)    {$-0.24 \: (0.15)$};
\node [above, thick ] at (11,7)    {$1.14 \: (0.14)$};

\node [left, thick ]  at (-14,-11) {$-1.76 \: (0.15)$};
\node [left, thick ]  at (-14,-3)  {$-0.27 \: (0.15)$};
\node [above, thick ] at (-2,-7)   {$-0.78 \: (0.12)$};
\node [right, thick ] at (15,1)    {$0.24 \: (0.27)$};
\node [above, thick ] at (6,4)     {$0.62 \: (0.20)$};
\node [right, thick ] at (15,5)    {$0.74 \: (0.32)$};

\end{tikzpicture}
\caption{Estimates from cryptocurrency example when assuming a general rectangular structure model (standard errors in parentheses)}
\label{fig:bitnonlatt}
\end{figure}}

Another aspect illustrated by the thresholds in Figure \ref{fig:bitnonlatt} regarding the decision making process is that individual decisions can be modeled  using a binary decision tree, where each node represents a decision based on a single latent process. This structure can be termed a \textit{hierarchical non-lattice model}, a specific subset of non-lattice models.\footnote{Any  model with a general rectangular structure can be represented by a decision tree, but typically, each node may involve multiple or all latent processes.} Hierarchical non-lattice models maintain coherence, as each node in the decision tree further refines the partitioning of the latent space. Figure \ref{fig:correctedhierarchicaltree} in the Online Appendix depicts a binary decision tree outlining the estimated hierarchical decision making process for this cryptocurrency application.\\

\noindent \textit{Sequential Timing Model.}
As an interim model between the lattice and the general rectangular structure, we estimate a sequential timing model in which familiarity is determined first, with thresholds functionally independent of optimism, while thresholds for optimism are then free to depend on familiarity. This is strictly narrower than the hierarchical non-lattice structure described above, as it imposes independent thresholds in the familiarity dimension while leaving the optimism dimension unrestricted. The sequential model is coherent by construction.

The estimated correlation coefficients (Table~\ref{tabREGbit}) follow a monotone pattern with $\hat{\rho}_{\text{latt}} = 0.09 < \hat{\rho}_{\text{seq}} = 0.26 < \hat{\rho}_{\text{non-latt}} = 0.75$. The explanation for this ordering is as follows. The lattice model misses this shifting threshold and infers from fewer people crossing into the highly optimistic category than a pure positive correlation would predict that $\hat{\rho}$ must be low. But the non-lattice model instead allows a high threshold into the highly optimistic category and, once that is accounted for, correctly interprets the residual covariation, given thresholds as a larger correlation in unobservables. The sequential model goes some way to fixing this but the fixed thresholds in familiarity dimension block the full increase in $\rho$ as in the fully flexible model.

\begin{footnotesize}
\begin{table}[t!]
\centering
\begin{threeparttable}
\caption{Estimation coefficients: bitcoin familiarity and optimism} \label{tabREGbit}
\centering
\begin{tabular}{lccccccccccccccc}
\toprule 
\textbf{Variable} && \textbf{Non-lattice} && \textbf{Lattice} && \textbf{Sequential} \\
\hline
\textit{Familiarity with Bitcoin}\\
\hline
\textsc{Low Income}         && 0.02 (0.05)    && 0.01 (0.06) && 0.01 (0.05)  \\
\textsc{Age}                 && -0.02 (0.00)  && -0.02 (0.00) && -0.02 (0.00)  \\
\textsc{Male}                && 0.46 (0.06)  && 0.51 (0.06) && 0.51 (0.05) \\
\textsc{Low Education}        && -0.44 (0.09)   && -0.48 (0.09) && -0.48 (0.08)  \\
\\
\textit{Bitcoin ``optimism''}\\
\hline
\textsc{Low Income}       && 	 0.05 (0.06) && 0.04 (0.06) && 0.03 (0.05)   \\
\textsc{Age}              && 	 -0.01 (0.00)  && -0.01 (0.00) && -0.01 (0.00) \\
\textsc{Male}             && 	 0.12 (0.07)   && -0.08 (0.05) && -0.02 (0.09) \\
\textsc{Low Education}    && 	 -0.10 (0.08)   && 0.08 (0.07) && 0.02 (0.08)   \\
\hline 
$\rho$  && 0.75 (0.14) && 0.09 (0.03) && 0.26 (0.21)  \\
\bottomrule 
\end{tabular} \vspace*{-0.1cm}
\begin{tablenotes}[flushleft]
\footnotesize
\item  Notes:  The ``Non-lattice'' column provides estimates from using non-lattice bivariate ordered probit model. The ``Lattice'' column assumes a lattice structure. The ``Sequential'' column assumes that thresholds in the familiarity dimension are lattice in structure, but thresholds in the optimism dimension are free (see Appendix Figure \ref{fig:sequential}).
\end{tablenotes}
\end{threeparttable}
\end{table}
\end{footnotesize}

The general rectangular structure appears to reject sequential timing in the familiarity direction. Figure~\ref{fig:bitnonlatt} shows that familiarity thresholds vary systematically across opinion outcomes. In particular, individuals who already hold strong opinions (either bullish or bearish about bitcoin) require a meaningfully different level of latent familiarity to self-report as, say, ``moderately familiar'' than do those who are agnostic about price movements. This two-way feedback is inconsistent with any  sequential timing model. Economically, it suggests that stated familiarity and price beliefs are formed jointly, consistent with belief-consistent information acquisition, which means that people who are already bullish may selectively engage with the asset and thereby report higher familiarity, while those who are agnostic remain less engaged. The lattice and sequential timing models suppress this channel entirely.

The male coefficient reinforces this conclusion. In the lattice, the coefficient on male in the optimism equation is $-0.08$, in the sequential timing model it attenuates to $-0.02$, and then only in the general rectangular structure does the coefficient become positive ($+0.12$). The lattice misspecification of the familiarity thresholds contaminates the optimism coefficients, and the sequential timing restriction is insufficient to correct that.

\subsection{Parental investments in children}
\label{sec:psid_cds_applications}
\noindent \textit{Motivation.} Our third and final application considers parental investment in their children, which is a central topic in the economics of human capital. The recent empirical literature is framed around how families allocate their time and money towards their children's educational development. Recent notable work (e.g., \cite{CunhaHeckman2007} and \citealp{CunhaHeckmanSchennach2010}) focuses on the dynamic technology that converts parental investments into the cognitive and non-cognitive skills of the children, stressing the importance of early investments.\footnote{A more foundational contribution is \cite{BeckerTomes1986}, which describes the partial inheritance of endowments and resources across generations.} 

Our application uses a general rectangular structure framework to study the \textit{interdependence} of families' decisions on (i) financial investments into their child's education, and (ii) the non-financial cognitive environment they create for their child, the latter of which relates to their time investment into their child's education. Structural studies on this topic, most notably \cite{DelBocaFlinnWiswall2014}, estimate production technologies with parental time and money as inputs. Both parental utility and production technologies typically take the Cobb Douglas form, which imposes complementarity of time and monetary investments on the margin. The argument is that a dollar of education spending yields more when the child is already in a cognitively stimulating home environment, and hours of stimulating parental time yield more when schooling investments reinforce them. However, parents may view time and monetary investments as substitutable, or they may narrow bracket their time and money investments into their child's education, as a lattice model would imply. 

Furthermore, in line with Beckerian tradition, there is a budgetary constraint between time investments and financial investments, arising from the labor supply decision of parents. Households in which parents work longer hours have more monetary resources available for education spending but less time and attention available for cognitive stimulation at home. A lattice model bundles all these forces into the correlation between unobservables. But a general rectangular structure model allows the data to inform on the complementarity/substitutability of parents' financial and time investments into their children's education through correlation in latent outcome unobservables and the threshold structure.\\

\noindent \textit{Data and Model Specification.} To estimate a general rectangular structure model in this context, we use the Panel Study of Income Dynamics (PSID) Child Development Supplement (CDS).\footnote{See \cite{tuzel2026} for another paper using the CDS data.} The CDS sampled approximately 3{,}500 children of PSID household members, aged 0--12 in 1997. We pool the second and third waves of the survey, which took place in 2002 and 2007, as they contain all requisite variables. The model we consider has financial and non-financial investments in their child's cognitive environment as jointly determined:  $$Y^{*}_{\text{spend}} = x_{\text{spend}}\beta_{\text{spend}} + \varepsilon_{\text{spend}} \qquad Y^{*}_{\text{stim}}=x_{\text{stim}}\beta_{\text{stim}} + \varepsilon_{\text{stim}}$$ with $Y_{\text{spend}}$ reflecting ordinal categories of financial investments into the child's education, and $Y_{\text{stim}}$ as an ordinal variable reflecting the quality of the child's cognitive environment. More specifically, for illustration, $Y_{\text{spend}}$ denotes annual education spending on the child, coded into four ordered categories split by quartile of educational spending. The quality of the child's cognitive home environment, $Y_{\text{stim}}$ is measured by the HOME Cognitive Stimulation scale, coded into terciles: low, medium, and high, for illustration. The HOME scale is constructed from interviewer-reported and parent-reported items on play with educational toys, stimulating activities, and parent-child reading and conversation.\footnote{The analysis sample restricts to children aged 10 or older so that the HOME-10+ scale is consistently applied.} It is a measure of parental time and environmental investment, not monetary investment. Education spending and HOME investment reflect the canonical inputs of money and time, respectively, in the child-quality production function of \cite{DelBocaFlinnWiswall2014}.

The index vectors $x_{\text{spend}}$ and $x_{\text{stim}}$ share a common set of demographic and socioeconomic controls: log family income, family size, child age, child gender, head's race, head's years of schooling, and three region indicators. We include an indicator for if the child has been previously classified as gifted and talented (\texttt{ever-gifted}) in both equations, and an indicator for if the primary caregiver works a non-standard shift as an exclusive covariate in the cognitive environment equation. The argument is that, conditional on income, if the primary caregiver works a non-standard shift, it will not affect financial investments into the child's education, but the nature of their job will render it more difficult to, for example, read to children and create a favorable cognitive environment for their child.\footnote{Two points are worth noting. First, a specification excluding \texttt{ever-gifted} from the HOME equation (so that each equation has an exclusive covariate) yields the same qualitative conclusions. Second, there may be a concern that parents working non-standard shifts may compensate for their lack of time by spending more on their child's education, invalidating the exclusion. However, because our measure of financial investment focuses mainly on direct schooling costs, it is unlikely to capture the direct substitution of paid daytime care for parental time.}  \\

\noindent \textit{Results and Interpretation.} 
Figure \ref{fig:gridpanel2} shows the full threshold structure in the latent space. At higher HOME levels, a lower latent index is required to enter the higher education-spending category, consistent with complementarity of the decisions between home inputs and financial investment in child production. The thresholds for the cognitive environment shift less in the same way. The estimate of the correlation coefficient is negative, $\widehat{\rho}=-0.27$ (0.12). This is not inconsistent with the complementarity displayed by the thresholds as the threshold layer captures the production complementarity emphasized by \cite{DelBocaFlinnWiswall2014}, while the error correlation captures dependence in unobservables. Its negative sign is consistent with labor supply constraints, whereby households with a higher latent propensity for monetary investment may have a lower latent propensity for time- and attention-intensive home investment, conditional on observed income and covariates.

These results illustrate why separating the two layers matters. By contrast, the lattice model estimates a positive correlation, $\widehat{\rho}=0.17$ (0.02). We interpret this as a consequence of forcing threshold dependence and residual dependence into a single correlation coefficient, where the complementarity embedded in the threshold structure dominates the negative trade-off in unobservables, giving a positive estimate.

{\tiny\begin{figure}[!t]
\centering
\begin{tikzpicture}[scale=0.25]

\draw [very thick, dotted ] (-15,-16) -- (-15,16);
\draw [very thick, dotted ] (16,-16) -- (16,16);
\draw [very thick, dotted ] (-15,16) -- (16,16);
\draw [very thick, dotted ] (-15,-16) -- (16,-16);

\draw [very thick, dotted ] (-10,13) -- (-10,16);
\draw [very thick, dotted ] (0,13) -- (0,16);
\draw [very thick, dotted ] (6,13) -- (6,16);
\draw [very thick, dotted ] (-2,-13) -- (-2,-16);
\draw [very thick, dotted ] (5,-13) -- (5,-16);
\draw [very thick, dotted ] (10,-13) -- (10,-16);
\draw [very thick, dotted ] (-13,-3) -- (-15,-3);
\draw [very thick, dotted ] (-13,9) -- (-15,9);
\draw [very thick, dotted ] (14,-10) -- (16,-10);
\draw [very thick, dotted ] (14,5) -- (16,5);

\draw [very thick ] (-2,-13) -- (-2,-3);    
\draw [very thick ] (-8,-3)  -- (-8,9);      
\draw [very thick ] (-10,9)  -- (-10,13);    
\draw [very thick ] (5,-13)  -- (5,-3);      
\draw [very thick ] (0,-3)   -- (0,13);      
\draw [very thick ] (10,-13) -- (10,5);      
\draw [very thick ] (6,5)    -- (6,13);      

\draw [very thick ] (-13,-3) -- (10,-3);     
\draw [very thick ] (10,-10) -- (14,-10);    
\draw [very thick ] (-13,9)  -- (0,9);       
\draw [very thick ] (0,5)    -- (14,5);      

\node at (-2,-3)  [circle,fill,inner sep=1.5pt]{};
\node at (-8,-3)  [circle,fill,inner sep=1.5pt]{};
\node at (-8,9)   [circle,fill,inner sep=1.5pt]{};
\node at (-10,9)  [circle,fill,inner sep=1.5pt]{};
\node at (0,-3)   [circle,fill,inner sep=1.5pt]{};
\node at (0,9)    [circle,fill,inner sep=1.5pt]{};
\node at (0,5)    [circle,fill,inner sep=1.5pt]{};
\node at (5,-3)   [circle,fill,inner sep=1.5pt]{};
\node at (6,5)    [circle,fill,inner sep=1.5pt]{};
\node at (10,-3)  [circle,fill,inner sep=1.5pt]{};
\node at (10,-10) [circle,fill,inner sep=1.5pt]{};
\node at (10,5)   [circle,fill,inner sep=1.5pt]{};
\node at (-13,-3) [circle,fill,inner sep=1.5pt]{};
\node at (-13,9)  [circle,fill,inner sep=1.5pt]{};
\node at (-2,-16) [circle,fill,inner sep=1.5pt]{};
\node at (5,-16)  [circle,fill,inner sep=1.5pt]{};
\node at (10,-16) [circle,fill,inner sep=1.5pt]{};
\node at (-10,16) [circle,fill,inner sep=1.5pt]{};
\node at (0,16)   [circle,fill,inner sep=1.5pt]{};
\node at (6,16)   [circle,fill,inner sep=1.5pt]{};
\node at (16,-10) [circle,fill,inner sep=1.5pt]{};
\node at (16,5)   [circle,fill,inner sep=1.5pt]{};

\draw[very thick,-> ] (-17,-20) -- (16,-20) node[right] {$Y^{*}_{\text{spend}}$};
\draw[very thick,-> ] (-17,-16) -- (-17,16) node[above] {$Y^{*}_{\text{stim}}$};

\node [below, thick ] at (-2,-16) {$2.77 \: (0.28)$};
\node [below, thick ] at (5,-17.5) {$3.37 \: (0.28)$};
\node [below, thick ] at (10,-16) {$3.86 \: (0.28)$};
\node [above, thick ] at (-10,16) {$2.02 \: (0.33)$};
\node [above, thick ] at (0,16) {$2.92 \: (0.31)$};
\node [above, thick ] at (8,16) {$3.51 \: (0.33)$};
\node [left, thick ] at (-4,-4) {$2.20 \: (0.30)$};   

\node [right, thick ] at (16.5,-3) {$1.65 \: (0.41)$};
\node [right, thick ] at (16.5,-10) {$0.92 \: (0.51)$};
\node [right, thick ] at (16.5,9) {$3.02 \: (0.41)$};
\node [right, thick ] at (16.5,5) {$2.58 \: (0.47)$};

\end{tikzpicture}
\caption{Threshold structure: parental investment example}
\label{fig:gridpanel2}
\end{figure}}

Finally, the index coefficients (reported in Table \ref{tabREGpsid}) differ across specifications in some places: in the education spending equation, the positive effect of child age is exacerbated in the non-lattice model, whereas the positive effects of family income, child gender (female), parental race (white), and having a gifted child are muted.

Our results have implications for the literature on investments in children's human capital. First, the complementarity assumption between financial and non-financial investment into a child's human capital as adopted in the literature is supported by our threshold structure. Second, a key finding of \cite{DelBocaFlinnWiswall2014} is that non-financial investments are more productive than financial ones, and cash subsidies to households with children have small impacts on child quality due to the relatively low impact of money investments on child outcomes. Our model does not speak to the productivity of each of the investment types. Instead, we point out that while cash subsidies themselves may have a muted effect, they may relax budget constraints. Furthermore, even if financial investments into education are less productive than non-financial ones, the complementarity between financial and non-financial investments can make financial subsidies foster non-financial investments, rather than lead to a substitution between them.

\section{Conclusion} 
\label{sec:conclusion} 
This paper develops a class of multivariate ordered discrete response models with general rectangular structures that substantially extends the reach of traditional lattice ordered choice models. The central insight is that the thresholds governing decisions in one dimension need not be functionally independent of outcomes in other dimensions. Allowing thresholds to shift as a function of the full vector of outcome indices introduces a second layer of dependence through the structure of decision rules, which is conceptually distinct from the correlation of latent unobservables. Standard lattice models collapse these two layers into one, conflating behavioral interdependence (narrow/broad bracketing) with unobserved heterogeneity and thereby misattributing the sources of covariation across outcomes.

By formalizing coherency, deriving  microfoundations, and proving identification,  our framework enables realistic modeling of complex multidimensional choices. The approach can reveal sign-changing partial effects, indirect covariate influences, and varying complementarity or substitutability, all of which are absent in lattice models. This  expands the econometric toolkit for studying multidimensional decisions and improves interpretability. As our empirical examples demonstrate, important economic contexts, such as selection markets, can be revisited, and new environments explored with the models we have introduced.

Several directions for future research are natural. Allowing for endogeneity of covariates   would broaden the range of empirical settings in which the model is applicable. A second direction concerns scalability as the number of dimensions $D$ grows.

More broadly, our framework makes a testable restriction for any lattice model. Wherever researchers have used bivariate ordered probit, bivariate probit, or correlated ordered logit models to study joint discrete decisions, the lattice assumption can now be potentially tested rather than maintained, and the two dependence layers can be estimated rather than conflated. We hope the framework opens new opportunities to revisit important questions in applied economics where the joint structure of ordered decisions carries the economic content of interest.

\bibliographystyle{ecta}
\bibliography{biblio_restud}

\clearpage 
\appendix

\setcounter{page}{1}
\setcounter{section}{0}

\begin{center}
    {\LARGE \textbf{Online Appendix}}
\end{center}

\section{Appendix A: Coherency} 

Coherency plays a critical role in both our identification and estimation routines, as well as in the conceptual framing of the model. Specifically, we interpret the model as representing the behavior of a single decision maker, for whom incoherent (i.e., logically inconsistent) choices would be implausible.

Although the literature on strategic interaction often distinguishes between incoherency and incompleteness, we subsume both under the broader notion of incoherency. From a technical standpoint, we define a model with the given set of thresholds in the latent space as \textit{coherent} (or \textit{coherent in the latent space}) if the  rectangles $R_{j_1, \ldots, j_D}$ defined in (\ref{rectangleR}) form a partition of $\mathbb{R}^D$; that is, they are mutually exclusive and collectively exhaustive over $\mathbb{R}^D$.

An equivalent way to characterize this notion of coherence is to ask: under what conditions on the latent thresholds does the observed coherency in choice probabilities and reflected in the fact that $\sum_{j_1=1}^{M_1} \cdots \sum_{j_D=1}^{M_D} P\left(Y=\left(y_{j_1}^{(1)}, \ldots, y_{j_D}^{(D)}\right) | x\right) = 1$ translate into coherence in the latent space? These conditions must be generic; that is, they should not depend on the distributional assumptions regarding observables or unobservables. 

\subsection{Bivariate case} 

We begin by examining the bivariate case, $D=2$. The main result for this case is presented in Proposition \ref{prop:coherencyD2}, which states that a bivariate model with a general rectangular structure is generically coherent in the latent space if and only if it is locally hierarchical in every local configuration. Specifically, this involves examining each  of four outcomes $(y^{(1)}_{j_1}, y^{(2)}_{j_2})$, $(y^{(1)}_{j_1+1}, y^{(2)}_{j_2})$, $(y^{(1)}_{j_1}, y^{(2)}_{j_2+1})$ and $(y^{(1)}_{j_1+1}, y^{(2)}_{j_2+1})$, where we consider incremental moves from a given point 
$(j_1,j_2)$ along one or both dimensions. The model must satisfy local hierarchies across all such configurations to ensure overall coherency. 

\smallskip

\textit{Proof of Proposition \ref{prop:coherencyD2}.}  \textit{Sufficiency.} First, observe that an incremental move from \((j_1, j_2)\) in only one dimension -- either to \((j_1+1, j_2)\) or to \((j_1, j_2+1)\) -- cannot by itself generate incoherency. This is because the definition of the rectangles \( R_{j_1, j_2} \) in equation ~(\ref{rectangleR}) ensures continuity at the thresholds along single-dimensional moves. Specifically, in dimension 1, the rectangle \( R_{j_1, j_2} \) ends at the threshold \( \alpha^{(1)}_{j_1, j_2} \), which simultaneously serves as the lower bound for the adjacent rectangle \( R_{j_1+1, j_2} \). In other words, incoherency patterns such as the one illustrated in Panel 1 of Figure ~\ref{fig:coherencyviolate} are ruled out by construction. A similar argument holds in dimension 2: the rectangle \( R_{j_1, j_2} \) terminates at threshold \( \alpha^{(2)}_{j_1, j_2} \), which also acts as the starting threshold for \( R_{j_1, j_2+1} \). Thus, the design of the threshold structure inherently prevents discontinuities along single-dimensional moves. 

To show that condition \eqref{eq:coherencybiv} makes these local alignments form a partition of the entire latent space, fix any point $(u,v)\in\mathbb R^2$.

For this argument, write $A_{ij}=\mathbf 1\{u\leq\alpha^{(1)}_{i,j}\}$, $B_{ij}=\mathbf 1\{v\leq\alpha^{(2)}_{i,j}\}$. The number of rectangles containing $(u,v)$ is
\[
\begin{aligned}
\sum_{i=1}^{M_1}\sum_{j=1}^{M_2}
\mathbf 1\{(u,v)\in R_{i,j}\}
&=
\sum_{i=1}^{M_1}\sum_{j=1}^{M_2}
(A_{ij}-A_{i-1,j})(B_{ij}-B_{i,j-1})\\&=
1+\sum_{i=1}^{M_1-1}\sum_{j=1}^{M_2-1}
(A_{ij}-A_{i,j+1})(B_{ij}-B_{i+1,j}).
\end{aligned}
\]
The second equality follows by expanding and telescoping,
using $A_{0j}=B_{i0}=0$ and
$A_{M_1,j}=B_{i,M_2}=1$.
By condition \eqref{eq:coherencybiv}, for each $(i,j)$ either
$\alpha^{(1)}_{i,j}=\alpha^{(1)}_{i,j+1}$
or $\alpha^{(2)}_{i,j}=\alpha^{(2)}_{i+1,j}$. 
Hence every summand in the last expression vanishes.
Exactly one rectangle therefore contains $(u,v)$.
Since the point was arbitrary, the rectangles are
mutually exclusive and collectively exhaustive.


\begin{figure}[!t]
\centering
\caption{Potential violations of coherency.}
\begin{tikzpicture}[scale=0.65]

\fill [opacity = 0.8, pattern=north west lines, pattern color=green] (2,2) rectangle (0,0);
\draw [very thick] (0,2) -- (2,2);
\draw [very thick] (2,0) -- (2,2);

\fill [opacity = 0.3, pattern=crosshatch, pattern color=red] (1.5,2) rectangle (5,0);
\draw [very thick] (1.5,2) -- (5,2);
\draw [very thick] (1.5,2) -- (1.5,0);

\fill [opacity = 0.3, pattern=crosshatch, pattern color=blue] (3,2) rectangle (0,5);
\draw [very thick] (3,2) -- (3,5);

\fill [opacity = 0.3, pattern=crosshatch, pattern color=black] (3,2) rectangle (5,5);

\node [below=1cm, align=flush center,text width=2cm]
        {
            Panel 1
        };
		
\end{tikzpicture}
\hskip 0.3in 
\begin{tikzpicture}[scale=0.65]
	
\fill [opacity = 0.8, pattern=north west lines, pattern color=green] (2,3) rectangle (0,0);
\draw [very thick] (0,3) -- (2,3);
\draw [very thick] (2,0) -- (2,3);

\fill [opacity = 0.3, pattern=crosshatch, pattern color=red] (5,2.5) rectangle (2,0);
\draw [very thick] (2,0) -- (2,2.5);
\draw [very thick] (2,2.5) -- (5,2.5);

\fill [opacity = 0.3, pattern=crosshatch, pattern color=blue] (2.5,5) rectangle (0,3);
\draw [very thick] (2.5,5) -- (2.5,3);
\draw [very thick] (2.5,3) -- (0,3);

\fill [opacity = 0.3, pattern=crosshatch, pattern color=black] (2.5,2.5) rectangle (5,5);
\draw [very thick] (2.5,2.5) -- (2.5,5);

\node [below=1cm, align=flush center,text width=2cm]
        {
            Panel 2
        };
\end{tikzpicture}
\hskip 0.3in 
\begin{tikzpicture}[scale=0.65]
	
\fill [opacity = 0.8, pattern=north west lines, pattern color=green] (2,2) rectangle (0,0);
\draw [very thick] (0,2) -- (2,2);
\draw [very thick] (2,0) -- (2,2);

\fill [opacity = 0.3, pattern=crosshatch, pattern color=red] (5,2.5) rectangle (2,0);
\draw [very thick] (2,0) -- (2,2.5);
\draw [very thick] (2,2.5) -- (5,2.5);

\fill [opacity = 0.3, pattern=crosshatch, pattern color=blue] (2.5,5) rectangle (0,2);
\draw [very thick] (2.5,5) -- (2.5,2);
\draw [very thick] (2.5,2) -- (0,2);

\fill [opacity = 0.3, pattern=crosshatch, pattern color=black] (2.5,2.5) rectangle (5,5);
\draw [very thick] (2.5,2.5) -- (2.5,5);

\node [below=1cm, align=flush center,text width=2cm]
        {
            Panel 3
        };

\end{tikzpicture}

\caption*{Notes: Violation of coherency in Panel 1 is ruled out by the thresholds structure in (\ref{rectangleR}). Violations of coherency in Panels 2 and 3 are not immediately ruled out by (\ref{rectangleR}). }
\label{fig:coherencyviolate}
\end{figure} 
 \smallskip 

 \noindent \textit{Necessity}. Suppose the model is coherent but condition~\eqref{eq:coherencybiv} fails.

 Fix a height $v$. For each category $i$ in dimension~1 there is exactly one $j$ with
$v\in(\alpha^{(2)}_{i,j-1},\alpha^{(2)}_{i,j}]$, and the cross-section of $R_{i,j}$ at height $v$ is the
nonempty interval $(\alpha^{(1)}_{i-1,j},\alpha^{(1)}_{i,j}]$. So the horizontal slice at $v$ consists of
$M_1$ intervals, one per $i$, and coherency means that these intervals partition $\mathbb{R}$ for every $v$.

Call $(i,j)$ \emph{failing} if both factors in condition~\eqref{eq:coherencybiv} are nonzero. Among failing pairs choose $(i,j)$ such that
$t:=\min\{\alpha^{(2)}_{i,j},\alpha^{(2)}_{i+1,j}\}$ is smallest and, among those, $j$ is largest. Then:

\begin{quote}
($\star$) If $\alpha^{(2)}_{i',k}\neq\alpha^{(2)}_{i'+1,k}$ and their smaller value is either below $t$, or equal to $t$ with $k>j$, then
$\alpha^{(1)}_{i',k}=\alpha^{(1)}_{i',k+1}$.
\end{quote}

Take heights $v^-<t<v^+$ such that $t$ is the only  threshold value in the second dimension  in $[v^-,v^+]$. All thresholds equal to $t$ are crossed simultaneously. At $v^-$, category $i$
lies in category $p_i$ of dimension~2, where $\alpha^{(2)}_{i,p_i-1}<t\le\alpha^{(2)}_{i,p_i}$. Call $i$
\emph{moving} if $\alpha^{(2)}_{i,p_i}=t$. At $v^+$ a moving category lies in $p_i+1$ and every other category
stays in $p_i$. Call the interval of $i$ in the slice at $v^-$ its \emph{preceding} interval and in the slice at $v^+$
its \emph{succeeding} interval; for non-moving $i$ they coincide. By coherency, both the preceding and the succeeding intervals
partition $\mathbb{R}$.

\emph{Step 1: the preceding intervals are consecutive.} Let $i<M_1$ and $\min\{p_i,p_{i+1}\}\le k<\max\{p_i,p_{i+1}\}$.
The category with the smaller $p$ has $\alpha^{(2)}_{\cdot,k}\ge t$, the other has $\alpha^{(2)}_{\cdot,k}<t$.
So the two thresholds differ and the smaller is below $t$, and ($\star$) gives
$\alpha^{(1)}_{i,k}=\alpha^{(1)}_{i,k+1}$. Chaining over $k$, we obtain that 
\[
\alpha^{(1)}_{i,k}\ \text{is the same for all $k$ between $p_i$ and $p_{i+1}$.}
\]
In particular the left endpoint of the preceding interval of $i+1$, $\alpha^{(1)}_{i,p_{i+1}}$, equals the right
endpoint $\alpha^{(1)}_{i,p_i}$ of the preceding interval of $i$.

\emph{Step 2: confinement.} Let $i$ be moving, and let $\{L,\dots,U\}$ be the maximal set of consecutive
categories containing $i$ with $p_h\le p_i$ throughout. Then the succeeding interval of $i$ lies in the union of the
preceding intervals of $L,\dots,U$. To see this, put $k=p_i$. If $U<M_1$, then $p_U\le k<k+1\le p_{U+1}$, so by
Step~1 $\alpha^{(1)}_{U,k+1}=\alpha^{(1)}_{U,p_U}$, the right end of that union. Since $i\le U$, the right end
$\alpha^{(1)}_{i,k+1}$ of the succeeding interval of $i$ is at most this. If $L>1$, the left end is symmetric, using
$\alpha^{(1)}_{L-1,k+1}=\alpha^{(1)}_{L-1,p_{L-1}}$ and $i-1\ge L-1$.
If $L=1$ or $U=M_1$, the corresponding bound is $-\infty$ or $+\infty$ by normalization.

\emph{Step 3: changes from categories above $j$ can be ignored.} A \emph{run} is a maximal set $\{r,\dots,s\}$ of consecutive
moving categories with a common $p=h$. Its preceding intervals tile $P=(\alpha^{(1)}_{r-1,h},\alpha^{(1)}_{s,h}]$ and
its succeeding intervals tile $W=(\alpha^{(1)}_{r-1,h+1},\alpha^{(1)}_{s,h+1}]$. If $h>j$ then $W=P$. Indeed, if $s<M_1$, maximality of the run gives
$\alpha^{(2)}_{s+1,h}\ne t=\alpha^{(2)}_{s,h}$, so ($\star$) implies
$\alpha^{(1)}_{s,h+1}=\alpha^{(1)}_{s,h}$. Likewise, if $r>1$,
$\alpha^{(2)}_{r-1,h}\ne t=\alpha^{(2)}_{r,h}$, and ($\star$) gives
$\alpha^{(1)}_{r-1,h+1}=\alpha^{(1)}_{r-1,h}$.
At an exterior boundary, the corresponding preceding and succeeding endpoints coincide by normalization to $-\infty$ or $+\infty$.
Replace the succeeding intervals of all runs with $h>j$ by their preceding
intervals. The result, the \emph{modified slice}, still partitions $\mathbb{R}$.

\emph{Step 4: runs changing from category $j$ preserve their unions.} Let the runs with $p=j$ be $\rho_1,\dots,\rho_m$ from
left to right, with preceding unions $P_1<\dots<P_m$ and succeeding unions $W_1<\dots<W_m$. Call every other category an
\emph{outsider}.
\begin{enumerate}
\item[(a)] \emph{The $W_k$ are separated.} Between $\rho_k=\{r_k,\dots,s_k\}$ and $\rho_{k+1}$ there is an
outsider, so $s_k<r_{k+1}-1$ and
$\alpha^{(1)}_{s_k,j+1}<\alpha^{(1)}_{r_{k+1}-1,j+1}$: a positive gap separates $W_k$ and $W_{k+1}$.
\item[(b)] \emph{Outsiders remain outside the preceding unions.} An outsider's interval in the modified slice is either its preceding
interval or, if it moves from category $k_0<j$ in the second dimension, lies by Step~2 in the union of preceding intervals of a consecutive set with all
$p\le k_0<j$. Such a set contains no category in a run with $p=j$, so the interval remains in the same space between the $P_k$, or beyond the first or last $P_k$, as its preceding interval. It is therefore disjoint from every $P_k$.
Moreover, if $s_k<M_1$, the outsider $s_k+1$ has an interval starting exactly
at the right end of $P_k$: this is clear for a preceding interval, and if it moves from second-coordinate category $k_0<j$ its left end is
$\alpha^{(1)}_{s_k,k_0+1}=\alpha^{(1)}_{s_k,j}$ by Step~1. Likewise, if $r_k>1$, the interval of $r_k-1$ ends exactly at the
left end of $P_k$.
\item[(c)] \emph{$P_k\subseteq W_k$.} By (b), every point of $P_k$ is covered by a run's succeeding interval, so
$P_k\subseteq\bigcup_\ell W_\ell$. By (a), the interval $P_k$ lies in a single $W_\ell$. No $W_\ell$ contains two
different $P$'s, since it would then contain the interval of an outsider lying between them (by (b)) and
overlap it. Hence $k\mapsto\ell$ is injective and increasing on $\{1,\dots,m\}$, so $\ell=k$.
\item[(d)] \emph{$W_k\subseteq P_k$.} If $W_k$ extended beyond a finite right endpoint of $P_k$, then, as $W_k\supseteq P_k$,
it would overlap the interval of $s_k+1$, which starts exactly there by (b). The argument at a finite left endpoint is symmetric; infinite endpoints coincide by normalization. So
$W_k=P_k$ for every $k$.
\end{enumerate}

\emph{Step 5: contradiction.} Return to the failing pair $(i,j)$. If
$\alpha^{(2)}_{i,j}=t<\alpha^{(2)}_{i+1,j}$, then $i$ is moving with $p_i=j$ and $i+1$ does not move from category $j$ in the second dimension,
so $i$ is the last member of its run. The right ends of $W$ and $P$ are $\alpha^{(1)}_{i,j+1}$ and
$\alpha^{(1)}_{i,j}$, which differ because $(i,j)$ is failing. If instead
$\alpha^{(2)}_{i+1,j}=t<\alpha^{(2)}_{i,j}$, then $i+1$ is the first member of its run and the left ends are
$\alpha^{(1)}_{i,j+1}\neq\alpha^{(1)}_{i,j}$. Either way $W\ne P$ for some run from second-coordinate category $j$, contradicting Step~4.
Hence condition~\eqref{eq:coherencybiv} is necessary. $\square$

\subsection{D Greater Than 2} 

For \( D > 2 \), we again examine each local configuration of the form \(\{(j_1+\ell_1, \ldots, j_D+\ell_D)\}_{\ell_d \in \{0,1\},\, d=1,\ldots,D}\). Coherency within every such local configuration ensures  global coherency of the  model. We proceed incrementally in deriving coherency conditions: starting with moves in one dimension, then in pairs of dimensions, and so on, up to moves in all \( D \) dimensions.

For concreteness, consider the case $D=3$. As in the bivariate case, any single-dimensional move within a local configuration (i.e., from \((t_1, t_2, t_3)\) to a neighboring point along one axis) cannot induce incoherency. This follows from the definition of the rectangles \( R_{t_1, t_2, t_3} \) in equation~(\ref{rectangleR}), which ensures continuity at threshold boundaries along each coordinate axis.

The next step is to consider moves that involve changes along two dimensions. For example, transitions such as \((j_1+1, j_2+1, j_3)\) to \((j_1, j_2+1, j_3+1)\), among others, must also preserve coherency. To verify this, we project the local configuration onto the relevant two-dimensional subspace, holding the remaining  coordinate fixed. For each such projection, we apply the bivariate condition from Proposition~\ref{prop:coherencyD2}. This yields conditions
\[
(\alpha_{j_1, j_2, t_3}^{(1)} - \alpha_{j_1, j_2+1, t_3}^{(1)}) 
(\alpha_{j_1+1, j_2, t_3}^{(2)} - \alpha_{j_1, j_2, t_3}^{(2)}) = 0, 
\quad t_3 \in \{j_3, j_3+1\}, 
\]
\[
(\alpha_{j_1, t_2, j_3}^{(1)} - \alpha_{j_1, t_2, j_3+1}^{(1)}) 
(\alpha_{j_1+1, t_2, j_3}^{(3)} - \alpha_{j_1, t_2, j_3}^{(3)}) = 0, 
\quad t_2 \in \{j_2, j_2+1\},
\]
\[
(\alpha_{t_1, j_2, j_3}^{(2)} - \alpha_{t_1, j_2, j_3+1}^{(2)}) 
(\alpha_{t_1, j_2+1, j_3}^{(3)} - \alpha_{t_1, j_2, j_3}^{(3)}) = 0, 
\quad t_1 \in \{j_1, j_1+1\}.
\]
This must hold for all pairs of dimensions and all fixed values of the remaining coordinate.

Finally, we must consider transitions involving simultaneous changes in all three dimensions. These correspond to movements between opposite orthants within the local configuration. To prevent incoherency, the thresholds where these orthants ``meet'' must align in at least one dimension.
For \(\boldsymbol{\ell}\in\{0,1\}^3\), define
\[
\mathbf j_d(\boldsymbol{\ell})
=
(j_1+\ell_1,\ldots,j_{d-1}+\ell_{d-1},
j_d,
j_{d+1}+\ell_{d+1},\ldots,j_3+\ell_3).
\]
Then the  coherency condition pooling all three dimensions is
\(
\prod_{d=1}^3
\left(
\alpha^{(d)}_{\mathbf j_d(\boldsymbol{\ell})}
-
\alpha^{(d)}_{\mathbf j_d(1-\boldsymbol{\ell})}
\right)
=0.
\)
It is enough to impose this for four choices of
\(\boldsymbol{\ell}\), one from each pair
\(\{\boldsymbol{\ell},1-\boldsymbol{\ell}\}\). This condition ensures that, for each pair of opposite orthants within a local configuration, the corresponding threshold surfaces meet along at least one boundary, thereby ruling out both gaps and overlaps in the latent space. 

An alternative way to impose coherency in the trivariate model is to combine pairwise disjointness with an exhaustiveness restriction. Fix any strictly increasing continuous bijection \(G:\mathbb R\to(0,1)\), such as the logistic c.d.f., with \(G(-\infty)=0\) and \(G(+\infty)=1\). This transformation is chosen by the researcher and imposes no restriction on the distribution of unobservables. For \(\mathbf j=(j_1,j_2,j_3)\), write \(\mathbf j-e_d\) for the tuple obtained by reducing its \(d\)-th coordinate by one, so that the lower and upper thresholds of \(R_{\mathbf j}\) in dimension \(d\) are \(\alpha^{(d)}_{\mathbf j-e_d}\) and \(\alpha^{(d)}_{\mathbf j}\), respectively. Pairwise disjointness is imposed by requiring, for every unordered pair of distinct cells \(\mathbf j,\mathbf k\),
\[
\prod_{d=1}^{3}
\left[
\min\left\{
G\!\left(\alpha^{(d)}_{\mathbf j}\right),
G\!\left(\alpha^{(d)}_{\mathbf k}\right)
\right\}
-
\max\left\{
G\!\left(\alpha^{(d)}_{\mathbf j-e_d}\right),
G\!\left(\alpha^{(d)}_{\mathbf k-e_d}\right)
\right\}
\right]_{+}=0,
\]
where \([a]_{+}=\max\{a,0\}\). Exhaustiveness is then imposed through
\[
\sum_{j_1=1}^{M_1}
\sum_{j_2=1}^{M_2}
\sum_{j_3=1}^{M_3}
\prod_{d=1}^{3}
\left[
G\!\left(\alpha^{(d)}_{j_1,\ldots,j_d,\ldots,j_3}\right)
-
G\!\left(\alpha^{(d)}_{j_1,\ldots,j_d-1,\ldots,j_3}\right)
\right]=1.
\]
Together, these restrictions are necessary and sufficient for coherency. 


The last approach straightforwardly extends for general $D > 3$. 

\section{Appendix B: Proofs} 
\label{sec:proofs}

\textit{\textbf{Proof of Theorem \ref{th:nonlattice_semiparametric}.}}  Fix $d$ and hold the remaining effective covariates fixed at any value for which condition (c) holds. Since $q(x)$ is the probability of a fixed ``corner'' event and
$x_{d,1:L_d}$ contains only covariates exclusive to process $d$,
$q(x)$ depends on these covariates only through $x_{d,1}+x_{d,2:L_d}\beta_{d,2:L_d}$, and is weakly monotone in this index, with the direction determined by
the chosen ``corner''.

By condition (c)(ii), there exists $p\in(0,1)$ belonging to the range
of $q(x)$ for every $x_{d,2:L_d}\in S_{d,2:L_d}$. By continuity and
strict monotonicity in condition (c)(i), for each
$x_{d,2:L_d}\in S_{d,2:L_d}$ there is a unique
$x^p_{d,1}(x_{d,2:L_d})$ such that $q(x)=p$. Equal probabilities must then correspond to equal index values.
Indeed, if two distinct index values generated the same probability
$p$, weak monotonicity would imply that the probability is constant
between them, contradicting condition (c)(i). Hence, for any  $x_{d,2:L_d},\widetilde x_{d,2:L_d}\in S_{d,2:L_d}$,
\[
x^p_{d,1}(x_{d,2:L_d})
-
x^p_{d,1}(\widetilde x_{d,2:L_d})
=
-
\left(
x_{d,2:L_d}-\widetilde x_{d,2:L_d}
\right)\beta_{d,2:L_d}.
\]
The left-hand side is identified from observable choice probabilities.
Since $S_{d,2:L_d}$ has full affine dimension $L_d-1$, these equations
uniquely identify $\beta_{d,2:L_d}$. Together with the normalization
$\beta_{d,1}=1$, this identifies $\beta_{d,1:L_d}$. Repeating the
argument for each $d$ completes the proof. $\square$

\vskip 0.1in

{\bf\textit{Proof of Theorem \ref{th:nonlattice_semiparametric2}.}} By Theorem \ref{th:nonlattice_semiparametric}, the
coefficients $\beta_{d,1:L_d}$ are identified for every
$d$. It remains to identify $\beta_{d,L_d+1:k_d}$ for
those equations with $L_d<k_d$.

Assumption \ref{assn:errors} and the definition of the ``corner'' give $q(x)=P\left(\bigcap_{h=1}^D
\{\varepsilon_h\,\kappa_h\,(a_h-x_h\beta_h)\}\right)$. Absolute continuity makes this probability continuous in
the exclusive covariates. The same is true of the probability
in condition (b), since its event is a finite union of cells.
For almost every $w\in S$, observationally equivalent models
therefore agree on both probabilities throughout the
conditional Cartesian support in (a).

Fix $d$.  Assumption \ref{assn:errors}, the ordering of thresholds and coherency give  $0\leq
F_{d,\kappa_d}(a_d-x_d\beta_d)-q(x)
\leq
1-P\left(
    \bigcap_{h\neq d}\{Y^{c_h}=y^{(h)}_{j_h}\}
    \,\middle|\,x
\right)$. Condition (b) makes the right-hand side tend to zero.
Exclusivity ensures that $x_d$ remains fixed along these
limits. Hence, 
\begin{equation}
\label{eq:th2-corner-limit}
q_d(x_d)=F_{d,\kappa_d}(a_d-x_d\beta_d).
\end{equation}
The conditional Cartesian support in (a) ensures that
the limits are taken along admissible covariate values. Convexity of the joint error support implies that each
marginal support is an interval and that every nonempty
open subinterval of its interior has positive probability.
This guarantees that each marginal c.d.f. is strictly increasing on its
support interior. It is continuous by  absolute continuity. 

Now fix $d$ with $L_d<k_d$. By (c), for almost every
$w\in S$ there is a unique value $x_{d,1}^*(w)$ in the
specified interval at which $q_d(x_d)=p_d$. Existence
follows from (c). Uniqueness follows from
\eqref{eq:th2-corner-limit}, the normalization
$\beta_{d,1}=1$, and strict monotonicity of the marginal
distribution at probability levels in $(0,1)$.

These crossings are identified from the limiting
probabilities. Since $p_d$ is common across the relevant
values of $w$, the argument of $F_{d,\kappa_d}$ at the
crossing is constant. Thus, $x_{d,1}^*(w)
+\sum_{\ell=2}^{L_d}x_{d,\ell}\beta_{d,\ell}
+x_{d,L_d+1:k_d}\beta_{d,L_d+1:k_d}
=\text{constant}$ for almost every $w\in S$, with the sum interpreted as
zero when $L_d=1$. The first two terms are identified with 
the first being the observed crossing, and the coefficients
in the second being identified by Theorem
\ref{th:nonlattice_semiparametric}.

Consider an observationally equivalent admissible
parameter system with coefficients $\widetilde\beta_d$.
Theorem \ref{th:nonlattice_semiparametric} implies $\widetilde\beta_{d,1:L_d}=\beta_{d,1:L_d}$. Subtracting its crossing equation from
the preceding equation then gives $x_{d,L_d+1:k_d}
\left(
\beta_{d,L_d+1:k_d}
-\widetilde\beta_{d,L_d+1:k_d}
\right)
=\text{constant}$ for almost every $w\in S$. If the coefficient difference were nonzero, this equality
would concentrate the conditional distribution of
$x_{d,L_d+1:k_d}$ on a proper affine hyperplane,
contradicting (a). Therefore, $\beta_{d,L_d+1:k_d}
=\widetilde\beta_{d,L_d+1:k_d}$. Repeating this argument for every equation with
$L_d<k_d$, and combining it with Theorem
\ref{th:nonlattice_semiparametric}, identifies all
$\beta_d$. $\square$

\vskip 0.1in 

{\bf\textit{Proof of Theorem \ref{th:nonlattice_semiparametric3}.}} By Theorem \ref{th:nonlattice_semiparametric2},
all $\beta_d$ and the functions $q_d(x_d)=F_{d,\kappa_d}(a_d-x_d\beta_d)$ are identified. Fix $w\in S$ for which the support
and range conditions hold.

Absolute continuity and convex support make each
marginal c.d.f. continuous and strictly increasing
on its support interior. The range condition therefore
gives a unique $x_{d,1}$ at which $q_d(x_d)=c_{0d}$
if $j_d=1$, or $q_d(x_d)=1-c_{0d}$ if $j_d=M_d$.
At this crossing, the normalization implies
$a_d=e_{0d}+x_d\beta_d$, identifying each ``corner'' threshold.

The range condition and strict monotonicity
ensure that $a_d-x_d\beta_d$ spans the entire marginal
support interior. The identified $q_d$ therefore
determines each marginal c.d.f., with endpoint and
tail values supplied by continuity and the usual
zero and one limits.

The ``corner'' probabilities satisfy
$q(x)=F_{\kappa_1,\ldots,\kappa_D}
(a_1-x_1\beta_1,\ldots,a_D-x_D\beta_D)$. Exclusivity and the conditional Cartesian support
in Theorem \ref{th:nonlattice_semiparametric2}(a)
allow these known arguments to vary jointly over
$\prod_d(\underline e_d,\overline e_d)$.
Thus the mixed distribution function is identified
on this product. Continuity and saturation limits
determine its remaining values and marginal
probabilities. Expanding complements in coordinates
with $\kappa_d$ being $>$ then recovers the ordinary joint
c.d.f., identifying the joint error distribution.
$\square$

\vskip 0.1in 

\noindent {\bf\textit{Proof of Theorem \ref{th:nonlattice_semiparametric4}. }} 

\noindent \textbf{Common argument.} We first establish a common argument for identifying
the remaining thresholds of a region $R_{\mathbf j}$.
For each $d\in A$, suppose its lower threshold is
already known if the marginal error support is bounded
below or equals $\mathbb R$, and its upper threshold
is already known otherwise. We show that all other
thresholds of the region are identified.

Note that Theorem \ref{th:nonlattice_semiparametric3} identified 
the index coefficients and $F$. Since absolute continuity
makes rectangular probabilities continuous in $t$,
then their almost surely identified values determine
them throughout the index support.

\smallskip
\noindent\textit{Limiting rectangular probabilities.} By coherence, dropping the restrictions in \(B\) changes the cell probability by at most the probability that a selected extreme outcome in \(B\) fails to occur. Condition (a) makes this difference tend to zero, so
\[
q(t_A)=P\left(
\alpha^{(d)}_{j_1,\ldots,j_d-1,\ldots,j_D}-t_d
<\varepsilon_d\leq
\alpha^{(d)}_{\mathbf j}-t_d,\quad d\in A
\right).
\]
The same argument identifies \(q_h(t_A,t_h)\), keeping \(t_h\) fixed and retaining the one-sided restriction in coordinate \(h\).

\smallskip
\noindent\textit{Thresholds in $C$.}
Let
\[
\Delta_d=
\alpha^{(d)}_{\mathbf j}
-\alpha^{(d)}_{j_1,\ldots,j_d-1,\ldots,j_D}>0,
\qquad d\in A,
\]
be the width of the outcome region in coordinate \(d\). For each \(d\in U\), the index \(t_d\) ranges over all of \(\mathbb R\). Holding the error fixed, the values of \(t_d\) satisfying its threshold restriction form an interval of length \(\Delta_d\). Integrating over these coordinates therefore gives
$$r(t_C)
=\int_{\mathbb R^{|U|}}q(t_A)\,dt_U =\left(\prod_{d\in U}\Delta_d\right)
P\left(
\alpha^{(d)}_{j_1,\ldots,j_d-1,\ldots,j_D}-t_d
<\varepsilon_d\leq
\alpha^{(d)}_{\mathbf j}-t_d,\quad d\in C
\right).$$
(Interchanging integration and expectation is valid by Tonelli’s theorem as the integrand is nonnegative.)
Suppose \(C\ne\varnothing\), and let \(\mathcal E_C\) be the support of \(\varepsilon_C\). The last expression defines \(r\) for all \(t_C\), including values outside the observed index range. 
The closure of the set where this expression is positive is 
\[
S_C=
\prod_{d\in C}
\left[
\alpha^{(d)}_{j_1,\ldots,j_d-1,\ldots,j_D},
\alpha^{(d)}_{\mathbf j}
\right]-\mathcal E_C,  
\]
with subtraction understood in Minkowski sense. This characterization follows from absolute continuity and the definition of support. The set \(S_C\) is convex.
Condition (b) provides \(|C|\) supporting hyperplanes: planes that touch \(S_C\) and leave the whole set on one side. Write their contact points as \(z^i\) and their outward normal vectors as \(v^i\). These contact points lie inside the observed index range, so the local boundary of the positive set of \(r\) identifies the planes. Any alternative thresholds producing the same observed probabilities must have the same local boundary. By convexity, these planes also support the alternative set globally.
At each contact point, \((v^i)^\top z^i\) is the largest value of \((v^i)^\top t_C\) over \(S_C\). Maximizing over the threshold rectangle selects its upper endpoint when \(v_d^i>0\) and its lower endpoint when \(v_d^i<0\). Thus,
\[
\sum_{\substack{d\in C\\v_d^i>0}}
v_d^i\alpha^{(d)}_{\mathbf j}
+
\sum_{\substack{d\in C\\v_d^i<0}}
v_d^i\alpha^{(d)}_{j_1,\ldots,j_d-1,\ldots,j_D}
=
(v^i)^\top z^i+
\inf_{e_C\in\mathcal E_C}(v^i)^\top e_C.
\]
The sign restrictions in condition (b) ensure that the infima are finite and that each equation involves the thresholds still to be identified. The right-hand sides are known. Since the normal vectors are linearly independent, these \(|C|\) linear equations uniquely determine the \(|C|\) remaining thresholds. Both thresholds in every coordinate of \(C\) are therefore identified.

\smallskip
\noindent\textit{Thresholds in $U$.}
If $U\ne\varnothing$ and $C\ne\varnothing$, choose
an observed $t_C$ with $r(t_C)>0$.
For $d\in U$, let $H_d$ be the conditional c.d.f.
of $\varepsilon_d$ given $\alpha^{(k)}_{j_1,\ldots,j_k-1,\ldots,j_D}-t_k
<\varepsilon_k\leq
\alpha^{(k)}_{\mathbf j}-t_k$, $ k\in C$,
and let $\eta_d$ be its median.
When $C=\varnothing$, use the unconditional marginal
c.d.f. and its median.

These distributions are identified as $F$
and the conditioning intervals are known.
They are absolutely continuous and have interval
support, by convexity of the error support.
Integration over the unrestricted indices gives
\[
\frac{
\displaystyle
\int_{\mathbb R^{|U|}}
\mathbf 1\left\{
t_d\geq
\alpha^{(d)}_{j_1,\ldots,j_d-1,\ldots,j_D}-\eta_d
\right\}q(t_A) \, dt_U
}{
\displaystyle
\int_{\mathbb R^{|U|}}q(t_A) \, dt_U
}
=
\frac{1}{\Delta_d}\int_0^{\Delta_d}
H_d(\eta_d+s)\,ds
=
\int_0^1H_d(\eta_d+s\Delta_d)\,ds.
\]
The denominator equals $\prod_{d\in U}\Delta_d$
times the positive probability of the conditioning
event, interpreted as one when $C=\varnothing$.
It is therefore positive and finite.
Since the lower threshold is known, the left-hand
side is identified. The right-hand side is strictly increasing in
$\Delta_d>0$. This is because  for any two different widths,
the c.d.f. values are strictly ordered for all
sufficiently small $s>0$, when both arguments lie
in the support interior.
Thus $\Delta_d$, and hence the upper threshold,
is identified for every $d\in U$.

\smallskip
\noindent\textit{Thresholds in $B$.}
All interval restrictions in $A$ are now known.
For each $h\in B$, the ratio $q_h(t_A,t_h)/q(t_A)$
is the conditional c.d.f. of $\varepsilon_h$
when $j_h=1$, and its conditional survival function
when $j_h=M_h$, evaluated at the remaining finite
threshold minus $t_h$.
The conditioning event is the known rectangular
restriction on $\varepsilon_A$.

Condition (c) gives an observed point where
this event has positive probability and the ratio
lies strictly between zero and one.
The conditional distribution is identified,
absolutely continuous, and has interval support.
Its c.d.f. is therefore continuous and strictly
increasing on its support interior, so inversion
identifies the remaining threshold.
When $A=\varnothing$, the same argument uses
the unconditional marginal distribution and $q=1$.

\paragraph*{Stage 1} Pick any ``corner'' rectangular region $R_{j_1,\ldots,j_D}$, $j_d \in \{1,M_d\}$ for each $d=1, \ldots, D$. It is described by $D$  unknown thresholds  $\alpha^{(d)}_{j_1,\ldots, j_{d-1}, r_d(j_d), j_{d+1},\ldots,j_D}$, 
where 
$$r_d(j_d)=\left\{ 
\begin{array}{l}
1, \text{ if  } j_d=1, \\
M_d-1, \text{ if } j_d=M_d. 
\end{array}
\right.$$
Our goal is to identify them. Once again, it is convenient to associate a certain direction for the distribution of $\boldsymbol{\varepsilon}$ with this ``corner''. If $j_d=1$ we take $\kappa_d $ to be $\leq$, and if $j_d=M_d$ we take $\kappa_d $ to be $>$,

For a ``corner'' region, $A=\varnothing$.
Condition (a) identifies each limiting marginal
probability $q_h(t_h)$.
Condition (c) makes it strictly between zero and one
at an observed index value.
Since the marginal error distribution is identified
by Theorem \ref{th:nonlattice_semiparametric3},
its strict monotonicity on its support interior
identifies the corresponding finite threshold.

Thus, this stage identifies thresholds shaping all the ``corner'' rectangular regions. 

\paragraph*{Stage 2}  In this stage we continue to consider  rectangular regions near the border. In Stage 1, each ``corner'' region had $D$ thresholds
fixed at $\infty$ or $-\infty$. At least one threshold will be finite and known from
Stage 1 or an earlier region in this stage.

Namely, fix $d$ and consider a border rectangular region $R_{j_1,\ldots,j_{d-1},q_d,j_{d+1},\ldots, j_D}$ where $j_h \in \{1,M_h\} $ for $h \neq d$ and $q_d=2, \ldots, M_d-1$. In the threshold expressions below, set $j_d=q_d$.  The region is described by $D+1$ thresholds $\alpha^{(d)}_{j_1,\ldots,j_{d-1},q_d-1,j_{d+1},\ldots, j_D}$, $\alpha^{(d)}_{j_1,\ldots,j_{d-1},q_d,j_{d+1},\ldots, j_D}$ (in dimension $d$) and $\alpha^{(h)}_{j_1,\ldots,j_{h-1},r_h(j_h),j_{h+1},\ldots,j_D}$, $h \neq d$. For $q_d=2$ and $q_d=M_d-1$ we only have at most $D$ unknown thresholds since $\alpha^{(d)}_{j_1,\ldots,j_{d-1},1,j_{d+1},\ldots, j_D}$ and $\alpha^{(d)}_{j_1,\ldots,j_{d-1},M_d-1,j_{d+1},\ldots, j_D}$ have been identified in Stage 1. The idea is then to proceed sequentially from $q_d=2$ in increasing order if the marginal error support is bounded below or equals $\mathbb R$, and from $q_d=M_d-1$ in decreasing order otherwise.

Within this stage, note that we can identify the remaining
finite threshold in dimension $d$ in the following way. Choosing, once again, $\kappa_h$ to be $\leq$ if $j_h=1$ and to be $>$ if $j_h=M_h$, $h \neq d$, obtain that  
\begin{multline*} P \left(Y^{c_d} = y^{(d)}_{q_d},  \, \cap_{h \neq d} \left(Y^{c_h}=y^{(h)}_{j_h} \right) | \, x \right)  =  P\bigg( \alpha^{(d)}_{j_1,\ldots,j_{d-1},q_d-1,j_{d+1},\ldots, j_D} <Y^{*c_d} \\ \leq 	\alpha^{(d)}_{j_1,\ldots,j_{d-1},q_d,j_{d+1},\ldots, j_D},   \cap_{h \neq d} \left(Y^{*c_h} \; \; \kappa_h \; \; \alpha^{(h)}_{j_1,\ldots,j_{h-1},r_h(j_h),j_{h+1},\ldots,j_D} \right) | \,x\bigg).  
\end{multline*}

Holding $t_d=x_d \beta_d$ fixed, send each $t_h=x_h \beta_h$, $h\neq d$,
to the lower endpoint of $I_h$ if $j_h=1$
and to its upper endpoint if $j_h=M_h$.
By condition (a), we identify 
\begin{multline*}
\lim_{\substack{
t_h\to\text{respective endpoint of }I_h\\h\neq d}}
P\left(
Y^{c_d}=y^{(d)}_{q_d},\,
\bigcap_{h\neq d}\left(Y^{c_h}=y^{(h)}_{j_h}\right)
| t
\right) =  \\	F_{d,\leq}\big(
\alpha^{(d)}_{j_1,\ldots,j_{d-1},q_d,j_{d+1},\ldots, j_D}- x_d\beta_d\big) - F_{d,\leq}\big(\alpha^{(d)}_{j_1,\ldots,j_{d-1},q_d-1,j_{d+1},\ldots, j_D} -x_d\beta_d\big). 
\end{multline*}
The opposite threshold is already identified.
Applying the interior-threshold part of the common
argument with $A=\{d\}$ identifies the remaining threshold.
Proceeding sequentially in the specified direction
identifies both finite thresholds of every region
considered in this stage.

Thus, in each rectangular region $R_{j_1,\ldots,j_{d-1},q_d,j_{d+1},\ldots, j_D}$ under consideration in this Stage 2 we now only have at most \(D-1\) unknown thresholds $\alpha^{(h)}_{j_1,\ldots,j_{h-1},r_h(j_h),j_{h+1},\ldots,j_D}$, $h \neq d$. 

Now we also fix one $h$ such that $h \neq d$ and show the threshold $\alpha^{(h)}_{j_1,\ldots,j_{h-1},r_h(j_h),j_{h+1},\ldots,j_D}$ is identified. For this, we consider $P (Y^{c_d} = y^{(d)}_{q_d}  \, \cap_{h \neq d} \left(Y^{c_h}=y^{(h)}_{j_h} \right) | \, x )$ again (as above) but now hold $t_d,t_h$ fixed and take limits only
in coordinates $\tilde h\notin\{d,h\}$.
For each such $\tilde h$, send $t_{\tilde h}$
to the lower endpoint of $I_{\tilde h}$ if
$j_{\tilde h}=1$ and to its upper endpoint if
$j_{\tilde h}=M_{\tilde h}$.
By condition (a), we identify 
\begin{multline} F_{d,h; \leq, \kappa_h} \left(\alpha^{(d)}_{j_1,\ldots,j_{d-1},q_d,j_{d+1},\ldots, j_D}- x_d\beta_d, \alpha^{(h)}_{j_1,\ldots,j_{h-1},r_h(j_h),j_{h+1},\ldots,j_D} - x_{h} \beta_{h}\right) \\ 
- F_{d,h; \leq, \kappa_h} \left(\alpha^{(d)}_{j_1,\ldots,j_{d-1},q_d-1,j_{d+1},\ldots, j_D}- x_d\beta_d, \alpha^{(h)}_{j_1,\ldots,j_{h-1},r_h(j_h),j_{h+1},\ldots,j_D} - x_{h} \beta_{h}\right), 
\label{eq-ident-threshold-nonlattice1}
\end{multline}	
where $F_{d,h; \leq, \kappa_h} (a_1,a_2) \equiv P(\varepsilon_d \leq a_1, \varepsilon_h \; \kappa_h \;  a_2)$. Function $F_{d,h; \leq, \kappa_h} $ is identified as an implication of Theorem \ref{th:nonlattice_semiparametric3}. Thus, in the known limit (\ref{eq-ident-threshold-nonlattice1}) there is only one unknown threshold $\alpha^{(h)}_{j_1,\ldots,j_{h-1},r_h(j_h),j_{h+1},\ldots,j_D}$. 

Condition (c) provides an observed point at which
the probability in \eqref{eq-ident-threshold-nonlattice1}
lies strictly between zero and the corresponding
one-dimensional interval probability.
Their ratio is the conditional c.d.f. or survival
function of $\varepsilon_h$, given the known interval
restriction on $\varepsilon_d$.
Its continuity and strict monotonicity on its support
interior identify the remaining threshold.

\paragraph*{Stage 3} In this stage we continue to consider  rectangular regions near the border. Compared with Stages 1 and 2, now we will have only $D-2$ known thresholds  fixed at $\infty$ or $-\infty$. Two other thresholds will be finite and known from
Stage 2 or earlier regions in this stage.

Consider a rectangular border region $R_{j_1,\ldots,j_{d_1-1},q_{d_1},j_{d_1+1},\ldots, j_{d_2-1},q_{d_2},j_{d_2+1}, \ldots,  j_D}$ for some $d_1$, $d_2$ such that $d_1<d_2$  and 
where $j_h \in \{1,M_h\} $ for $h \neq d_1, d_2$ and $q_{d_j}=2, \ldots, M_{d_j}-1$, $j=1,2$. In this border region we allow discrete processes in dimensions $d_1$ and $d_2$ to  take any of their interior categories whereas in all the other dimensions they take their boundary values. Without loss of generality, take $d_1=1$ and $d_2=2$. 

Region $R_{q_1, q_2, j_3, \ldots,  j_D}$, where $j_h \in \{1,M_h\} $ for $h \geq 3$ is described by $D+2$ thresholds $\alpha^{(1)}_{q_1-1,q_2, j_3,\ldots, j_D}$, $\alpha^{(1)}_{q_1,q_2, j_3,\ldots, j_D}$, $\alpha^{(2)}_{q_1,q_2-1, j_3,\ldots, j_D}$, $\alpha^{(2)}_{q_1,q_2, j_3,\ldots, j_D}$  (in dimensions 1 and 2) and  for $h\geq3$ (other dimensions) $\alpha^{(h)}_{q_1,q_2,j_3,\ldots,j_{h-1},r_h(j_h),
j_{h+1},\ldots,j_D}$. We proceed sequentially in each coordinate $d=1,2$. 
If the marginal error support is bounded below or
equals $\mathbb R$, start with $q_d=2$ and increase
$q_d$. Otherwise, start with $q_d=M_d-1$ and decrease
$q_d$. Within this stage, order the regions by the
sum of their positions in these coordinatewise orders.

For concreteness, suppose the coordinatewise orders
in dimensions 1 and 2 are both increasing.
We first consider $q_1=2$, $q_2=2$.
Using results of Stage 2, we find that we only deal
with at most $D$ unknown thresholds
$\alpha^{(1)}_{2,2,j_3,\ldots,j_D}$,
$\alpha^{(2)}_{2,2,j_3,\ldots,j_D}$, and
$\alpha^{(h)}_{2,2,j_3,\ldots,j_{h-1},r_h(j_h),
j_{h+1},\ldots,j_D}$, $h\geq3$.
Notice that, due to coherency requirements,
strictly fewer than $D$ of these thresholds may
be unknown. However, our identification argument
allows all $D$ to be unknown.

Consider the observed probability
\[
P\left(
Y^{c_1}=y^{(1)}_2,\,
Y^{c_2}=y^{(2)}_2,\,
\bigcap_{h\geq3}
\left(Y^{c_h}=y^{(h)}_{j_h}\right)
\mid x
\right).
\]
Holding $t_1,t_2$ fixed, send each $t_h$, $h\geq3$,
to the lower endpoint of $I_h$ when $j_h=1$
and to its upper endpoint when $j_h=M_h$.
By condition (a), this identifies
\[
\sum_{\ell_1=0}^1\sum_{\ell_2=0}^1
(-1)^{\ell_1+\ell_2}
F_{1,2;\leq\leq}
\left(
\alpha^{(1)}_{1+\ell_1,2,j_3,\ldots,j_D}
-x_1\beta_1,\,
\alpha^{(2)}_{2,1+\ell_2,j_3,\ldots,j_D}
-x_2\beta_2
\right)
\in[0,1]
\]
for $(t_1,t_2)\in I_1\times I_2$.

Apply the common argument established at the
beginning of the proof with $A=\{1,2\}$.
The opposite thresholds
$\alpha^{(1)}_{1,2,j_3,\ldots,j_D}$ and
$\alpha^{(2)}_{2,1,j_3,\ldots,j_D}$
are already known from Stage 2.
Condition (b) identifies the remaining thresholds
in coordinates 1 and 2, and condition (c) identifies
the finite thresholds in coordinates $h\geq3$.
Thus, all the thresholds of $R_{2,2,j_3,\ldots,j_D}$
are identified.

Building on this result, we next look at the
rectangles $R_{3,2,j_3,\ldots,j_D}$ and
$R_{2,3,j_3,\ldots,j_D}$, whenever these indices
are interior, and identify their thresholds in
the same manner. We then consider the remaining
rectangles in increasing order of $q_1+q_2$.
The opposite thresholds are known from Stage 2
or an earlier rectangle in this order.
We continue until all thresholds of the rectangles
considered in this stage are identified.

If either coordinatewise order is decreasing,
start with $q_d=M_d-1$ in that coordinate and
proceed downward, using its already identified
upper threshold as the opposite threshold.
The same argument applies, with rectangles ordered
by the sum of their positions in the respective
coordinatewise orders.

\paragraph*{Stage 4} In this stage we build on the results of previous stages and  consider rectangular border regions where we allow discrete processes in three dimensions $d_1$, $d_2$, $d_3$ to take their interior categories whereas in all the other dimensions they take their boundary values. To analyze the identification of threshold considered in this stage, without a loss of generality we can take $d_1=1$, $d_2=2$, $d_3=3$. 

Consider $R_{q_1,q_2,q_3,j_4,\ldots,j_D}$, where
$2\leq q_d\leq M_d-1$ for $d=1,2,3$ and
$j_h\in\{1,M_h\}$ for $h\geq4$.
Proceed in the coordinatewise orders specified above.
The opposite threshold in each of the first three
coordinates is known from Stage 3 or an earlier
region in this stage.

Condition (a) identifies the limiting
three-dimensional rectangular probability.
The common argument identifies its remaining
thresholds, first in $C$ and then in $U$.
Condition (c) identifies the remaining thresholds
in coordinates $h\geq4$.
Repeating this procedure identifies every region
considered in this stage.

\paragraph*{Stages 5 to $D+1$.} 
Stage $m+1$ considers regions with
$m$ interior outcome coordinates.
In each coordinate, proceed in increasing outcome
order if the marginal error support is bounded below
or equals $\mathbb R$, and in decreasing order otherwise. Within a stage, order regions by the sum of their positions in these coordinatewise orders.

The opposite threshold in each interior coordinate
is identified from the adjacent region earlier in
this order, or from the preceding stage when that
adjacent outcome is a boundary category.
The common argument above identifies the remaining
interior thresholds and then the boundary thresholds.
This completes the induction. $\square$


\section{Appendix C: Estimation in semiparametric model}

\label{sec:semiest}

The majority of existing estimation approaches for univariate semiparametric ordered response models  (either under full stochastic independence of the unobservable from covariates or under a slightly more general formulation with a multiplicative scale function like in \citealp{chenkhan2003})  do not extend to multivariate models with general rectangular structures.\footnote{Most of these methods, however, can be extended to multivariate lattice models, as discussed in \cite{km2025lattice}.} For example, in the two-stage approach of \cite{klein2002}, which first estimates the index parameter using kernel density estimates of the conditional choice probabilities and then identifies the threshold parameters through shift restrictions, adapting the shift restrictions to non-lattice settings proves challenging. The same limitation applies to \cite{liu2019}. Similarly, the approaches proposed in \cite{lewbel2000, lewbel2003} do not generalize to non-lattice frameworks. Moreover, the strategy of \cite{chenkhan2003} cannot be directly applied to models with general rectangular structures to estimate all finite-dimensional parameters of interest. For example, under the
strict monotonicity condition in Theorem \ref{th:nonlattice_semiparametric},  equality of lower ``corner'' probabilities 
$ P(Y^{(1)}=y^{(1)}_{1}, Y^{(2)}=y^{(2)}_{1} \ | \ x_1,x_2) = P(Y^{(1)} = y^{(1)}_{1}, Y^{(2)}=y^{(2)}_{1} \ | \ \tilde{x}_1, \tilde{x}_2)$ implies that $\tilde{x}_d \beta_d = x_d \beta_d$ when $x_{-d} =\tilde{x}_{-d}$, for $d=1,2$. This comparison can exploit variation in exclusive covariates, but does not by itself identify the coefficients on shared covariates.

If we were interested in the estimation of parameters on exclusive covariates, we could proceed in many ways. We could combine pairwise differences \citep{honore2005} with maximum rank correlation (MRC) estimation \citep{han87} or any other method suitable for single-index models to estimate $\beta_{d,1:L_d}$ for exclusive regressors. Pairwise differencing would restrict attention to comparisons where shared covariates and all other dimensions are similar, while MRC would exploit the stochastic dominance, which is behind the result in Theorem \ref{th:nonlattice_semiparametric}. 

We found that the only existing approach that can potentially be directly extended to non-lattice models and which permits the estimation of all the index parameters as well as all the thresholds and the unknown joint distribution of unobservables is  \cite{coppejans2007}. It is a spline-based sieve likelihood approach
and, analogously to this paper, relies on the assumption of independence of the unobservable from covariates as well as enough variation in covariates. Focusing on the bivariate case for clarity, we describe how to extend \cite{coppejans2007} to our setting. 

Consider a random sample $\left\{(y^{(1)(i)}, y^{(2)(i)}, x_1^{(i)}, x_2^{(i)}) \right\}_{i=1}^N$. First, create an estimate of the probability of the bivariate latent process falling into the rectangle $R_{j_1,j_2}$: 
\begin{align*}
\widehat{\ell}^{(i)}_{j_{1},j_{2}} =& \, \sum_{\ell_1=0}^1 \sum_{\ell_2=0}^1 (-1)^{\ell_1+\ell_2} \widehat{F}\left(a_{j_{1}-\ell_1,j_{2}}^{(1)} - x_1^{(i)} b_1, a_{j_{1},j_{2}-\ell_2}^{(2)} - x_2^{(i)} b_2\right) 
\end{align*} 
\cite{coppejans2007} deals with the univariate $\widehat{F}$ and models it using a quadratic \textit{B}-spline whose coefficients are estimated jointly with the index and threshold parameters. In the multivariate case, we can model $\widehat{F}$ using tensor-product \textit{B}-splines and estimate their coefficients jointly with thresholds and index parameters.  The estimation proceeds by 
\[
\max_{\theta, \widehat{F}} \mathcal{L}(\theta) = \frac{1}{N} \sum_{i=1}^{N} \sum_{j_{1}=1}^{M_{1}} \sum_{j_{2}=1}^{M_{2}} 1\left[(y^{(1)(i)}, y^{(2)(i)}) = (y_{j_{1}}^{(1)}, y_{j_{2}}^{(2)})\right] \log(\widehat{\ell}^{(i)}_{j_{1},j_{2}}),
\]
where $\theta$ combines all the index and threshold parameters. In the bivariate setting, the basis functions in tensor-product representation for $\widehat{F}$ consist of $S_1 \cdot S_2$ products 
$\mathcal{R}_{1;s_1,S_1}(e_1;q_1) \cdot \mathcal{R}_{2;s_2,S_2}(e_2;q_2)$,  $s_1 = 1,\dots,S_1$,  $s_2 = 1,\dots,S_2$,
of univariate \textit{B}-splines 
evaluated at specific $(e_1,e_2)$. Here $q_d$ is the degree in dimension $d=1,2$. There is a system of knots in each dimension which is not explicitly  incorporated in our notation. The full tensor-product \emph{B}-spline for $\widehat{F}(e_1,e_2)$ is a linear combination of these basis functions:
\[
\sum_{s_1=1}^{S_1} \sum_{s_2=1}^{S_2} h_{s_1 s_2} \mathcal{R}_{1;s_1,S_1}(e_1;q_1) \mathcal{R}_{2;s_2,S_2}(e_2;q_2),
\]
with coefficients $\{h_{s_1 s_2}\}$  constrained to impose c.d.f.-type shape restrictions on the spline domain. Namely, (a) 
 \textit{monotonicity} in each dimension is enforced by
    \begin{align*}
    h_{s_1 s_2} &\leq h_{s_1+1, s_2}, \quad s_1 = 1,\dots,S_1-1,\; s_2 = 1,\dots,S_2, \\
    h_{s_1 s_2} &\leq h_{s_1, s_2+1}, \quad s_2 = 1,\dots,S_2-1,\; s_1 = 1,\dots,S_1;
    \end{align*}
(b) c.d.f. \textit{bounds} are maintained by $0 \leq h_{s_1 s_2} \leq 1 \quad \text{for all } s_1, s_2$ and (c) the 2-increasing property is enforced through $$h_{s_1+1, s_2+1} - h_{s_1+1, s_2} - h_{s_1, s_2+1} + h_{s_1, s_2} \geqslant 0
$$ for $s_1=1,\dots,S_{1}-1$ and $s_2=1,\dots,S_{2}-1$.\footnote{For more details on shape constraints in tensor-product B-splines, see \cite{bk2022}.} Additional boundary normalizations are
imposed to obtain a proper c.d.f. normalization. Coherency requires additional constraints on thresholds. In the bivariate case, these can be imposed via a penalty term added to the objective, for example, $-\lambda_{N} \left(\alpha_{j_{1}+1,j_{2}}^{(2)} - \alpha_{j_{1},j_{2}}^{(2)}\right)^2 \cdot \left(\alpha_{j_{1},j_{2}+1}^{(1)} - \alpha_{j_{1},j_{2}}^{(1)}\right)^2$, 
for a large $\lambda_N > 0$ to ensure the constraints are imposed in practice.

This construction defines a proposed sieve likelihood estimator. Its consistency and limiting distribution require a separate analysis, including conditions on sieve growth and the local geometry of the coherency restrictions. These results are not established here.

\section{Appendix D: Additional simulation results}
\label{AdditionalSims}
In Table \ref{tabSIM1CUT}, we provide the results for the thresholds in simulation Design 2. It is evident how poorly the lattice model does in this case, relative to essentially no bias in the non-lattice model.

\begin{table}[tbp]
\caption{Simulation results: Design 2 thresholds} \label{tabSIM1CUT}
\centering
\begin{threeparttable}
\begin{tabular}{cccccccccc}
\toprule 
\textbf{Parameter} & \textbf{Truth} &&& \multicolumn{1}{c}{\textbf{Non-lattice model}}  &&&\multicolumn{1}{c}{\textbf{Lattice model}} \\
\hline 
$\alpha_{1,1}^{(1)}$ & -3.25 &&& -3.26 (0.11) &&& \multirow{3}{*}{-1.48 (0.04)} \\
$\alpha_{1,2}^{(1)}$ &  -3.25 &&& -3.25 (0.10) &&&                                  \\
$\alpha_{1,3}^{(1)}$ & -0.5  &&& -0.50 (0.07) &&&                                   \\
\hline
$\alpha_{2,1}^{(1)}$ & 0.5   &&& 0.51 (0.09)  &&& \multirow{3}{*}{1.59 (0.04)}  \\
$\alpha_{2,2}^{(1)}$ & 1     &&& 0.99 (0.12)  &&&                                   \\
$\alpha_{2,3}^{(1)}$ & 5     &&& 5.02  (0.14)   &&&                                   \\
\hline
$\alpha_{3,1}^{(1)}$ & 8     &&& 8.03 (0.19)  &&& \multirow{3}{*}{5.12 (0.09)}  \\
$\alpha_{3,2}^{(1)}$ &  8    &&& 8.03 (0.19)  &&&                                   \\
$\alpha_{3,3}^{(1)}$ &   8   &&& 8.03 (0.19)  &&&                                   \\
\hline
$\alpha_{1,1}^{(2)}$ & -4    &&& -3.98 (0.21) &&& \multirow{4}{*}{-1.10 (0.04)} \\
$\alpha_{2,1}^{(2)}$ & -2    &&& -2.02 (0.11) &&&                \\
$\alpha_{3,1}^{(2)}$ &  -2   &&& -1.99 (0.08)  &&&                                   \\
$\alpha_{4,1}^{(2)}$ & 0     &&& -0.01 (0.09) &&&                                   \\
\hline
$\alpha_{1,2}^{(2)}$ & 0.5   &&& 0.50 (0.05)  &&& \multirow{4}{*}{0.90 (0.04)}  \\
$\alpha_{2,2}^{(2)}$ &  0.5  &&& 0.50 (0.05)  &&&                                   \\
$\alpha_{3,2}^{(2)}$ &  0.5  &&& 0.50 (0.05)  &&&                                   \\
$\alpha_{4,2}^{(2)}$ & 4     &&& 4.00 (0.15)  &&&   \\
\bottomrule 
\end{tabular} \vspace*{-0.1cm}
\begin{tablenotes}[flushleft]
\footnotesize
\item Notes: Sample means and sample standard deviations (in parentheses) of the estimates of the Design 2 threshold parameters, over 250 repeated samples. 
\end{tablenotes}
\end{threeparttable}
\end{table}

\section{Appendix E: Additional applications exhibits}

\subsection{Cryptocurrency}

{\tiny\begin{figure}[ht]
\centering
\begin{tikzpicture}[scale=0.3]

\draw [very thick, dotted ] (-12,-14) -- (-12,8);
\draw [very thick, dotted ] (15,-14) -- (15,8);
\draw [very thick, dotted ] (-12,8) -- (15,8);
\draw [very thick, dotted ] (15,-14) -- (-12,-14);

\draw [very thick, dotted ] (-4,-12) -- (-4,-14);
\draw [very thick, dotted ] (-4,6.5) -- (-4,8);
\draw [very thick, dotted ] (4,-12) -- (4,-14);
\draw [very thick, dotted ] (4,6.5) -- (4,8);
\draw [very thick, dotted ] (10,-12) -- (10,-14);
\draw [very thick, dotted ] (10,6.5) -- (10,8);

\draw [very thick, dotted ] (-10,-10) -- (-12,-10);
\draw [very thick, dotted ] (-10,3) -- (-12,3);
\draw [very thick, dotted ] (13,-2) -- (15,-2);
\draw [very thick, dotted ] (13,5) -- (15,5);

\draw [very thick ] (-4,-12) -- (-4,6.5);
\draw [very thick ] (4,-12) -- (4,6.5);
\draw [very thick ] (10,-12) -- (10,6.5);

\draw [very thick ] (-10,-10) -- (-4,-10);       
\draw [very thick ] (-4,-6) -- (4,-6);            
\draw [very thick ] (4,-4) -- (10,-4);            
\draw [very thick ] (10,-2) -- (13,-2);           

\draw [very thick ] (-10,3) -- (-4,3);            
\draw [very thick ] (-4,-0.5) -- (4,-0.5);        
\draw [very thick ] (4,0.5) -- (10,0.5);          
\draw [very thick ] (10,5) -- (13,5);             

\node at (-12,-10) [circle,fill,inner sep=1.5pt]{};
\node at (-12,3) [circle,fill,inner sep=1.5pt]{};
\node at (-4,-10) [circle,fill,inner sep=1.5pt]{};
\node at (-4,-6) [circle,fill,inner sep=1.5pt]{};
\node at (-4,3) [circle,fill,inner sep=1.5pt]{};
\node at (-4,-0.5) [circle,fill,inner sep=1.5pt]{};
\node at (4,-6) [circle,fill,inner sep=1.5pt]{};
\node at (4,-4) [circle,fill,inner sep=1.5pt]{};
\node at (4,-0.5) [circle,fill,inner sep=1.5pt]{};
\node at (4,0.5) [circle,fill,inner sep=1.5pt]{};
\node at (10,-4) [circle,fill,inner sep=1.5pt]{};
\node at (10,-2) [circle,fill,inner sep=1.5pt]{};
\node at (10,0.5) [circle,fill,inner sep=1.5pt]{};
\node at (10,5) [circle,fill,inner sep=1.5pt]{};
\node at (15,-2) [circle,fill,inner sep=1.5pt]{};
\node at (15,5) [circle,fill,inner sep=1.5pt]{};
\node at (-4,-14) [circle,fill,inner sep=1.5pt]{};
\node at (4,-14) [circle,fill,inner sep=1.5pt]{};
\node at (10,-14) [circle,fill,inner sep=1.5pt]{};
\node at (-4,8) [circle,fill,inner sep=1.5pt]{};
\node at (4,8) [circle,fill,inner sep=1.5pt]{};
\node at (10,8) [circle,fill,inner sep=1.5pt]{};

\draw[very thick,-> ] (-14,-17.75) -- (15,-17.75) node[right] {$Y^{*\text{Familiarity}}$};
\draw[very thick,-> ] (-14,-14) -- (-14,8) node[above] {$Y^{*\text{Opinion}}$};

\node [below, thick ] at (-4,-14) {$-0.55 \: (0.10)$};
\node [below, thick ] at (4,-15.5) {$0.39 \: (0.10)$};
\node [below, thick ] at (10,-14) {$1.02 \: (0.10)$};

\node [left, thick ] at (-14.5,-10) {$-1.40 \: (0.24)$};
\node [above, thick ] at (0,-6) {$-0.85 \: (0.12)$};
\node [right, thick ] at (15.5,-4) {$-0.60 \: (0.25)$};
\node [right, thick ] at (15.5,-2) {$-0.29 \: (0.40)$};

\node [left, thick ] at (-14.5,3) {$0.08 \: (0.28)$};
\node [above, thick ] at (0,-0.5) {$0.01 \: (0.10)$};
\node [above, thick ] at (7,0.5) {$0.01 \: (0.22)$};
\node [right, thick ] at (15.5,5) {$0.09 \: (0.38)$};
\end{tikzpicture}
\caption{Estimates from cryptocurrency example when assuming a sequential model.}
\label{fig:sequential}
\end{figure}}

\begin{figure}[ht]
\centering
\rotatebox{90}{ 
\begin{forest}
for tree={
  l sep+=.05cm,
  s sep+=.05cm,
  shape=rectangle,
  rounded corners,
  draw,
  align=center,
  top color=white,
  bottom color=gray!20,
  font=\bfseries,
  content format={\strut\forestoption{content}}
}
[{$Y^{*\text{Familiarity}}\leq 0.36$}
  [{$Y^{*\text{Opinion}}\leq -0.27$}, edge label={node[midway,above left]{$\checkmark$}}
    [{$Y^{*\text{Familiarity}}\leq -0.80$}, edge label={node[midway,left]{$\checkmark$}}
      [{$Y^{*\text{Opinion}}\leq -1.76$}, edge label={node[midway,left]{$\checkmark$}}
        [{(-1, -1)}, edge label={node[midway,left]{$\checkmark$}}]
        [{(-1, 0)}, edge label={node[midway,right]{$\times$}}]
      ]
      [{$Y^{*\text{Opinion}}\leq -0.78$}, edge label={node[midway,right]{$\times$}}
        [{(0, -1)}, edge label={node[midway,left]{$\checkmark$}}]
        [{(0, 0)}, edge label={node[midway,right]{$\times$}}]
      ]
    ]
    [{$Y^{*\text{Familiarity}}\leq -0.24$}, edge label={node[midway,right]{$\times$}}
      [{(-1, 1)}, edge label={node[midway,left]{$\checkmark$}}]
      [{(0, 1)}, edge label={node[midway,right]{$\times$}}]
    ]
  ]
  [{$Y^{*\text{Opinion}}\leq 0.24$}, edge label={node[midway,above right]{$\times$}}
    [{$Y^{*\text{Familiarity}}\leq 0.73$}, edge label={node[midway,left]{$\checkmark$}}
      [{(1, -1)}, edge label={node[midway,left]{$\checkmark$}}]
      [{(2, -1)}, edge label={node[midway,right]{$\times$}}]
    ]
    [{$Y^{*\text{Familiarity}}\leq 1.14$}, edge label={node[midway,right]{$\times$}}
      [{$Y^{*\text{Opinion}}\leq 0.62$}, edge label={node[midway,left]{$\checkmark$}}
        [{(1, 0)}, edge label={node[midway,left]{$\checkmark$}}]
        [{(1, 1)}, edge label={node[midway,right]{$\times$}}]
      ]
      [{$Y^{*\text{Opinion}}\leq 0.74$}, edge label={node[midway,right]{$\times$}}
        [{(2, 0)}, edge label={node[midway,left]{$\checkmark$}}]
        [{(2, 1)}, edge label={node[midway,right]{$\times$}}]
      ]
    ]
  ]
]
\end{forest}
}
\caption{Binary decision tree describing the hierarchical decision process for general rectangular model}
\label{fig:correctedhierarchicaltree}
\end{figure}

\FloatBarrier
\subsection{Parental Investment in Children's Education}
\begin{footnotesize}
\begin{table}[ht]
\begin{threeparttable}
\caption{Estimation coefficients: parental investments in children} \label{tabREGpsid}
\centering
\begin{tabular}{lccccc}
\toprule 
\textbf{Variable} && \textbf{Non-lattice} && \textbf{Lattice} \\
\hline
\textit{Education Spending ($Y_1$)}\\
\hline
\textsc{Log family income}      && 0.21 (0.03)     && 0.24 (0.03)  \\
\textsc{Family size}            && $-$0.05 (0.02)  && $-$0.04 (0.02)  \\
\textsc{Child age}              && 0.04 (0.01)     && 0.02 (0.01) \\
\textsc{Child female}           && 0.08 (0.04)     && 0.12 (0.04)  \\
\textsc{Head white}             && 0.06 (0.05)     && 0.10 (0.05)  \\
\textsc{Head education yrs}     && 0.02 (0.01)     && 0.04 (0.01) \\
\textsc{Ever gifted}            && 0.03 (0.05)     && 0.10 (0.04) \\
\\
\textit{HOME Cognitive Stimulation ($Y_2$)}\\
\hline
\textsc{Log family income}      && 0.15 (0.03)     && 0.18 (0.03)  \\
\textsc{Family size}            && 0.05 (0.02)     && 0.04 (0.02)  \\
\textsc{Child age}              && $-$0.10 (0.01)  && $-$0.09 (0.01) \\
\textsc{Child female}           && 0.16 (0.04)     && 0.18 (0.04)  \\
\textsc{Head white}             && 0.23 (0.05)     && 0.25 (0.05)  \\
\textsc{Head education yrs}     && 0.11 (0.01)     && 0.11 (0.01) \\
\textsc{Ever gifted}            && 0.41 (0.05)     && 0.42 (0.05) \\
\textsc{Non-standard shift} && $-$0.10 (0.07) && $-$0.12 (0.06) \\
\bottomrule 
\end{tabular} \vspace*{-0.1cm}
\begin{tablenotes}[flushleft]
\footnotesize
\item Notes: The ``Non-lattice'' column provides estimates from the non-lattice bivariate ordered probit model. The ``Lattice'' column assumes a lattice structure, estimated jointly. Both specifications include region and wave fixed effects (not reported). Standard errors in parentheses. 
\end{tablenotes}
\end{threeparttable}
\end{table}
\end{footnotesize}

\end{document}